\documentclass[twocolumn]{aastex63}

\usepackage{amsmath}
\usepackage{graphicx}

\usepackage{savesym}
\savesymbol{tablenum}
\usepackage{siunitx}
\restoresymbol{SIX}{tablenum}

\usepackage{xspace}

\usepackage[normalem]{ulem}

\DeclareSIUnit\parsec{pc}
\newcommand{\msun}{\textrm{M}_{\odot}}

\newcommand{\flash}{{\tt FLASH}\xspace}
\newcommand{\enzo}{{\tt ENZO}\xspace}
\newcommand{\phfour}{{\tt ph4}\xspace}
\newcommand{\seba}{{\tt SeBa}\xspace}
\newcommand{\amuse}{{\tt AMUSE}\xspace}
\newcommand{\multiples}{{\tt multiples}\xspace}

\begin{document}


\title{Modelling of the Effects of Stellar Feedback during Star
  Cluster Formation Using a Hybrid Gas and N-Body Method}


\author[0000-0003-2128-1932]{Joshua E. Wall}
\affiliation{Drexel University, Department of Physics and Astronomy,
  Disque Hall, 32 S 32nd St., Philadelphia, PA 19104, USA}

\author[0000-0003-0064-4060]{Mordecai-Mark Mac Low}
\affiliation{Department of Astrophysics, American Museum of Natural
	History, 79th St at Central Park West, New York, NY 10024,
        USA}
\affiliation{Drexel University, Department of Physics and Astronomy,
	Disque Hall, 32 S 32nd St., Philadelphia, PA 19104, USA}
\affiliation{Center for Computational Astrophysics, Flatiron
	Institute, 162 Fifth Ave., New York, NY 10010, USA}
      \correspondingauthor{Mordecai-Mark Mac Low}
      \email{mordecai@amnh.org}

\author[0000-0001-9104-9675]{Stephen L. W. McMillan}
\affiliation{Drexel University, Department of Physics and Astronomy,
	Disque Hall, 32 S 32nd St., Philadelphia, PA 19104, USA}

\author[0000-0002-0560-3172]{Ralf S. Klessen}
\affiliation{Heidelberg University, Center for Astronomy, Institute for Theoretical
	Astrophysics, Albert-Ueberle-Str.\ 2, 69120 Heidelberg, Germany }
\affiliation{Heidelberg University, Interdisciplinary Center for Scientific Computing, INF 205, 69120, Heidelberg, Germany}

\author[0000-0001-5839-0302]{Simon Portegies Zwart}
\affiliation{Leiden Observatory, Leiden University, Niels Bohrweg 2, 2333 Leiden, Netherlands}

\author[0000-0001-9353-149X]{Andrew Pellegrino}
\affiliation{Drexel University, Department of Physics and Astronomy,
	Disque Hall, 32 S 32nd St., Philadelphia, PA 19104, USA}


\date{\today}

\begin{abstract}
  Understanding the formation of stellar clusters requires following
  the interplay between gas and newly formed stars accurately.  We
  therefore couple the magnetohydrodynamics code \flash to the N-body
  code \phfour and the stellar evolution code \seba using the Astrophysical Multipurpose Software Environment
  (\amuse) to model stellar dynamics, evolution, and collisional
  N-body dynamics and the formation of binary and higher-order
  multiple systems, while implementing stellar feedback in the form of
  radiation, stellar winds and supernovae in \flash. \added{We here
    describe the algorithms used for each of these processes.  We
    denote this integrated package Torch.} We
  \added{then} use this novel
  numerical method to simulate the formation and early evolution of
  \added{several examples of}
  open clusters of $\sim 1000$ stars formed from clouds with a mass
  range of \SIrange{e3}{e5}{\msun}. Analyzing the effects of stellar
  feedback on the gas and stars of the natal cluster\added{s}, we find
  that \replaced{our}{in these examples, the stellar}
  clusters are resilient to disruption, even in the presence of
  intense feedback. This can even slightly increase the amount of dense, Jeans
  unstable gas by sweeping up shells; thus, a stellar wind strong enough to
  trap its own H~{\sc ii} region shows modest triggering of star
  formation. Our clusters are born moderately mass segregated, an
  effect enhanced by feedback, and retained after the ejection of their
  natal gas, in agreement with observations.  
\end{abstract}

\section{Introduction} \label{intro2}

Star cluster formation is a nonlinear, attenuated, feedback problem:
initially cold and dense gas starts forming stars, and the more dense
gas is available, the more vigorous the star formation becomes
\citep{Mac_Low_Star_Formation}. However as the number of stars
increase, so do the chances of producing multiple OB stars that
can prevent further star formation by injection of kinetic energy, momentum, and
radiation in surrounding regions extending out to parsecs. This
feedback can not only make the gas Jeans stable, thereby halting star
formation, but can also eject the gas from the natal cluster
altogether. If the gas dominates the gravitational potential of the
cluster at the time of ejection, the cluster may be disrupted by the
gas expulsion entirely 
\citep{Baumgardt_cluster_disruption_gas_expulsion_2007}, perhaps
providing an explanation for the 90\% destruction rate of
all young clusters found by \citet{Lada_and_Lada_2003}. 

Stellar feedback in the context of entire clusters has been studied by
several authors. Early work included radiation
\citep{Peters_2010_starve_ind_frag, krumholz_orion_I_2011} and winds
\citep{krumholz_orion_II_2012}, as well as studying the effects of
clustered supernovae \citep{Joung_SN_driven_turb}.  In a seminal
series of papers Dale and coauthors used a smooth particle
hydrodynamics code \citep{Monaghan_SPH_1992} to study radiation
\citep{Dale_HII_disrupting_bound_MCs,Dale_HII_disrupting_unbound_MCs}
and winds \citep{Dale_winds_2013} independently as well as in
combination \citep{Dale_Winds_and_H2,Dale_early_evo_clusters} for
their effects on natal gas clouds. Subsequent studies have
further investigated feedback in the form of radiation
\explain{added references to \citet{vazquez-semadeni2017} and \citet{kim2018}}
\citep{rosen_unstable_2016,Peters_SILCC_4_2017,vazquez-semadeni2017,kim2018}, winds
\citep{Gatto_Walch_SILCC3_2016}, and supernovae
\citep{Kim_Ostriker_2015ApJ,simpson_kinetic_2015,ibanez-mejia_gravitational_2015,Girichidis_SILCC_2_2016}
with increasing accuracy. However none of these works has included at
the same time
ionizing radiation, winds, supernovae, stellar evolution, and
collisional N-body dynamics capable of following the formation and
dynamical evolution of wide binaries and
multiple systems.

Our goal is to predict the initial conditions of a newly born star
cluster: A cluster formed star by star from magnetized gas and
remaining gravitationally bound during the gas expulsion process
driven by stellar feedback from evolving massive stars.
The gravitationally bound state of young star clusters has been supported by
both observations
\citep{Tobin_sub_virial_ONC_2009,Karnath_gas_disperal_bound_2019} and
previous simulations of young clusters
\citep{Offner_sub_virial_clusters_2009}, although more recent studies
have suggested many young stellar groups may actually be supervirial
or unbound
\citep{Gouliermis_unbound_systems_2018,Kuhn_supervirial_2019}. We are
cautious with regard to observational claims of supervirial ratios,
though, as mass segregation has been shown to bias virial measures towards smaller group masses and more unbound configurations \citep{Fleck_varying_eta_2006}. 

\defcitealias{paper1}{Paper~I}
Our simulations bridge the gap between gas-dominated, star-formation
simulations and gas-free, N-body, star cluster simulations.  In a
previous paper \citep[][hereafter \citetalias{paper1}]{paper1}, we explained how
we use the \amuse framework \citep{AMUSE,AMUSE13,Mcmillan_Art_of_AMUSE} to couple
the adaptive mesh refinement magnetohydrodynamics (MHD) 
code \flash \citep{FLASH}, the N-body code \phfour \citep{ph4}, 
the stellar evolution code \seba \citep{Portegies_SeBa}, and a
treatment of tight multiple systems \citep[using \multiples][]{Mcmillan_Art_of_AMUSE}. In the current study
we focus on describing the numerical methods developed to implement
stellar feedback of the stars acting on gas, and their consequences.
The structure of this study is as follows: In Sect.~\ref{methods} we
describe our particular implementations of radiation, stellar winds
and supernovae. Further we discuss our modifications to \flash for far
ultraviolet and cosmic-ray background heating as well as our atomic,
molecular and dust cooling approach. In Sect.~\ref{fb_examples} we
describe four \added{proof-of-concept} simulations conducted with our
code\replaced{, and use them in}{; while in}
Sect.~\ref{fb_results} \added{we show how such models can be used} to
examine the effects of feedback on star 
formation, cluster structure, and the possibility of triggering star
formation through stellar feedback. Finally we close in
Sect.~\ref{fb_conclusions} with a summary.

The source code for our method, including our new and revised routines
for \flash and the bridge script to couple \flash and \amuse, are
available at \url{https://bitbucket.org/torch-sf/torch/}.
Documentation available is summarized at
\url{https://torch-sf.bitbucket.io/}. We invite community use of this method and
participation in its further development.

\section{Feedback Implementation} \label{methods}

In \citetalias{paper1} we described our star formation
method. \added{Because we do not resolve the full collapse process of
  individual stars, we must choose a subgrid approximation to sample
  the initial mass function.} \replaced{The method we choose is based on the use
of}{  In short,} sink
particles \citep{Bate_Sinks_1995,Krumholz_Sinks,Federrath_Sink_Particles} \added{that}
form in regions of dense, gravitationally bound gas.  As soon as a
sink forms, a list of stars is drawn by Poisson sampling from a
\citet{kroupa_IMF_2001} initial mass function (IMF) with a minimum and
maximum mass, using the same method as
\citet{Sormani_sinks_particles}.  As the sink accumulates enough mass
to form the next star on the list, that star is immediately formed with
position and velocity chosen based on distributions around the sink
values. In the initial models described here we used Gaussians with
width given by the local sound
speed and the sink radius. The sink then moves to the nearest local density maximum and continues
accreting.

\added{The choice of how to implement a subgrid model for star
  formation is a sensitive one. Two issues are of particular
  significance. First, different models for the choice of the initial
  mass function were examined by \citet{grudic2019} in models of
  cluster formation run at 1--2 orders of magnitude lower mass
  resolution than the proof-of-concept models presented here.  They
  found that implementing a stochastic model such as proposed by
  \citet{Sormani_sinks_particles} to determine the luminosity of sink
  particles representing multiple stars makes a significant difference
  compared to using an average over the initial mass function, even
  without following the dynamics of the single stars as our method
  does.
  Indeed our model even provides some responsiveness to the local
  environment, as the most massive stars will only form in regions
  where sinks can accrete substantial mass, although the connection
  remains stochastic.
  Second,
  our choice of the position of newly formed stars in phase space with
  respect to their parent sink particle represents a further
  approximation.  Although we have made a single choice in this paper,
  ongoing work in our group is studying how important this choice is
  to the structure of the resulting clusters, as we plan to describe
  in future publications.  Nevertheless, our star formation method remains an
  approximation that limits the accuracy of our model of cloud
  evolution.}

The minimum mass is chosen in the models presented to be the
hydrogen-burning limit of 0.08$\msun$. The maximum mass of a star is
correlated with the mass of the final cluster, a result found in
observations and parameterized by \citet{Weidner_IGIMF_2010}. We
preserve this correlation by choosing the maximum mass that a star can
obtain using their integrated galactic IMF. We
calculate the maximum mass of a star assuming a given star formation
efficiency $\epsilon_{\rm sfe}$ for conversion of gas into stars, taken to
be unity in our present work, from our initial clouds of mass
\SIrange{e3}{e5}{\msun}. This gives us a maximum mass for our
\SI{e3}{\msun} runs of $\sim \SI{30}{\msun}$ and a maximum mass for
the \SI{e5}{\msun} runs of $\sim \SI{110}{\msun}$.

\subsection{Radiation}
\label{RT}

\subsubsection{Photoionization}

For radiation transport we use the \flash module {\tt FERVENT}
\citep{baczynski_fervent:_2015}. This module follows the ray tracing
algorithm implemented in \enzo by \citet{wise_enzo+moray:_2011}. The
method creates rays from point sources along directions defined using
the HEALPIX \citep{HealPix} tiling of a sphere, then traces these rays
through the block-structured, adaptively refined grid.  The number of
rays from a specific source hitting each individual block (usually
$8^3$ or $16^3$ cells) is kept constant by splitting rays when
necessary. As each ray intersects a cell the number of photons is
reduced by absorption while the gas in the cell is ionized and
heated accordingly.

The {\tt FERVENT} package calculates the ionization fraction due to ultraviolet (UV) radiation \citep{baczynski_fervent:_2015}
\begin{eqnarray}
\frac{dx_{\rm H^+}}{dt} = C_{\rm cl} n_{\rm e} x_{\rm H^0} + k_{\rm ion} x_{\rm H^0} - \alpha_{\rm B} n_{\rm e} x_{\rm H^+}. \label{ioneq}
\end{eqnarray}
Here $x_n$ is the fraction of species $n$, $C_{\rm cl}(T)$ is the
collisional ionization rate, $\alpha_{\rm B}(T)$ is the case B
recombination coefficient, $n_{\rm H}$ is the number density of
neutral hydrogen,
$n_{\rm e}$ is the number density of electrons and
\begin{eqnarray}
k_{\rm ion} = \frac{N_{\gamma}}{n_{\rm H}(1-x)V \delta t}
\end{eqnarray}
is the rate of photon ionization, specifically formulated to be photon
conservative \citep{baczynski_fervent:_2015}. Here $N_{\gamma}$ is the
number of photons, $V$ is the volume of a cell, $x_{\rm H^+}$ is the hydrogen ionization
fraction, shortened hereafter to $x$, and $\delta t$ is an ionization time step. 

Equation~(\ref{ioneq}) was originally solved explicitly.  Since we
only follow hydrogen ionization, we have modified this method of {\tt FERVENT} to implicitly solve the ionization evolution, which allows for a solution with a much longer time step. First we rewrite Equation \ref{ioneq} with substitutions for the fractional ionization $x$ everywhere
\begin{eqnarray}
f(x) =\frac{dx}{dt} &=& C_{\rm cl} x n_{\rm H} - C_{\rm cl} x^2 n_{\rm H} + \\
& & + k_{\rm ion} -k_{\rm ion}x - \alpha_{\rm B} x^2 n_{\rm H}, \nonumber 
\end{eqnarray}
then approximate it with a forward finite difference equation
\begin{equation}
\frac{x_1-x_0}{\delta t} ={}k_{\rm ion} (C_{\rm cl} n_{\rm H} - k_{\rm
                             ion})x_1 - (C_{\rm cl} + \alpha_{\rm B}) n_{\rm H} x^2_1,  
\end{equation}
which is quadratic in the ionization $x_1$ at time $t+\delta t$,
leading to an algebraic solution for $x_1$.
%

The error in this method is given by the next term in the Taylor
expansion of the method,
%
\begin{equation}
x_1 - x_0 = f(x_1) \delta t + \frac{f'(x_1)}{2} \delta t^2,
\end{equation}
which can be used to derive a more accurate estimate of the time step than the original method:
\begin{equation}
\delta t_{\rm ion}= \frac{c}{C_{\rm cl} n_{\rm H}-2 n_{\rm H} x_1\left(C_{\rm cl}+\alpha_{\rm B}\right)},
\end{equation}
where $c$ is a tunable safety parameter that we usually set to
0.8. (Our solutions do not seem sensitive to the exact value of $c$.)
Since ionization depends strongly on the radiation field and the
temperature, and rapidly converges once these fields find steady states, integration of the ionization differential equation can capture the proper timescale for each of these events. Basing the time step on the rate of change of ionization allows us to do this, while the implicit solution guarantees stability. 

We have implemented subcycling on the ray tracing, ionization and
heating/cooling within a single MHD time step to allow the gas
dynamical time steps to be as large as possible. This results
in an overall speedup of \flash of about an order of magnitude
compared to calling the (expensive) gravity and MHD solvers during
the short ionization time steps.

\subsubsection{Radiation Sources}
To calculate stellar radiation fluxes (and winds) based on mass we
only consider stars with $M > $\SI{7}{\msun}. These stars are evolved
using the stellar evolution code \seba \citep[][with
updates from \citealt{2016ComAC...3....6T}, who included triple evolution]{Portegies_SeBa}, which is integrated into the \amuse
framework.  The luminosity $N_{\gamma}$ and average energy
$\nu_{\gamma}$ of ionizing photons from these stars is calculated from
their mass and age-dependent surface temperature interpolation of the
{\tt OSTAR2002} grid from \citep{lanz_grid_2003} if the star has a surface
temperature $T_* > $\SI{27.5e3}{K}; or else just an estimate from
integrating the blackbody emission $B(\lambda)$ as a function of
wavelength $\lambda$ for the star if $T_* < $\SI{27.5e3}{K}
\citep[e.g.][]{stahler2004formation}.
%
Then the cross section $\sigma$ for the photons is calculated from $\nu_{\gamma}$ \citep{osterbrock2006gaseousnebula}
\begin{equation}
\sigma_{\rm H} = \sigma_0\left(\frac{\nu_{\gamma}}{\nu_{\rm
      \SI{13.6}{eV}}} \right)^{-3},
\end{equation}
%
with $\sigma_0= \SI{6.304e-18}{ cm^{2}}$ \citep{draine2011physics}.


\subsubsection{Photoelectric Effect from Far Ultraviolet}

\label{subsub:PE}

As a second energy bin for radiation we also include far ultraviolet
(FUV) radiation (\SIrange{5.6}{13.6}{eV}) which is absorbed by dust,
ejecting photoelectrons in the process. Especially for lower mass
stars ($7 \le M/\msun \le 13$), the power in this radiation bin
approaches that of photoionizing radiation. Since the cross section of
photoelectric photons is smaller than those that ionize hydrogen, these rays
penetrate farther into the gas, heating it farther from the source
star than the ionizing radiation.

Although only about one in ten FUV photons ejects an
electron from the dust, all impart momentum to the gas that can be
important in clearing dense gas around newly formed massive
stars. Indeed, at our typical numerical resolution, this radiation pressure
is the main process acting to clear gas out of zones with
densities $n_{\rm H} \gtrsim \SI{e6}{cm^{-3}}$ surrounding early B and late O type
stars with weak stellar winds.  If this process is not implemented,
the ionizing radiation from the star is trapped in the
zone, producing an unphysical ultracompact H~{\sc ii} region.

We limit dust to gas with temperatures less than $T_{\rm sputter} = \SI{3e6}{K}$ to ensure that gas does not cool unphysically in regions
where dust would have previously been destroyed. This is an estimate
based on the temperatures at which dust would sputter within the
period that we model \citep[][eqs.\ 25.13 and 25.14]{draine2011physics}.


To compute the attenuation of radiation in the FUV with luminosity
$N_{\gamma}$, we follow the method of the original {\tt FERVENT} paper
\citep{baczynski_fervent:_2015}, calculating the optical depth
$\tau_{\rm d} = n_{\rm H} \Delta r \sigma_{\rm d}$ as a function of
the path length of the ray $\Delta r$ and the dust cross-section
$\sigma_{\rm  d} = 10^{-21}$~cm$^{-2}$~H$^{-1}$ \citep[e.g.][]{draine_dustyHII_2011}. The attenuated radiation $N_{\rm d} = N_{\gamma} \left(1
  - \exp \left(\tau_{\rm d} \right) \right)$, and the flux through a
cell is then
\begin{equation}
G = \frac{N_{\rm d} E_{\gamma}}{G_0 dx^2 \delta t} \label{Gflux},
\end{equation}
where the \citet{Habing_1968BAN....19..421H} flux $G_0 =
\SI{1.6e-3}{erg \ cm^{-2}}$.

We then add the momentum $E_{\gamma}/c$ from the photons absorbed by each cell
to the gas, while we follow
\citet{weingartner_photoelectric_2001} to calculate the heating from
the photoelectric electrons ejected into the gas as detailed in
Sect.~\ref{HandC}.


Finally, we also allow for EUV photons to be absorbed by dust. Since
the cross section for EUV photons is so much larger for hydrogen
\citep[$\sigma_{\rm H} \sim$
\SI{6e-18}{cm^2}]{osterbrock2006gaseousnebula} compared to that of
dust \citep[$\sigma_{\rm d} \sim$
\SI{1e-21}{cm^2}]{draine_2003ARA&A_dust_review}, we first compute how
many UV photons are absorbed by the gas, then any remaining photons
are subject to absorption by dust.  As a result, most dust absorption occurs
inside H~{\sc ii} regions where the gas is completely ionized,
leaving the only the dust optically thick to radiation.
   Although the dust will eventually be destroyed by sputtering, this
   occurs on timescales long compared to the expansion time of H~{\sc
     ii} regions \citep{arthur2004,draine_dustyHII_2011}.   
Similar to
previous studies \citep{draine_dustyHII_2011,kim_dustHII_2016} we find
that dust absorption leads to density gradients within our H~{\sc ii} regions. 

\subsection{Supernovae}
 \label{SN}
We include the possibility of explosions from Type II supernovae
from massive stars formed in our molecular clouds, as well as Type Ia supernovae from white dwarfs
in the field.
Supernovae were previously implemented in \flash by pure thermal
energy injection to study the large-scale driving of turbulence
\citep{Joung_SN_driven_turb}. However recently several authors have
demonstrated that more accurate results can be achieved using
injection of both kinetic and thermal energy in a mixture that depends
on numerical resolution
\citep{simpson_kinetic_2015,Kim_Ostriker_2015ApJ,Gatto_modelling_supernova_driven_2015}. \citet{simpson_kinetic_2015}
derived an analytic expression for the kinetic fraction $f_{\rm kin}$
based on how well a simulation resolves the pressure-driven snowplow
(PDS) as 
\begin{equation}
f_{\rm kin} = \SI{3.97e-6}{} \mu n_o R_{\rm PDS}^7 t_{\rm PDS}^{-2}
E_{51}^{-1}\Delta x^{-2},
\label{eq:fkin}
\end{equation}
where $\Delta x$ is the width of a grid cell, $\mu$ is the mean
molecular weight, $n_o$ is the background number density, $E_{51}$ is
the supernova energy in units of \SI{e51}{erg}, and $R_{\rm PDS}$ and
$t_{\rm PDS}$ are the radius and time of the supernova transition into
the PDS \citep{draine2011physics}.

\begin{figure}
	\centering
	\includegraphics[width=\linewidth]{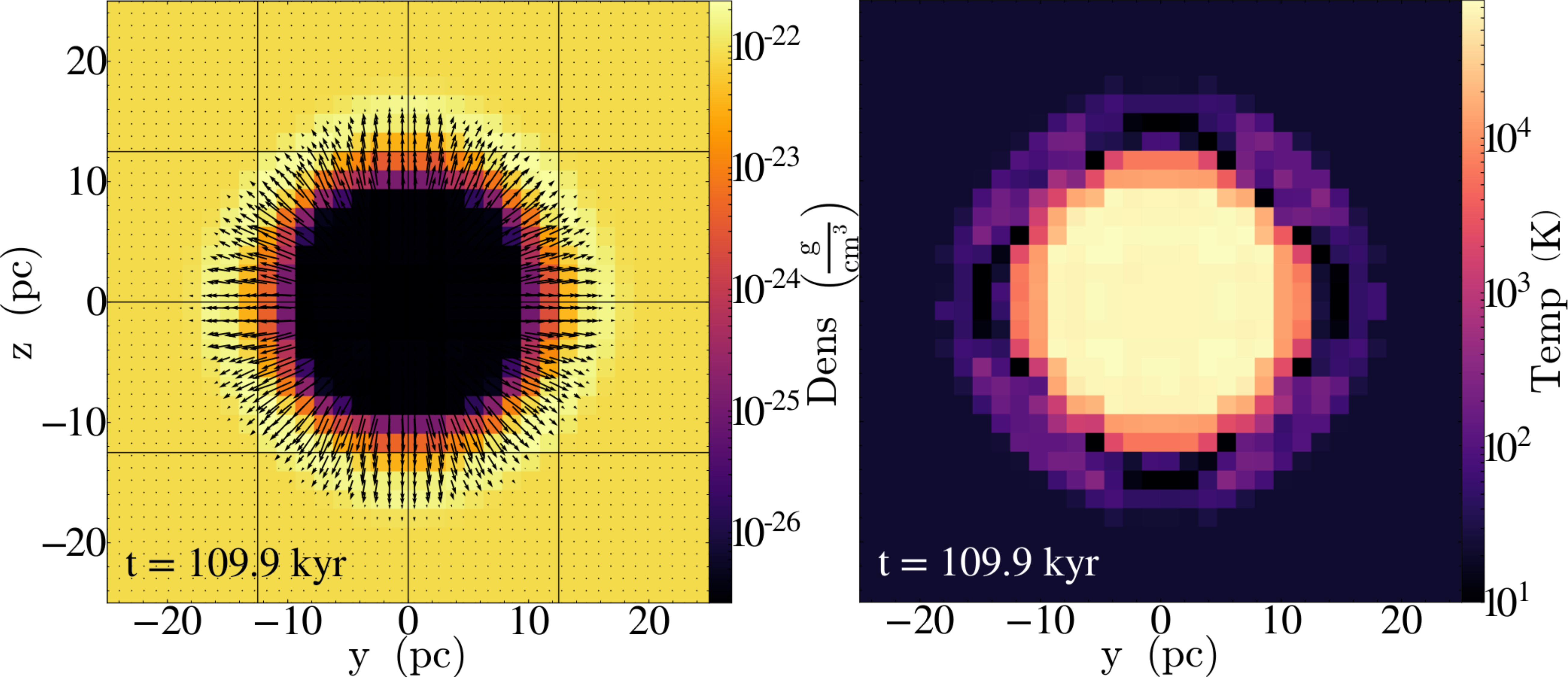} 
	\caption{Left: Density plot with overlaid velocity vectors for
          the supernova injection method of Simpson
          et. al. 2015. Right: Temperature plot. The kinetic fraction
          (Eq.~[\ref{eq:fkin}]) $f_{\rm kin} = 0.23$ and $\Delta x = 0.8$~pc.\label{fig:dens_vel0700}}
\end{figure}

We have implemented the method of \citet{simpson_kinetic_2015} for
supernova injection into our version of \flash: cloud-in-cell (CIC)
linear interpolation is used to map the energy input into the grid
from the supernova onto a $3^3$ cube centered at the supernova
location. Thermal energy and mass are equally divided among the 27
cells. Kinetic energy is also equally divided and injected in the form
of momentum into all but the center cell, where instead the kinetic
energy is converted into thermal energy and added to the thermal
energy already present. The contents of each zone in the cube are then mapped
onto grid zones they overlap
\citep[Fig. 1]{simpson_kinetic_2015}. Initial testing shows that even at low resolution, the supernova remnants are nearly spherical and exhibit the proper transitions from Sedov-Taylor to PDS \citep{draine2011physics}, as shown in Figure~\ref{fig:dens_vel0700}.


\begin{figure}
	\centering
	\includegraphics[width=0.95\linewidth]{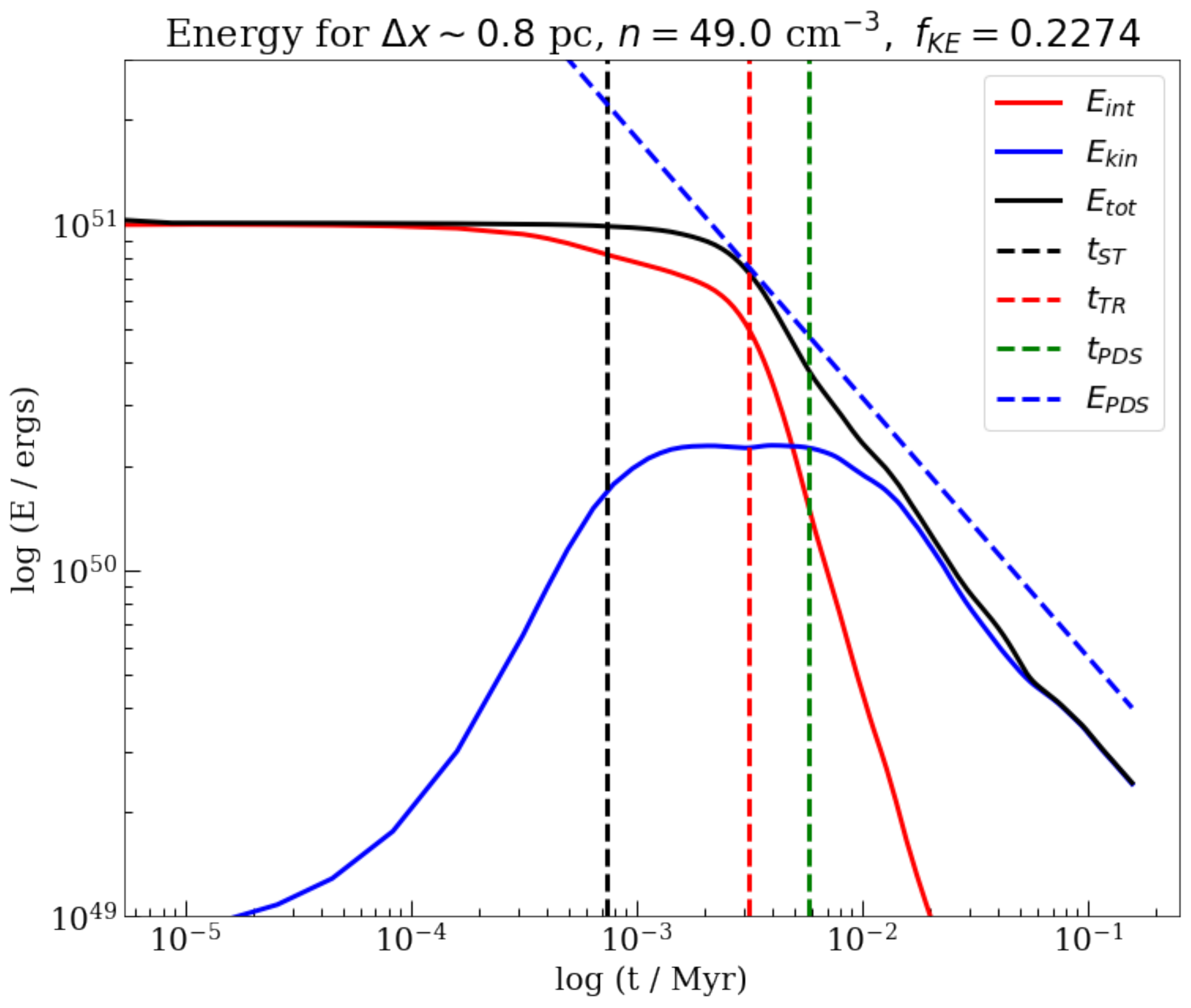}
	\caption{Energy of the supernova shown above. The Sedov-Taylor
          ($t_{\rm ST}$), thermal ($t_{\rm TR}$), and snow plow
          ($t_{\rm PDS}$) transition times \citep{draine2011physics} are shown as vertical dashed lines, and the analytic form of the energy during the pressure dominated snow plow phase is shown as a blue dashed line above the energy during this phase.\label{fig:SNenergy}}
\end{figure}

An energy plot for the same run is shown in
Fig. \ref{fig:SNenergy}. Here we note that the transitions between the
Sedov-Taylor solution \citep[when the kinetic energy fraction $\sim
0.25$]{draine2011physics}, transition
\citep{haid_supernova-blast_2016}, and PDS \citep{draine2011physics} all appear to match the analytic solutions very well. Also the slope of $E(t) \propto t^{-3/4}$ during the PDS is well recovered.

\subsection{Winds \label{winds}}


In recent years, the general strategy for injecting winds has been to inject
mass and velocity 
     in a region around the star to set the kinetic energy of the wind
     over the timestep
\citep{Pelupessy_embedded_SC,Gatto_Walch_SILCC3_2016, 2016ComAC...3....2R}. This is also
the method of \citet{simpson_kinetic_2015} for supernovae, who use
this energy to calculate the momentum for each cell. However,
adding momentum to a grid cell that already has mass does not
   add the same amount of kinetic energy as adding the momentum to an
   empty cell.
Therefore, after adding the momentum of the wind to
each cell in the source region, we compute the resulting kinetic
energy, and conserve total wind energy by injecting the missing energy as
thermal energy into the cell.

Stellar wind feedback is implemented using a method of momentum
injection of our own design which was inspired by inverting the method
of \citet{simpson_kinetic_2015}.
The amount of energy deposited by stellar winds onto the grid is given by the mechanical luminosity
\begin{eqnarray}
L_w = \frac{1}{2} \dot{M} v_w^2,
\end{eqnarray}
where $\dot{M}$ is the stellar mass loss rate, typically 
\SIrange{e-8}{e-6}{\msun~yr^{-1}}
for O and B stars, and $v_w$ is the wind terminal velocity, typically
\SIrange{3e2}{3e3}{km~s^{-1}} for the same stars.

We consider the update over time $\Delta t$ of an individual cell with volume $V_{\rm
  cell}$, density $\rho_{\rm old}$, specific internal energy $e_{\rm
  intold}$, and velocity $\mathbf{v}_{\rm old}$. We first calculate the overlap fraction $\phi$
between the spherical wind injection region and the cell itself using
a $20^3$ subgrid of sample points in each cell, following a routine by
D. Clarke included in ZEUS-MP \citep{hayes2006}. This value is
normalized to the full volume of the source region (i.e. $\sum_{\rm
  vol} \phi = 1$). The change in density of the cell
\begin{equation}
  \delta \rho = \phi \frac{\dot{M} \Delta t}{V_{\rm cell}}.
\end{equation}
The stellar wind input kinetic energy for this cell is
\begin{equation}
  \delta E_{\rm w}= \phi \frac{L_w \Delta t}{V_{\rm cell}}.
  \end{equation}
The final velocity of the cell can be computed from momentum
conservation to be
\begin{equation}
\mathbf{v} = \frac{\delta \rho \mathbf{v}_{\rm w} + \rho_{\rm old} \mathbf{v}_{\rm
old}}{\rho_{\rm old}+\delta \rho} < \mathbf{v}_{\rm w}.
\end{equation}
The final change in specific kinetic energy is then 
\begin{equation}
\delta e_{\rm kin} = \frac{\lvert \mathbf{v} \rvert^2}{2}  - \frac{
  \rho_{\rm old} \lvert \mathbf{v}_{\rm old} \rvert^2}{2 (\rho_{\rm old}
  + \delta \rho)},
\end{equation}
so the specific internal energy of the cell needs to be increased to
\begin{equation}
e_{\rm int} = \frac{\delta E_{\rm w}}{\rho} + e_{\rm intold} \frac{\rho_{\rm old}}{\rho} - \delta e_{\rm kin}.
\end{equation}

In determining the radius (e.g. the number of cells) across which to
inject the winds, both the physical radius of the wind and the ability
of the Cartesian grid to resolve the spherical input are
important. Here the analytic solution for a
stellar wind bubble is our guide.  The radius of
the wind termination shock \citep[][Eq.~{[12]}]{Weaver_1977} 
\begin{equation}
  R_1 = 
  0.74 \left(\frac{\dot{M}}{\rho_0}\right)^{\frac{3}{10}} v_w^{\frac{1}{10}} t_w^{\frac{2}{5}},
\end{equation}
where $t_w$ is the lifetime of the wind and $\rho_0$ is the background
density. Within this region the wind will be free streaming. If this radius
is resolved by more than a single cell, we directly inject the energy
and momentum calculated as above in the resolved region out to a
maximum radius of $6 \sqrt{3} \ \Delta x$, a radius at which we find
spherical winds to be well resolved.
If $R_1 < \Delta x$, we set the radius of the injection region to be
$\Delta x$.
Note that since the stars are Lagrangian particles not restricted to
the cell centers, even wind bubbles smaller than a single cell
generally inject not just thermal energy but also momentum and kinetic
energy into the grid by straddling cell boundaries. 
 
For each cell within this radius we determine the fractional overlap
of the cell with the wind injection region and add that fraction of
momentum and thermal energy into the cell, with the momentum and
energy evenly distributed throughout the sphere defined by the injection
radius. To guarantee that all cells are at maximum resolution in the
source region, we add a new criterion that enforces refinement of
all blocks that lie within the injection radius of the star. 

\begin{figure*}[htb]
	\includegraphics[width=\linewidth]{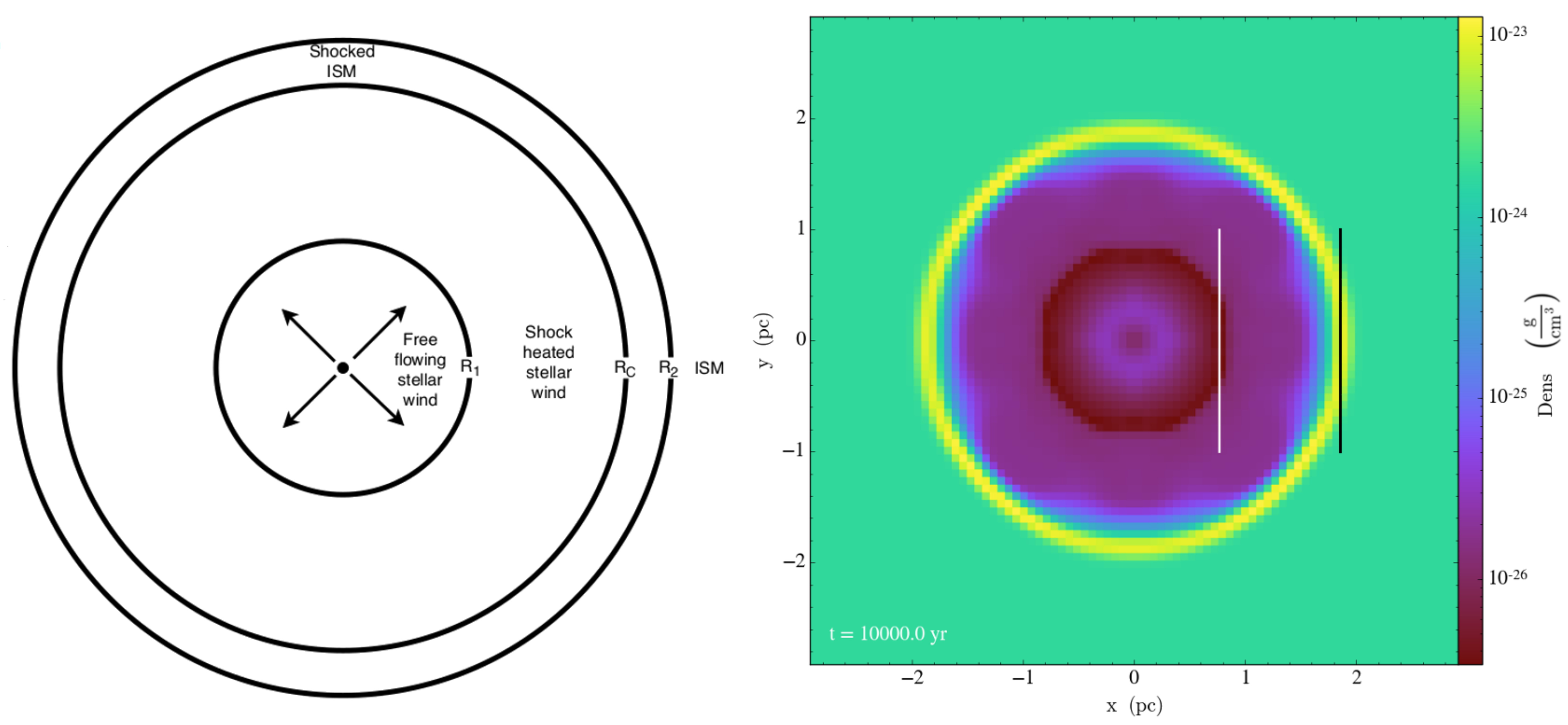}
	\caption{Comparison between Figure 1 of \citet{Weaver_1977} and our own
		test with $\dot{M}=$~\SI{e-6}{\msun~yr^{-1}} and
                $v=\SI{2e3}{km~s}^{-1}$, at time $t = 10^4$~yr. The inner white line
                indicates the analytic solution for the stellar wind
                termination shock radius $R_1$ and the outer black
                line the solution for the outer shock radius $R_2$ from \citet{Weaver_1977}. Note that the shell has cooled and collapsed at this point, with the distance between the contact discontinuity $R_c$ and $R_2$ now resolved by at most two cells.\label{fig:weavercomp}}
\end{figure*}

The mass of hot gas within real stellar wind bubbles is determined by
conductive evaporation \citep{Weaver_1977}, as well as turbulent mixing, across
the contact discontinuity at $R_{\rm C}$ (see Fig.~\ref{fig:weavercomp}) between the hot, rarefied, shocked stellar 
wind and the dense, radiatively cooled shell.  The extra density
injected by these mechanisms reduces the temperature in the hot region
between $R_1$ and $R_{\rm C}$, and thus the sound speed.  Capturing
this physics exactly is computationally challenging, as conductive
evaporation is a diffusive process with Courant timestep $\Delta
t_{\rm diff} \propto \Delta x^2$. However, the Courant timestep
$\Delta t_{\rm C} = C_{\rm CFL} \Delta x / \mbox{max}(v, c_s)$ is dominated by
the high sound speed $c_s$ in the hot region. Therefore, we introduce
the option to mass load the wind to bring its temperature to the
correct order of magnitude. We choose a mass-loaded temperature target
$T_{\rm ml} = $\SI{5e6}{K} in our simulations, set to lie at the low end of the quasi-stable hot gas
phase \citep{mckee1977}. We ensure this temperature
 by reducing the pre-shock velocity of the wind such that the
post-shock temperature \citep{draine2011physics} 
\begin{equation}
  T_s = \SI{1.38e7}{K}~\left(\frac{v_w }{\SI{e3}{km~s}^{-1}}\right)^2
  < T_{\rm ml}.
\end{equation}
We make up the lost energy by adding to the mass of the wind until we
recover the proper wind luminosity.  We have found this to be
sufficiently hot that the wind bubble in diffuse gas ($n \sim 1$~cm$^{-3}$) continues to conserve energy in
the shocked gas as expected, while
allowing significant gains in the size of the Courant time step. 

The combination of this radius with the division of kinetic and
thermal energy described above leads to bubbles in dense gas primarily
injected with thermal energy until the free streaming wind is resolved
(and with it the inner boundary of the hot wind bubble), at which
point the injection of energy shifts over towards kinetic. This method
shows excellent agreement with theoretical predictions for wind
bubbles given by \added{\citet{castor1975} and} \citet{Weaver_1977}. Comparison between their
analytic solution\deleted{s} and a plot from our test runs is shown in
Figure~\ref{fig:weavercomp}.

To calculate the mass loss rates for our stars we follow \citet{vink_2000_A&A} while for the velocities we use the
fitting formula of \citet{kudritzki_puls_2000_ARAA}. \replaced{Note that these}{These}
methods include the bi-stability jump in wind strength in late O and early B stars
produced by the higher absorption cross section of Fe~{\sc iii}
compared to Fe~{\sc iv} \citep{vink_2000_A&A}.  \added{However they do
  not include any correction for wind clumping, which observations
  suggest may produce a factor of three reduction in mass loss rates $\dot{M}$
  \citep{smith2014}.  Thus the effect of winds in the models shown
  here are likely an upper limit, although the effect is not great
  given that the radius of a stellar wind bubble depends on
  $\dot{M}^{1/5}$ \citep{castor1975}. A clumping correction should be
  considered in future work.}

\subsection{Heating and Cooling \label{HandC}}

\subsubsection{Ultraviolet Radiation Heating}

Heating from our stellar sources in the EUV and FUV is calculated by
converting the photons to energy fluxes, then applying those fluxes
weighted by the probability of an electron of a given energy actually being ejected from
an ion or dust grain when it absorbs a photon. For hydrogen
ionization this is done by simply differencing the energy of the
photon and the ionization potential of hydrogen.

For the background FUV we assume a constant flux of $1.7 G_0$ and
estimate the local visual extinction
\begin{equation}
  A_v \sim \frac{\lambda_J n_{\rm H}}{N_{\rm H}},
\end{equation}
where $N_{\rm H} = \SI{1.87e21}{cm^2}$ and we take the length scale to
be the local Jeans length \citep{Jeans_1902}
\begin{equation}
\lambda_J = \left[\pi c_s^2/ (G \rho)\right]^{1/2},
\end{equation}
following the methods of \citet{Seifried_2011MNRAS.417.1054S} and
\citet{Walch_SILLC1}. The
fraction of FUV radiation that can heat the gas is then $f_{\rm ext} =
\exp(-3.5 A_V)$.  

For the FUV flux we normalize to the Habing flux $G_0$ following
Equation~(\ref{Gflux}) to find $G$, and then calculate the heating function per
unit volume
\begin{equation}
  n_{\rm H} \Gamma_{\rm pe} =  n_{\rm H} \epsilon G
  \end{equation}
where $\epsilon$ is a heating efficiency function. We have implemented several
different approximations to this function, including detailed fits from
\citet{weingartner_photoelectric_2001} and \citet{wolfire_2003}, as
well as a simple adjustable parameter following
\citet{Joung_SN_driven_turb}. 

The \citet{weingartner_photoelectric_2001}
efficiency function is given by
\begin{equation} \label{WD01}
  \begin{aligned}
  & \frac{\epsilon}{10^{-26} \,\si{ erg \ s^{-1}}} = \\
   & = \frac{7.64 + 4.52T^{0.132}}{1 + \num{4.37e-2}
                G_{\rm f}^{0.452}   \left(1 + \num{5.57e-3}  G_{\rm
                    f}^{0.675} \right)}.
   \end{aligned}
\end{equation}
The coefficients are taken from their Table 2, using the case with a ratio of
visual extinction to reddening $R_V=3.1$, carbon abundance with
respect to hydrogen of
$b_c=6 \times 10^{-5}$, distribution A, which minimizes the amount of C and Si in
grains, and the stellar radiation field of a B0 star, corresponding to a blackbody
with $3\times 10^4K$ up to 13.6~eV.  The flux factor
\begin{equation}
G_{\rm f} =  \frac{G \sqrt{T}}{n_{\rm e}}.
\end{equation}
The \citet{wolfire_2003} function is given by 
\begin{equation}
  \begin{aligned}
  &\frac{\epsilon}{1.3 \times 10^{-24} \,\si{ erg \ s^{-1}}} = \\
&  =\frac{4.9 \times 10^{-2}}{1+4.0 \times 10^{-3}\left(G_{\rm f} /
    \phi_{\mathrm{PAH}}\right)^{0.73}} + \\
&+\frac{3.7 \times 10^{-2}\left(T / 10^{4}\right)^{0.7}}{1+2.0 \times
  10^{-4}\left(G_{\rm f} / \phi_{\mathrm{PAH}}\right)},
\end{aligned}
\end{equation}
with $\phi_{\mathrm{PAH}} = 0.5$ following their assumption. Finally,
the simplest assumption is to just follow \citet{Joung_SN_driven_turb}
and set a constant value of $\epsilon = 6.5 \times 10^{-26} \,\si{ erg \
  s^{-1}}$.  In the example models analyzed in
\citetalias{paper1} and here, we use the \citet{weingartner_photoelectric_2001}
approximation (Eq.~[\ref{WD01}]).

\subsubsection{Dust Temperature}
Any photons absorbed that do
not eject electrons contribute directly to heating the dust.
The dust density is assumed to be a constant 0.01 fraction of the gas
density \citep{draine2011physics}. When solving for the dust
temperature we use the radiative cooling rate from
\citet{goldsmith_molecular_2001}, assuming the dust is always
optically thin at our densities $n_{\rm H} < \SI{e6}{cm^{-3}}$, while applying photoelectric heating as previously described. To calculate the dust temperature we use Newton's root finding algorithm as in \citet{Seifried_2011MNRAS.417.1054S}, which generally converges in less than ten steps.

\subsubsection{Cosmic Ray Heating}
Cosmic ray heating is applied with an ionization rate of $\zeta =
\SI{e-17}{s^{-1}}$ and a heating rate of $\Gamma_{\rm cr} / n_{\rm H}
= (\SI{20}{eV}) \zeta = \SI{3e-27}{erg \ s^{-1}}$ as appropriate for
the dense regions we are attempting to simulate
\citep{galli_cosmic-ray_2015}.

\subsubsection{Gas Cooling}
For gas cooling we include contributions from atomic and molecular
species as well as dust grains.
For the atomic contribution we use the
piecewise power law in \citet[][Fig. 1]{Joung_SN_driven_turb}, itself
derived from equilibrium ionization values given by
\citet{sutherland_cooling_1993}. For molecular cooling we 
use tabulated values from \cite{neufeld_thermal_1995} which were
originally implemented in \flash by
\citet{Seifried_2011MNRAS.417.1054S}, while for dust we use
the method of \citet{goldsmith_molecular_2001} with the cooling
equation for dust from \citet{hollenbach_molecule_1989}. Note that
dust cooling is also limited to temperatures $T < T_{\rm sputter}$
(see Sect.~\ref{subsub:PE}).

\subsubsection{Numerical Solution}
To solve the implicit difference equation for the temperature of the
gas under all of these heating and cooling sources we have implemented
Brent's \citeyear{brent1973algorithms} method, which we find to be more
accurate and stable than the Euler method used by
\citet{Joung_SN_driven_turb} and \citet{baczynski_fervent:_2015}.
In each cell, all heating
$\Gamma_i(\epsilon)$ and cooling $\Lambda_j(\epsilon)$ rates are
combined to find the rate of change of specific internal energy
\begin{equation}
  \frac{de}{dt} = \Gamma_i(\epsilon)n - \Lambda_j(\epsilon)n^2.
\end{equation}
The cooling rate at the minimum allowed temperature
in the simulation (generally \SI{10}{K}, but could be as low as the
CMB background temperature) is calculated and subtracted from the
total cooling rate to set a temperature floor. Then,
the difference equation
\begin{equation}
  e^{i+1} - e^{i} - \Delta t \frac{de}{dt} =0
  \end{equation}
is solved for $e^{i+1}$ by the Brent method.
  
\begin{figure}
	\centering
	\includegraphics[width=0.8\linewidth]{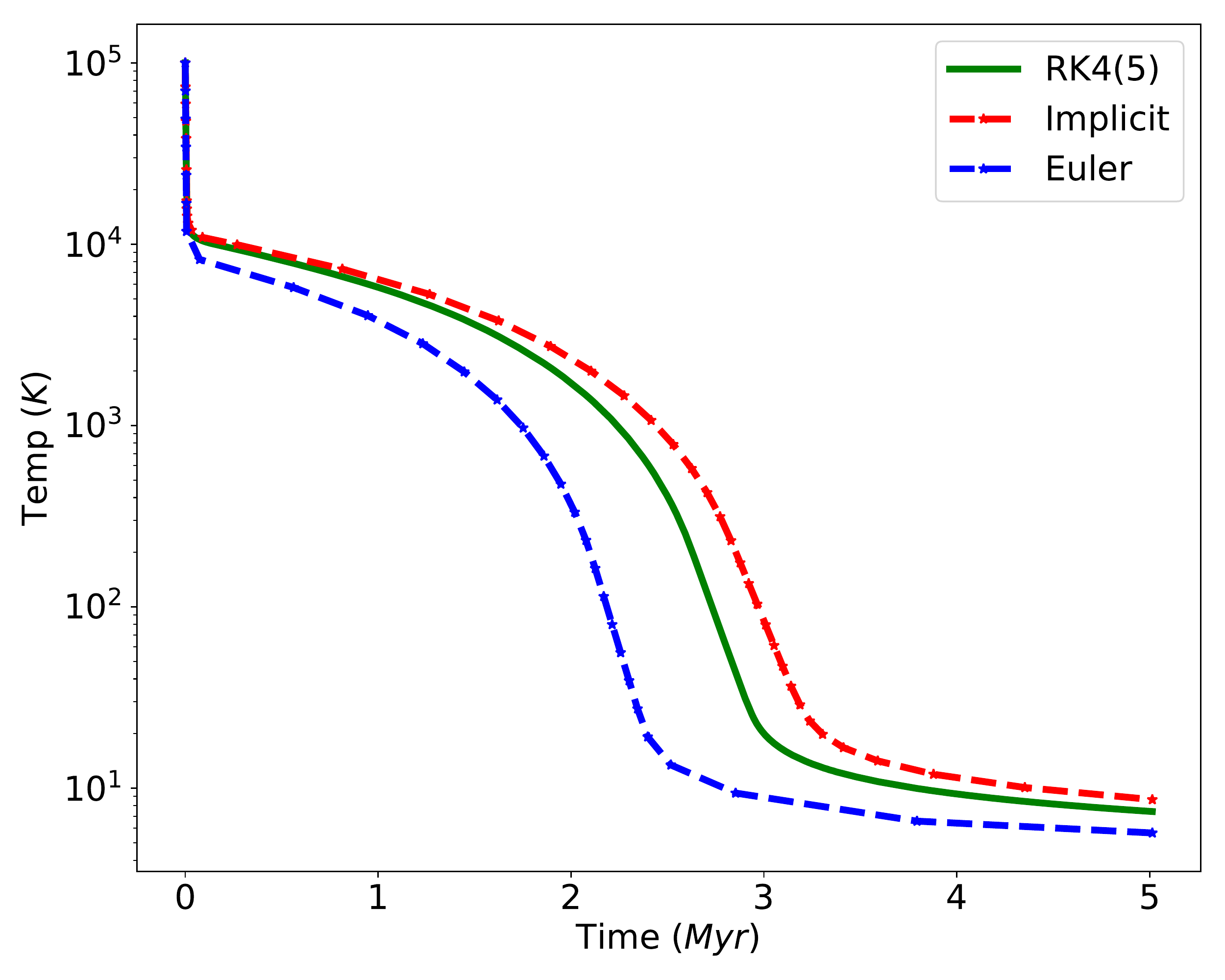}
	\caption{Temperature of a single gas cell with $n_{\rm H} = \SI{e4}{cm^{-3}}$ initially at \SI{e5}{K} as it evolves over \SI{5}{Myr} in the presence of a background UV field of $1.69 \ G_0$ integrated with our three different methods. \label{fig:cooling_one_cell}}
\end{figure}

Figure~\ref{fig:cooling_one_cell} shows the temperature for a cell
initially at $T=$ \SI{e5}{K} cooling over \SI{10}{Myr} using the
original Euler method of \citet{Joung_SN_driven_turb} and our implicit
method, as well as a more expensive Runge-Kutte 4(5) method with local
extrapolation \citep{press2007numerical}. For a given time step criterion, the implicit method is generally about twice as accurate as the Eulerian method and $\sim 30 \%$ faster than the Runge-Kutte method. Given that we call this solver on every iteration of the ray tracing method as we converge to an ionization solution, we have chosen to use the implicit method due to its combination of speed and accuracy.

\subsubsection{Tests}
In order for our simulations to resemble the real interstellar medium (ISM) as closely as
possible, our heating and cooling solutions should be able to
replicate the three-phase ISM
\citep{McKee_Ostriker_3phaseISM,Cox_3phase_revisited}, where we have
equal pressure solutions for a quasi-static hot medium and warm and
cold media. Being able to maintain the different thermal phases of the ISM in
pressure equilibrium is important generally.  It is particularly
important for the initial
conditions chosen for our proof-of-concept models, since the background
medium for 
some of our clouds is  warm
neutral medium, while the clouds themselves always consist of cold
neutral medium in pressure equilibrium with the background at the
initial time.

\begin{figure}
	\centering
	\includegraphics[width=0.85\linewidth]{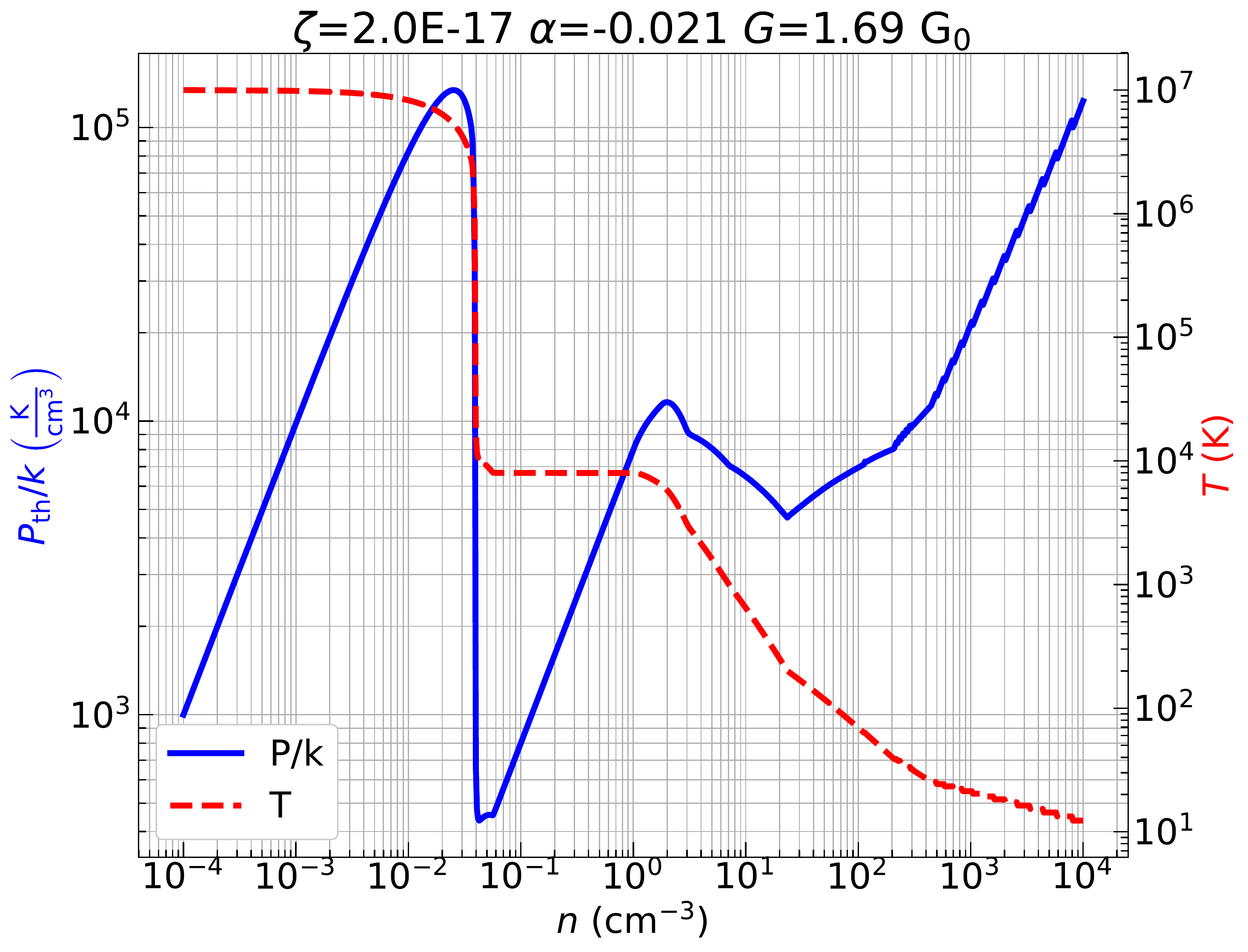}
	\caption{Temperature (blue, solid line) and pressure (red,
          dashed line) of a single gas cell with $10^{-4} \le n_{\rm
            H} \le 10^4 \,\si{cm^{-3}}$ evolved until equilibrium in the
          presence of a background UV flux of $G=1.69 \ G_0$ and
          cosmic ray background ionization rate of $\zeta =
          \SI{2e-17}{s^{-1}}$ integrated using the implicit
          method. The two-phase medium occurs for densities of roughly $10^{-1} \le
          n_{\rm H} < 10^3$~cm$^{-3}$, and a quasi-static third phase
          at high temperature and low density is then consistent with
          the pressure. \label{fig:cooling}}
\end{figure}

To test our heating and cooling methods we therefore created a single cell
simulation that iterates over hydrogen number densities with $10^{-4}
\le n_{\rm H} \le \SI{e4}{cm^{-3}}$, solving for the equilibrium
temperature and pressure with Milky Way-like background FUV and cosmic
ray values. The results are shown in Figure~\ref{fig:cooling}, where the
three-phase medium is shown to be stable in our simulations from $P
\sim$ \SIrange{4e3}{2e4}{K~cm^{-3}}, reproducing similar ranges found
in \citet[][Fig. 7]{wolfire_2003} for the solar neighborhood. Note
that we can adjust this range by increasing or decreasing our
background FUV, as discussed in \citet{Hill_three_phase_ISM_2018}, which also uses the atomic cooling model our method is based on.

\section{Example Runs}\label{fb_examples}

\begin{deluxetable*}{c c c c c c c c | c c c c}
	\tablecaption{Parameters for each of the four runs described
		here including cloud mass $M$, radius $R$, and central density $\rho_c$, in
                units of $\rho' =2.39 \times 10^{-23} \mbox{ g
                  cm}^{-3}$, normalization of the
                velocity perturbation spectrum $v_0$, the number of
                refinement levels $N_{\rm ref}$, cell size $\Delta x$
                at maximum refinement, and the
                domain size $D$.  We further give the 
 total number of stars $N_s$ \added{and total stellar mass $M_s$} \replaced{at end of
		run at time $t_{\rm end}$ when analysis was
                performed}{at the final time $t_{\rm end}$},
                as well as the time the first star formed $t_{\rm sf}$.
				Note that M3 and M3f used different random turbulent
		patterns initially, explaining their different values of $t_{\rm sf}$.
		\label{tab:table}}
	\tablehead{
		\colhead{Run\tablenotemark{a}} 
		& \colhead{$M$ ($\msun$)}
                & \colhead{$R$ (pc)}
                & \colhead{$\rho_c / \rho'$}
                & \colhead{$v_0$ (km~s$^{-1}$)}
                & \colhead{$N_{\rm ref}$}
                & \colhead{$\Delta x$ (pc)}
		& \colhead{$D$ (pc)}
		& \colhead{$N_{\rm s}$}
                & \colhead{$M_s$ (M$_\odot$)}
		& \colhead{$t_{\rm sf}$ (Myr)} 
		& \colhead{$t_{\rm end}$ (Myr)}
              }
	\startdata
	M3  & $10^3$  & 3  & 46 & 0.616 & 8 & 0.01 &  10 & 1100 & 514 & 2.86  & 4.38 \\ 
	M3f & $10^3$  & 3  & 46 & 0.616 & 7 & 0.02 &  10 & 1062 & 338 & 2.31  & 3.90 \\ 
	M3f2& $10^3$ & 5  & 10 & 0.616 & 7 & 0.01 &  14 & 52    & 43.5 & 5.22  & 5.60 \\ 
	M5f  & $10^5$ & 50 &1.0 & 1.58  & 8 & 0.2  & 110 & 1144 & 501 & 15.4  & 17.8
	\enddata
	\tablenotetext{a}{Runs ending in ``f'' include feedback due to
          radiation and stellar winds.}
\end{deluxetable*}

For testing \added{our models of} stellar feedback we present four proof-of-concept runs, three of which
include radiation, winds and supernova. However, we terminated these
runs for cost reasons before any massive star had exploded
as a supernova, so we restrict our discussion to radiation and
winds. In Table~\ref{tab:table} we show the initial cloud properties,
grid resolution and time of initial star formation for these runs. 

All four simulations use an initial density
field that is spherically symmetric and 
distributed radially as a Gaussian
\citep{Bate_Sinks_1995,goodwin-whitworth-W-T} with full width at half
maximum \replaced{of}{at} the cloud radius $R$.
\added{The central density was chosen in each case to be the characteristic
  density of regions at the mass scale chosen, following Table 3.1 of
  \citet{stahler2004formation}.  Because the length scales given were
  used as the radii of the initial clouds, the resulting surface densities of around
  10~$M_\odot$~pc$^{-2}$ are
  characteristic of Jeans unstable regions of the atomic ISM rather
  than the roughly 50~$M_\odot$~pc$^{-2}$ more typical of already formed
  molecular clouds that have undergone substantial collapse prior to
  molecule formation.}
The velocity field is initialized with a turbulent Kolmogorov
velocity spectrum $v(k) = v_0 k^{-5/3}$ from wavenumber $k = 2$ to
$k = 32$ \citep{WunschCloud} for the dense gas, where $k = 2 \pi / D$
for a simulation domain with side $D$.
We note that
choosing the initial conditions does determine a good deal about the
subsequent evolution \citep{goodwin-whitworth-W-T,Girichidis_ICs_and_statistics_1_inital_cloud_profiles}. The surrounding medium
is initialized with zero velocity and in pressure equilibrium with the cloud gas.  
Refinement and derefinement are controlled by the \citet{Jeans_1902} criterion as described in
\citet{Federrath_Sink_Particles}. We now describe individual
characteristics of these runs.

\begin{figure*}[h]
	\begin{tabular}{cc}
		\includegraphics[width=0.5 \textwidth]{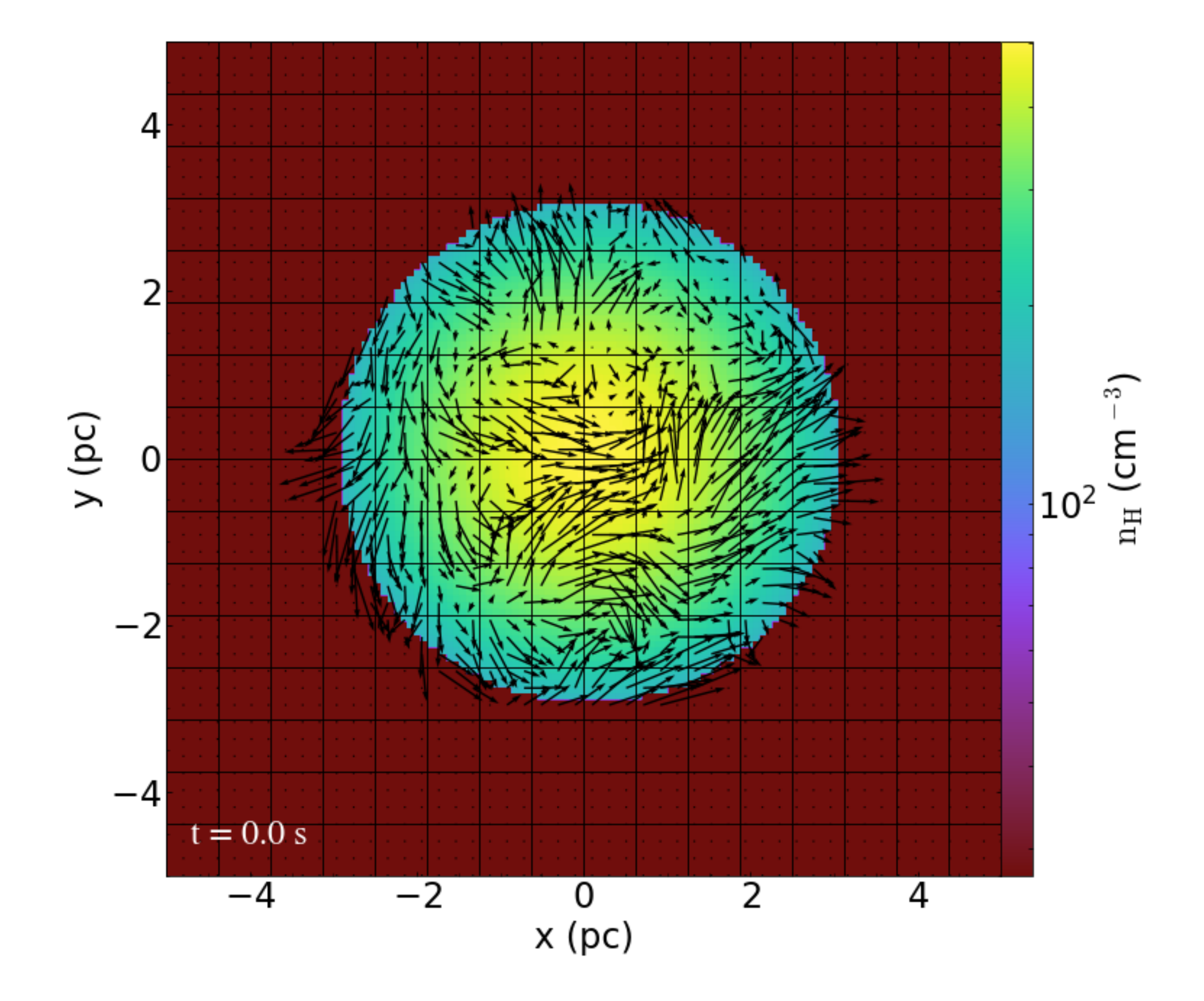}
          &
		\includegraphics[width=0.5 \textwidth]{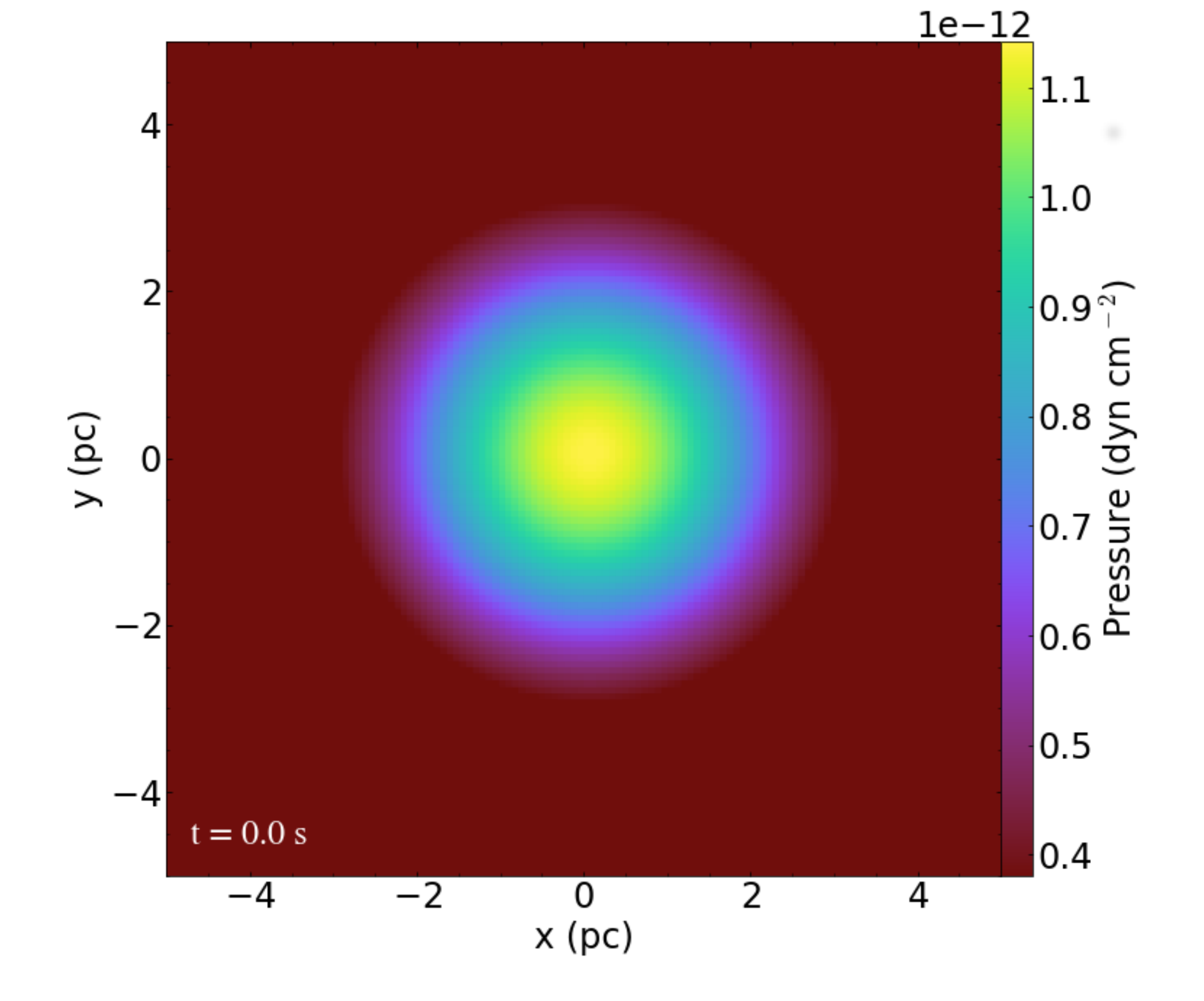}
          \\
		(a) & (b) \\[6pt]
	\end{tabular}
	\caption{Plots showing the initial conditions in (a) H nuclei density and (b) pressure for run M3f. Velocity vectors and initial grid blocks, which each contain $16^3$ cells, are annotated on the density plot. \label{fig:example_run}}
\end{figure*}




\subsection{M3}

In this \added{control} run, which \replaced{does not
  include}{includes no} feedback, two subclusters form
that subsequently merge. We highlight the merger event in the relevant
plots throughout by including a grey shaded box on each plot covering
the time of the merger event. This is \replaced{our}{the} only run among our
proof-of-concept models that features \added{such} a merger.

\subsection{M3f}\label{Sect:M3f}

In this run\added{, including feedback,} several stars form in the main cluster that are massive
enough to have ionizing and wind feedback. 
The growth of an H~{\sc ii} region is initially suppressed by dense gas accretion,
creating flickering H~{\sc ii} regions
\citep{peters2010,peters2010a,De_Pree_flickering_HII}. This continues
until the number of massive stars gets large enough for their combined wind and ionization feedback to clear an expanding H~{\sc ii} region
around the main cluster, near the end of the run. Around this same
time a second cluster forms in the simulation. While the two
clusters appear to be falling toward each other, they have yet to
merge at the end of the run.

\subsection{M3f2}

This run starts with the surrounding lower density gas in the warm,
neutral phase at \SI{4e3}{K}, as opposed to the cold phase at roughly
\SI{60}{K} in the other two \SI{e3}{\msun} runs. Therefore, the
\replaced{surrounding gas is less dense}{surface density is lower}
than \added{in} those runs, resulting in slower
accretion onto the star forming region. Similarly, feedback is more
effective in expelling the gas from the star forming region, since the
feedback sees a smaller surface density above it
\citep{Grudic_Hopkins_feedback_fails_surf_dens_to_force_2018}. The
more effective feedback rapidly shuts down star formation in this run,
which therefore only produces 52 stars.

\subsection{M5f}

In our final run we start with an initial sphere of \SI{e5}{\msun} and
a radius of \SI{50}{pc}. The gas outside the sphere is initially in
the warm ionized phase at $\sim \SI{8e3}{K}$. Once the gas collapses
and begins to form stars, a large central cluster appears, as well as
a secondary, smaller cluster. The central cluster rapidly grows until
it stochastically forms a \SI{97}{\msun} star, the most massive star
formed to date in any simulation we have run.
\added{Although the statistical chance of any individual cluster
  having a star of this size form is modest, its appearance does allow
  exercise of the full dynamical range of the feedback mechanisms that
  we have implemented, so we consider this a valuable example of the
  behavior of strong feedback.}
  
\begin{figure*}
  	\gridline{\fig{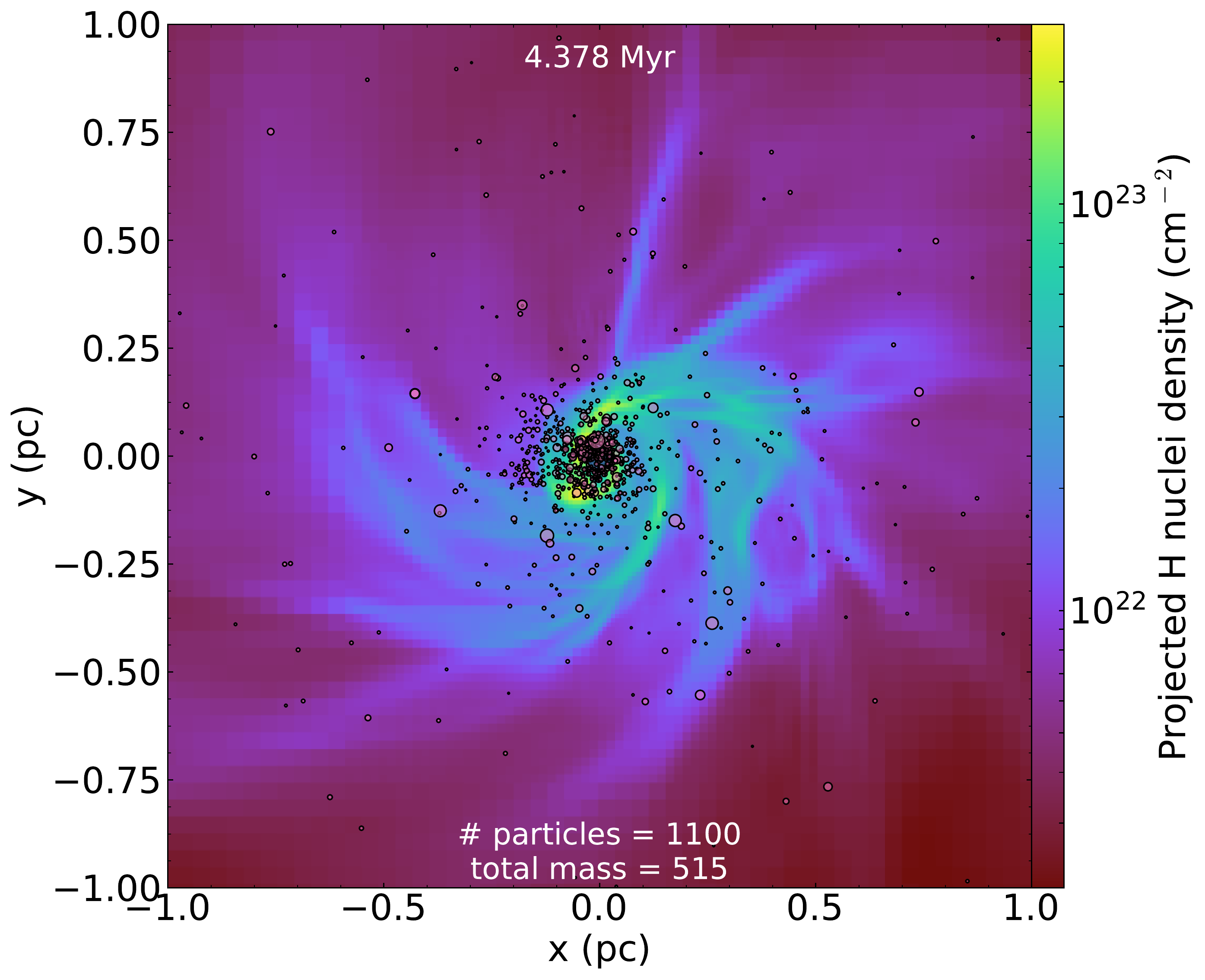}{0.5\textwidth}{(a)}
		       \fig{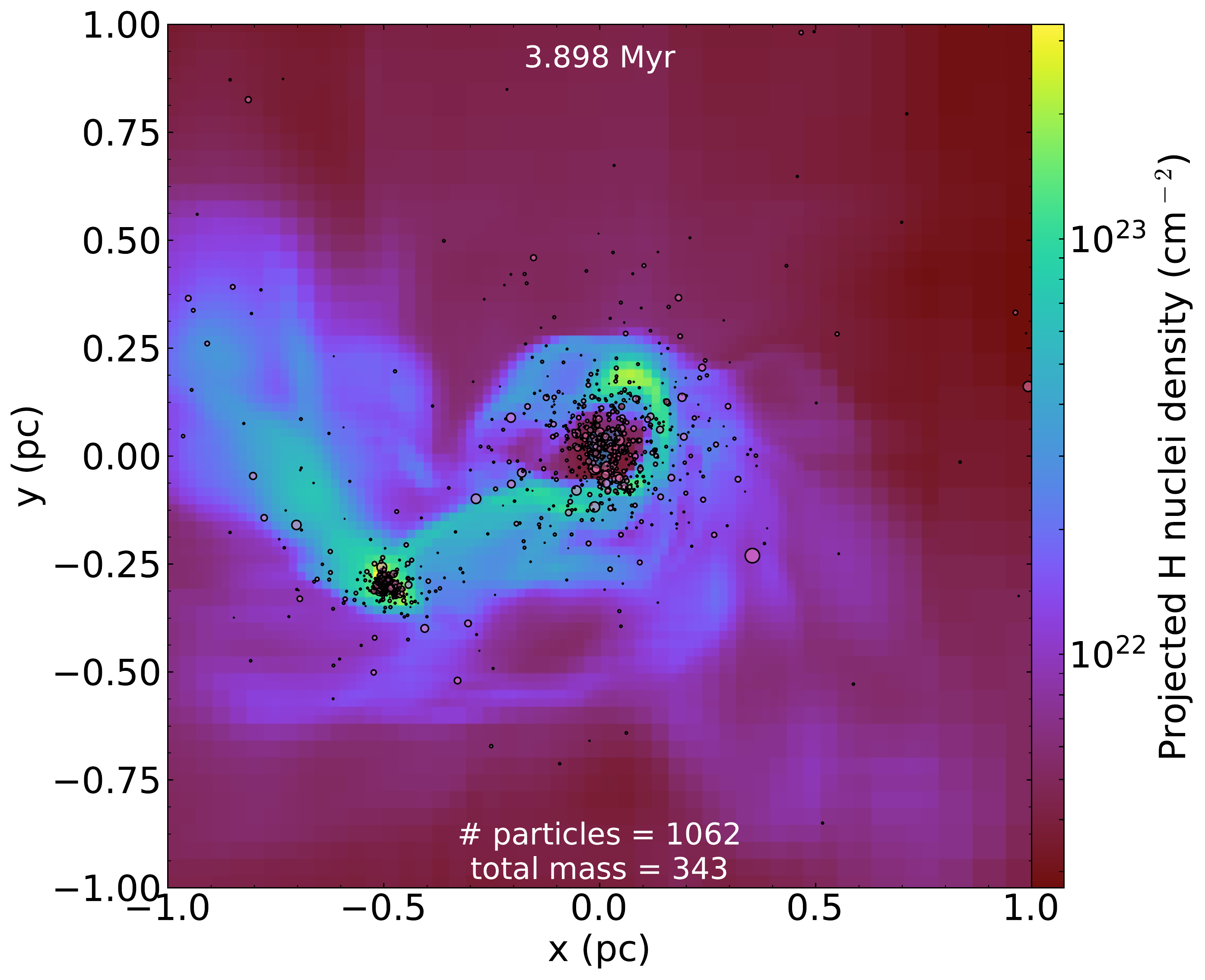}{0.5\textwidth}{(b)}
	}
	\gridline{\fig{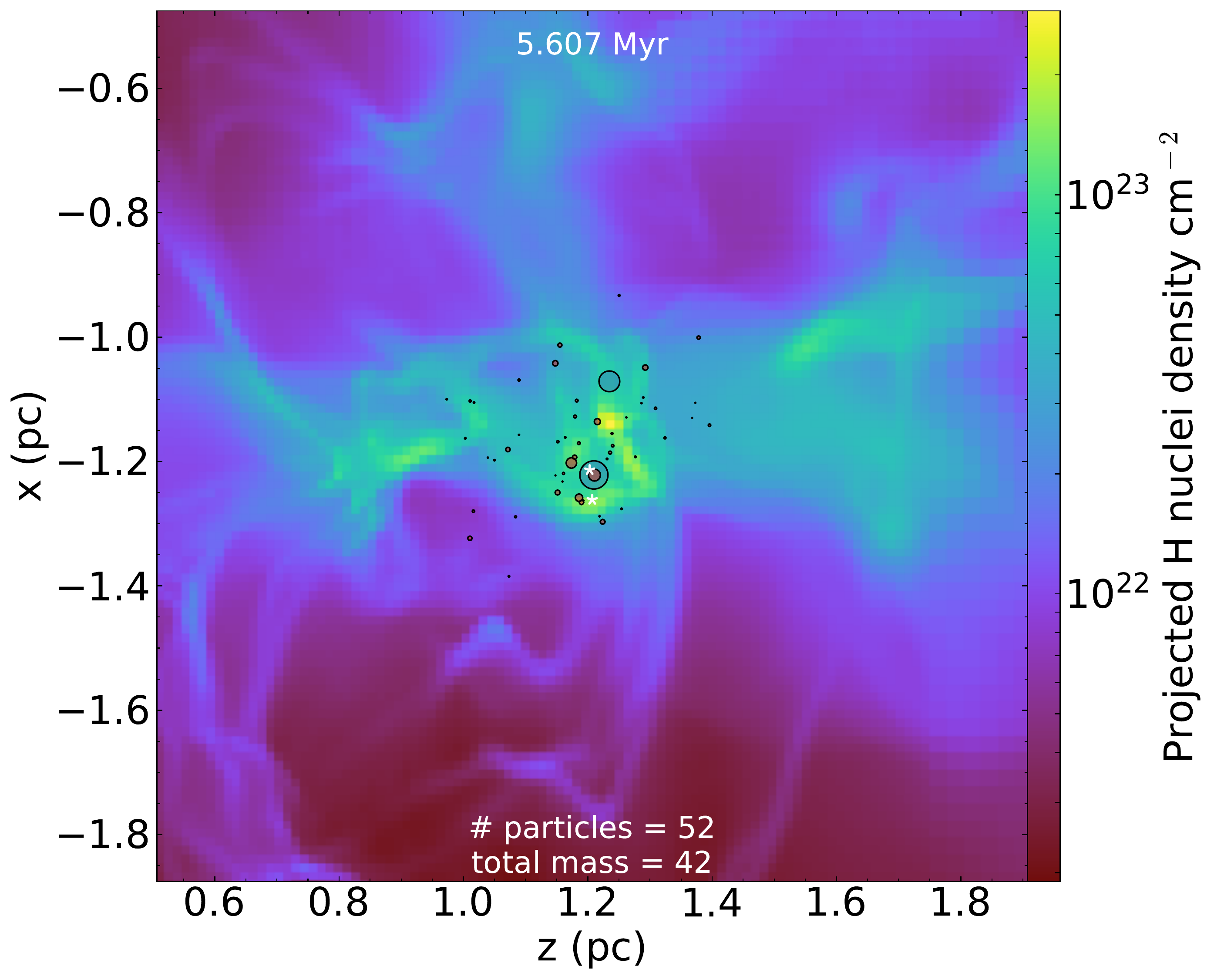}{0.5\textwidth}{(c)}
		       \fig{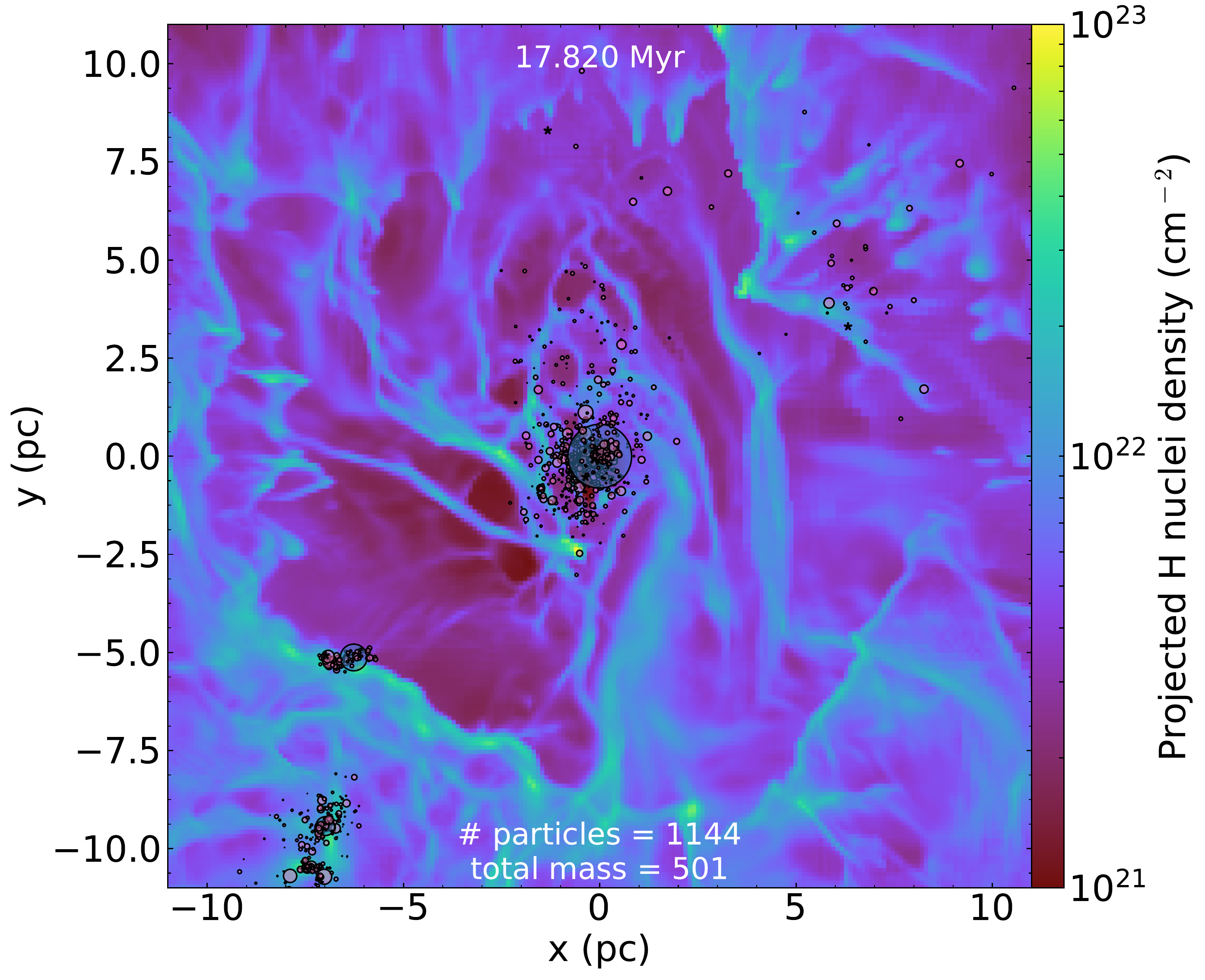}{0.5\textwidth}{(d)}
	}
	\caption{Projected number density along the $z$-axis for runs (a) M3 (b) M3f (c) M3f2
		and (d) M5f at the last data file from each run. The area of the circles
		representing stars are proportional to their mass,
                while the locations of sink particles are shown by
                white star symbols. Feedback is most effective in run
                (b) where multiple massive stars with strong feedback have sunk together to the center of the cluster and in (d) due to the $97~\msun$ star in the center of the image.\label{fig:ndens_grid_fb}}
\end{figure*}

This \added{massive} star rapidly expels
all remaining gas from the central cluster, \added{terminating star
  formation there and} leaving pillars of gas
surrounding the star forming region (see Fig.~\ref{fig:ndens_grid_fb} d) that
resemble the Eagle Nebula and similar formations
\citep[e.g.][]{hester1996,mccaughrean2002,mcleod2015}. 

A difficulty in performing simulations of large clouds from idealized initial density conditions stems from the initial free-fall time for the gas in the Gaussian sphere, which for this run is only \SI{8.6}{Myr}. Since turbulence decays within a free fall time $t_{\rm ff}$ \citep{Mac_Low_Kinetic_1998}, the velocity distribution of the gas became quite smooth by the time star formation commenced in this run at \SI{15.4}{Myr}. 
Similar concerns were discussed by \citet{krumholz_orion_II_2012}, who
noted that this affects both star formation rate and efficiency.  
%
Future models of high mass clouds will need to start with more
realistic initial conditions that better model the actual assembly of
such structures.


\subsection{Stellar Group Identification}

To identify stars as group members within the simulation we used two
methods, {\tt HOP} \citep{Eisenstein_HOP_1998}, which is included in \amuse,
and the {\tt Scikit-Learn} \citep{scikit-learn} implementation of
{\tt DBSCAN} \citep{Ester_DBSCAN_1996,Schubert_DBSCAN_revisited_2017}.
{\tt HOP} determines group membership by the following procedure:
\begin{enumerate}
	\item Calculate the local density at each particle using its
          $N_{\rm nn}$ nearest neighbors and the local density gradient at each particle
          using its $N_{\rm hop}$ nearest neighbors.
	\item From each particle, hop to the next particle of the
          $N_{\rm hop}$ neighbors in the direction of the highest
          density gradient. Continue until the current
          particle is the density maximum of the $N_{\rm hop}$ nearest
          particles.
          \item Identify this particle as a group core
          particle, with this particle's density as the group peak
          density $\rho_{\rm peak}$. All particles in the path of hops
          leading to this
          particle are added to this
          group as members. Repeat this process until all particles
          are in groups. 
	\item Identify particles that reside on the boundary between
          two groups by identifying particles where one of its
          $N_{\rm merge}$ nearest neighbors belongs to a different group. Record the density at these locations as the saddle density, $\rho_{\rm saddle}$, calculated as the average density between the two boundary particles.
	\item Merge any groups where $\rho_{\rm saddle}$ is either greater than an absolute saddle density $\delta_{\rm saddle}$, or whose ratio of saddle to peak densities is less than a given relative saddle density factor threshold $f_{\rm saddle}$. In mergers the group with the lower peak density $\rho_{\rm peak}$ is merged into the group with the higher peak density.
	\item Remove any group whose peak density is lower than the outer threshold density $\delta_{\rm outer}$.
\end{enumerate}
For physical parameters in {\tt HOP} we use an outer stellar mass density
threshold $\delta_{\rm outer}=\SI{1}{\msun~pc^{-3}}$, an order of
magnitude lower than the average stellar density of an open cluster
and an order of magnitude greater than the stellar density of the Milky Way \citep{binney2011galactic}. For the peak density we use $\delta_{\rm peak}=3\delta_{\rm outer}$ as suggested in the original paper. For the saddle density we use a relative saddle density threshold, where the boundary saddle density is compared to the minimum peak density of the two groups, defined by
\begin{eqnarray}
D = \frac{\delta_{\rm saddle}}{{\rm min}\left(\delta_{\rm peak,1};\delta_{\rm peak,2}\right)}.
\end{eqnarray}
If $D < f_{\rm saddle}$, where $f_{\rm saddle}$ is the saddle density
threshold factor, the two groups are merged. For our analysis here
we set $f_{\rm saddle}=0.5$, slightly more aggressively merging
groups than the default value of 0.8. The values ($N_{\rm merge}$, $N_{\rm
  hop}$, $N_{\rm nn}$) = (4,16,64), again as suggested in
\cite{Eisenstein_HOP_1998}.

{\tt DBSCAN} on the other hand determines group membership using a simpler procedure:
\begin{enumerate}
	\item Any particle with at least $N_{\rm min}$ neighbors within a distance $\xi$ is considered a core particle.
	\item Any particle that is within distance $\xi$ of at
          least one core particle, but has fewer than $N_{\rm min}$ neighbors, is considered a boundary particle.
	\item All connected core and boundary particles define a group.
	\item Any other particles are defined as noise.
\end{enumerate}
For {\tt DBSCAN} we set \deleted{the}  $N_{\rm min} = 16$ for a core particle
\deleted{to be 16} and we calculate $\xi$, the maximum neighboring particle
separation, to match our physical parameters in {\tt HOP}. Assuming an
average stellar mass of $M_{\rm avg} =\SI{0.56}{\msun}$ for the
initial mass function (IMF) of \citet{kroupa_IMF_2001} and using
a stellar density of $\rho_{\rm bg} = \SI{1}{\msun~pc^{-3}}$ we compute
\begin{equation}
\xi = \left(\rho_{\rm bg}/M_{\rm avg}\right)^{-1/3} = \SI{0.84}{pc},
\end{equation}
which we used for the one run we analyzed with {\tt DBSCAN} here, M5f.

Generally we prefer {\tt HOP} due to its physically motivated thresholds,
particularly its ability to compare the relative saddle density
between two groups to the minimum peak density of the groups
themselves to determine if the two groups should be merged. However
it was more practical to use the simpler {\tt DBSCAN} technique for
our largest data set.
As with any group-finding numerical technique, both
methods struggle to disentangle whether one or two groups exist just
before the point of merger. However, we only have one
run that experiences a merger of two nearly equal sized groups, while
others are either well separated or have mergers where one group is
clearly the more massive and dominates the potential. As a check, we
examined several times during M5f and verified
that we found similar results with {\tt HOP} and {\tt DBSCAN}. Both methods
provide similar and consistent grouping results when compared on the
same data over multiple grouping computations, with an agreement of $>
95\%$ on cluster members.


\section{Results}\label{fb_results}

\subsection{Star Formation}

\label{subsec:SF}
\begin{figure*}
	\begin{tabular}{cc}
          	\includegraphics[width=0.5 \textwidth]{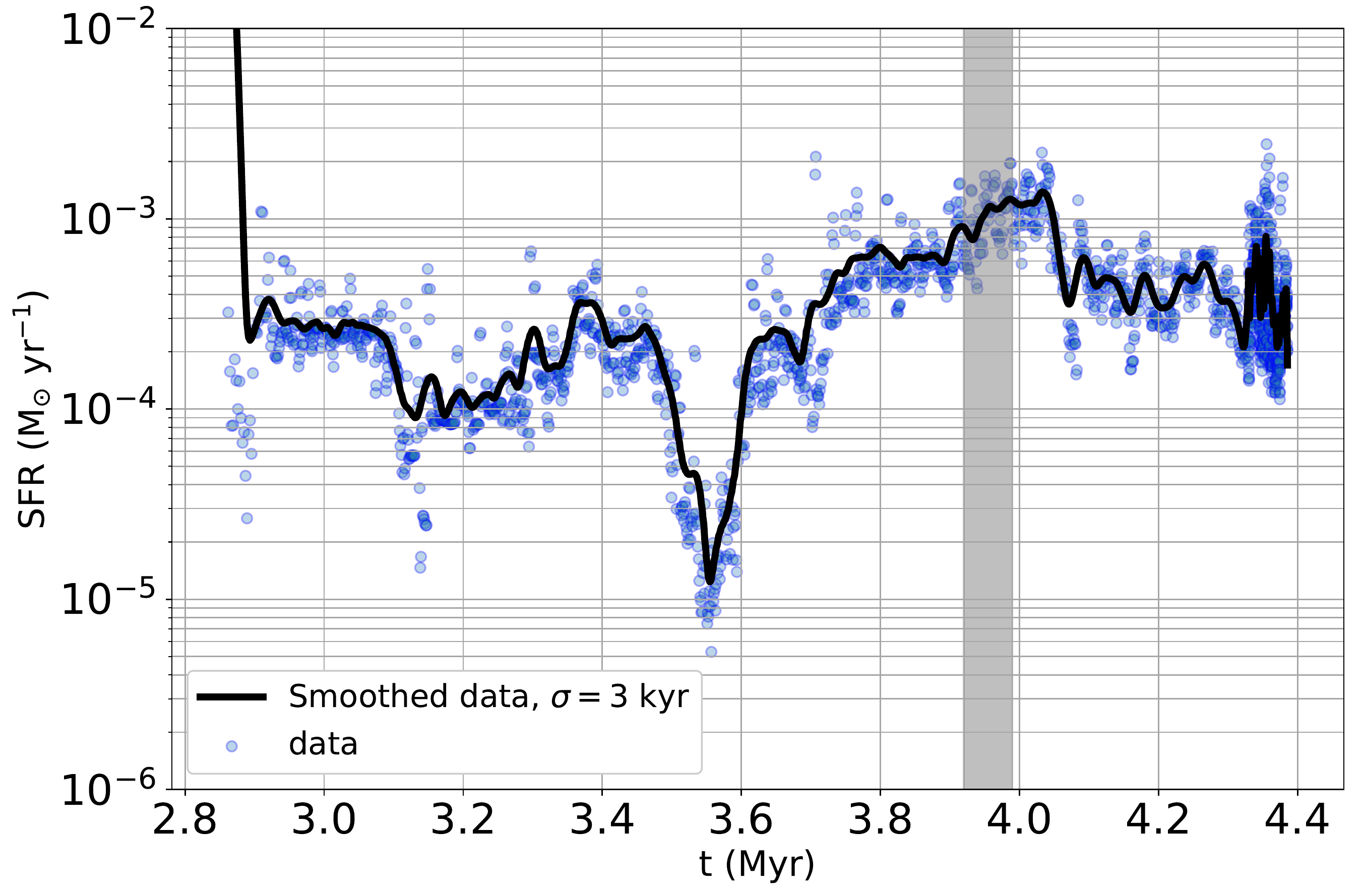} &
		\includegraphics[width=0.5 \textwidth]{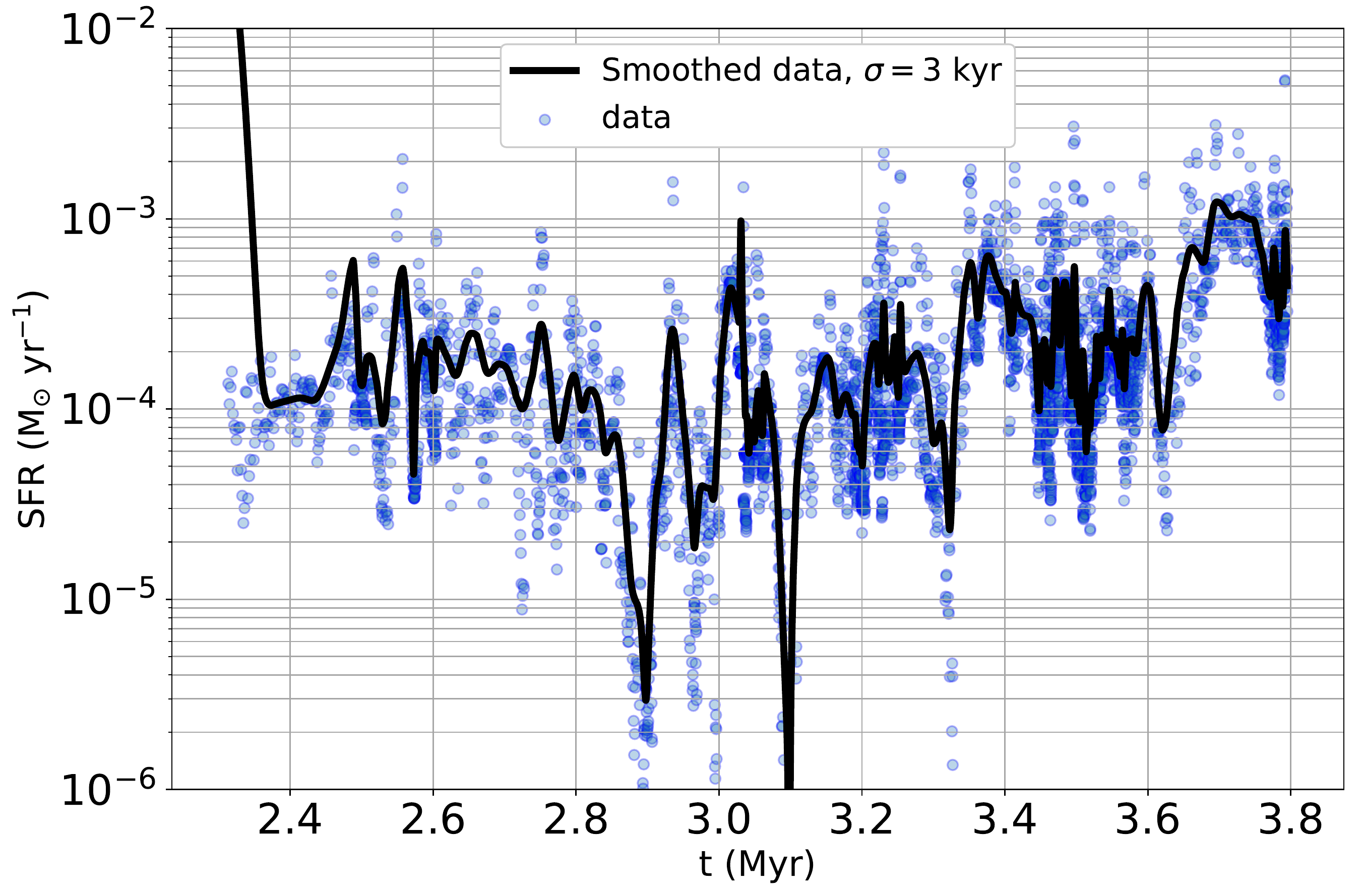} \\
		(a) & (b) \\[6pt]
		\includegraphics[width=0.5 \textwidth]{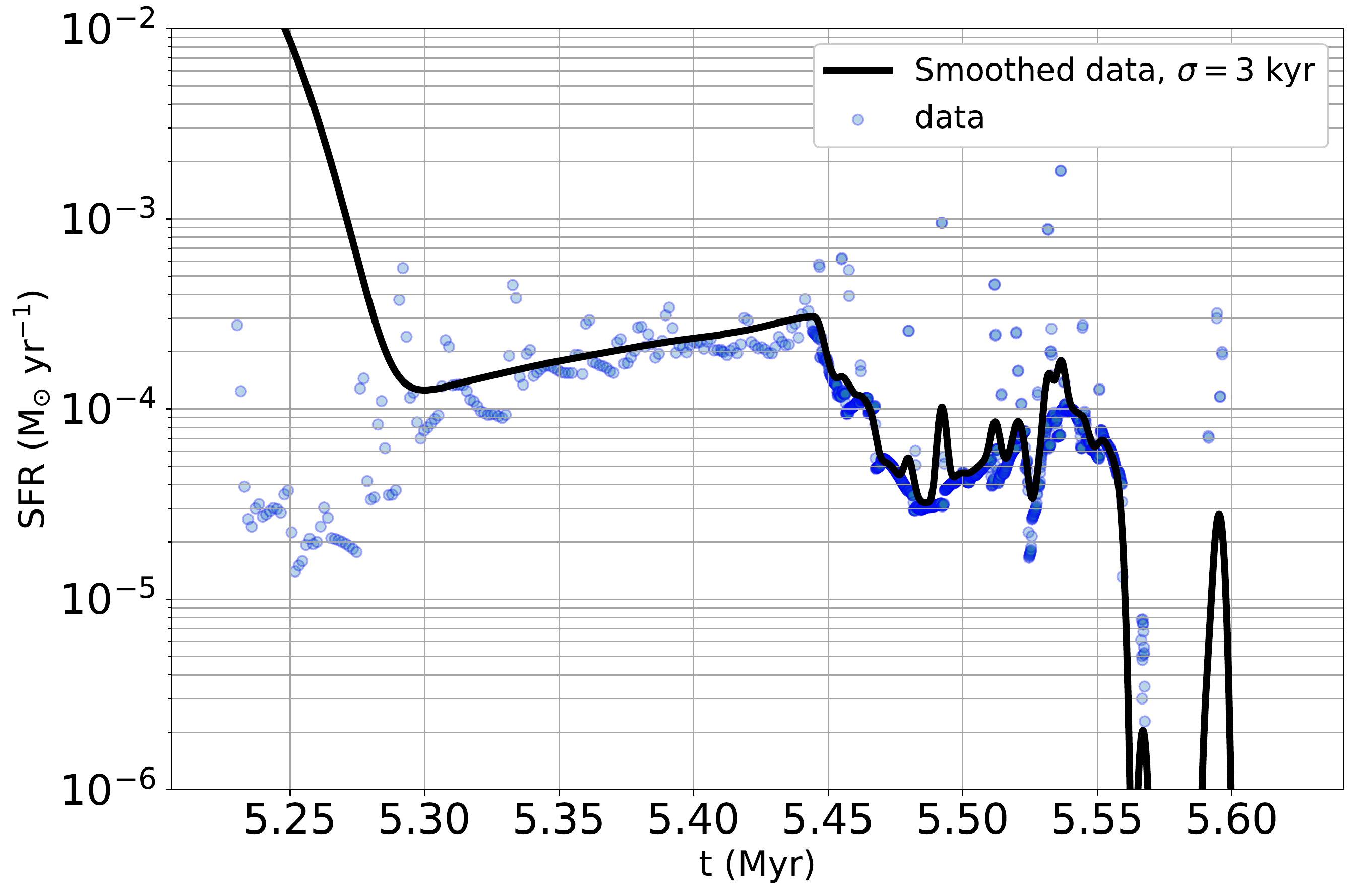} &
		\includegraphics[width=0.5 \textwidth]{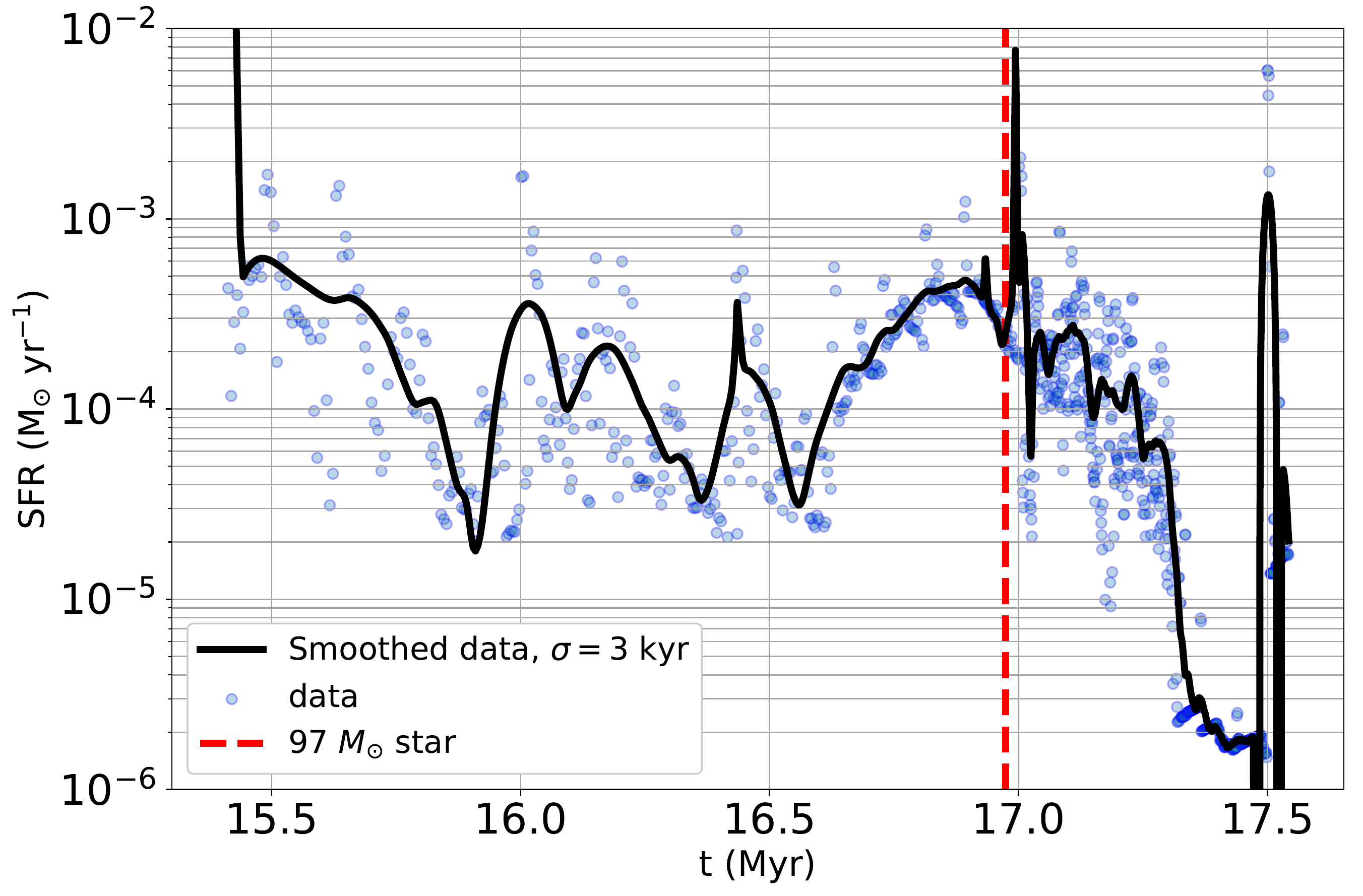} \\
		(c) & (d) \\[6pt]
	\end{tabular}
	\caption{Star formation rates for runs (a) M3 (b) M3f (c) M3f2
          and (d) M5f.  Blue dots show data points every timestep (which 
          during feedback can be as small as 100 yr),
          while black line shows the SFR smoothed with a Gaussian filter
          with $\sigma=\SI{3}{kyr}$.
          The grey shaded area in (a) shows time of subgroup merger, while
          the red dashed vertical line in (d) shows the formation of
          $A_*$, the \SI{97}{\msun} star.
          \label{fig:sfr}
        }
\end{figure*}

The star formation rate (SFR) as a function of time in our
\added{four proof of concept} simulations
is shown in Figure~\ref{fig:sfr}. The data shown as blue points is
initially calculated by a second-order central difference of the
stellar masses every timestep (as little as 100 yr),
which are generally quite noisy. We also show the
result of using a Savitzky-Golay (\citeyear{Savitzky_Golay_1964})
filter convolved with a window size of 51 and fit to a third-order
polynomial to smooth the data before we take the derivative (black
lines). We use this filter on all data hereafter presented with open 
circles, representing the data taken directly from the simulation, 
accompanied by line plots, which show the result of the smoothing.
For the SFR, we further smooth with a Gaussian filter with 3~kyr
variance.

In \replaced{M3}{the M3 run without feedback,} the SFR generally stays
high, only briefly decreasing due to a 
strong interaction in the main group that placed many of the massive
stars on wide orbits and slowed the overall accretion rate of the
region. In \replaced{M3f}{the first run with feedback, M3f,} the data
are much noisier due to the flickering H~{\sc 
  ii} regions (see Sect.\ \ref{Sect:M3f}), which heat the gas briefly
and inject turbulence, but never provide enough outward momentum to
the nearby gas to eject it nor ionize enough material to prevent it
from cooling again.
\replaced{Run}{The lower surface density run} M3f2 shows relatively
stable star formation until the first 
massive stars appear at \SI{5.45}{Myr}. Their feedback breaks up the
filament in which stars are forming, but cannot entirely disperse the
dense gas in the region. This allows star formation to continue for
another \SI{e4}{yr} until an interaction between two massive stars
expels one far enough out of the center of the group for a second
H~{\sc ii} region to form and expand out of the star-forming region.
\added{In all M3 runs a mild trend of increasing SFR over time can be
  seen, agreeing with \citet{gonzalez-samaniego2020}, though not
  accelerating at nearly the rate found by \citet{lee2015} in a more
  idealized model.}
In run M5f the SFR shows large variations as
filaments form and intersect in the first megayear and two separate groups
form.
\added{No clear trend of increasing SFR is seen in this
  large region, although the formation of multiple groups may obscure
  any signal.}
At around \SI{17}{Myr}, a \SI{97}{\msun} star forms in the more
massive group, emitting radiation and winds that travel throughout the
simulation domain, although star formation continues both near and far
from the star for another $\sim \SI{3e5}{yr}$ before effectively terminating.



\begin{figure*}[h]
	\begin{tabular}{cc}
                \includegraphics[width=0.5 \textwidth]{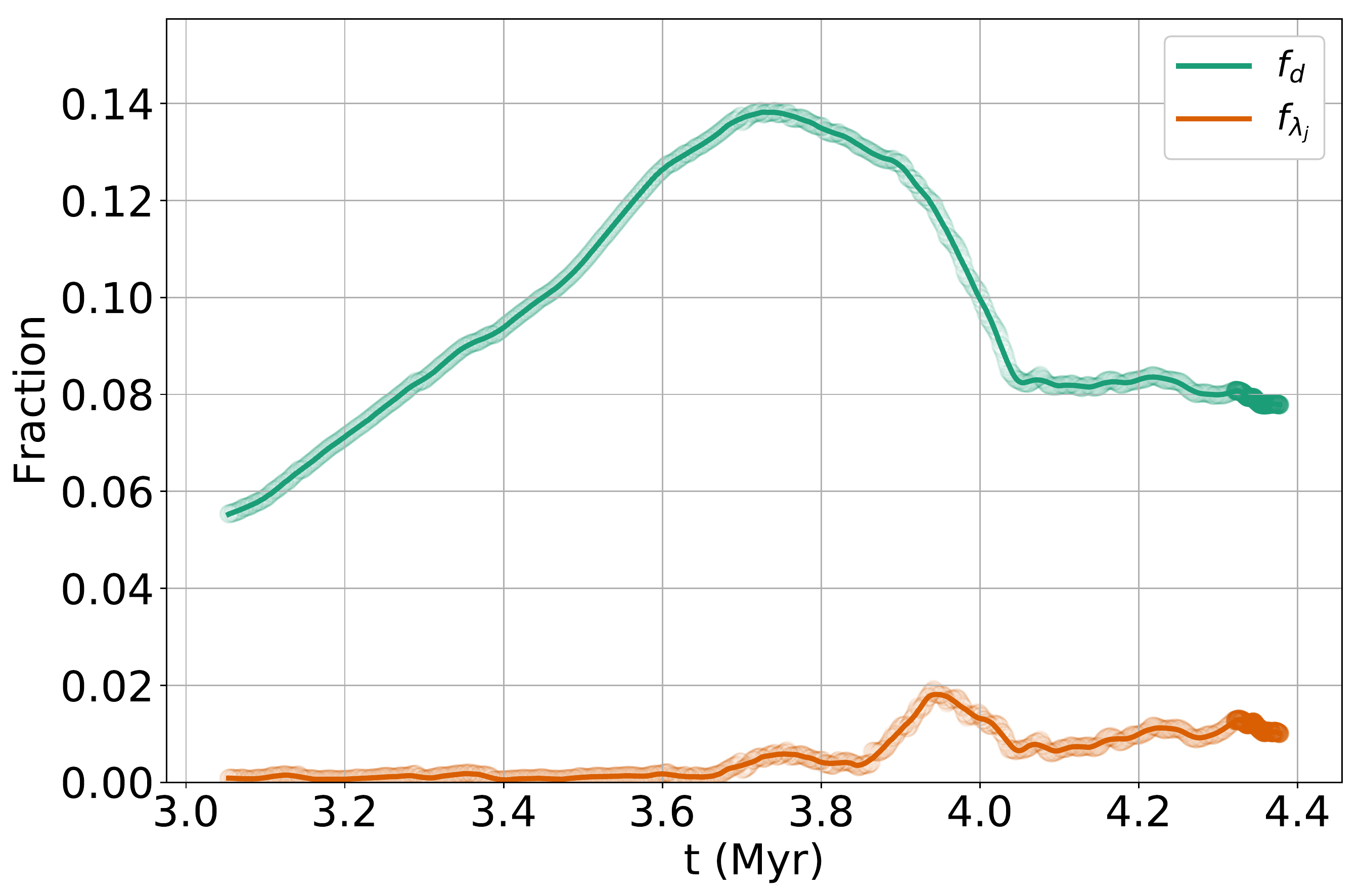} &
		\includegraphics[width=0.5 \textwidth]{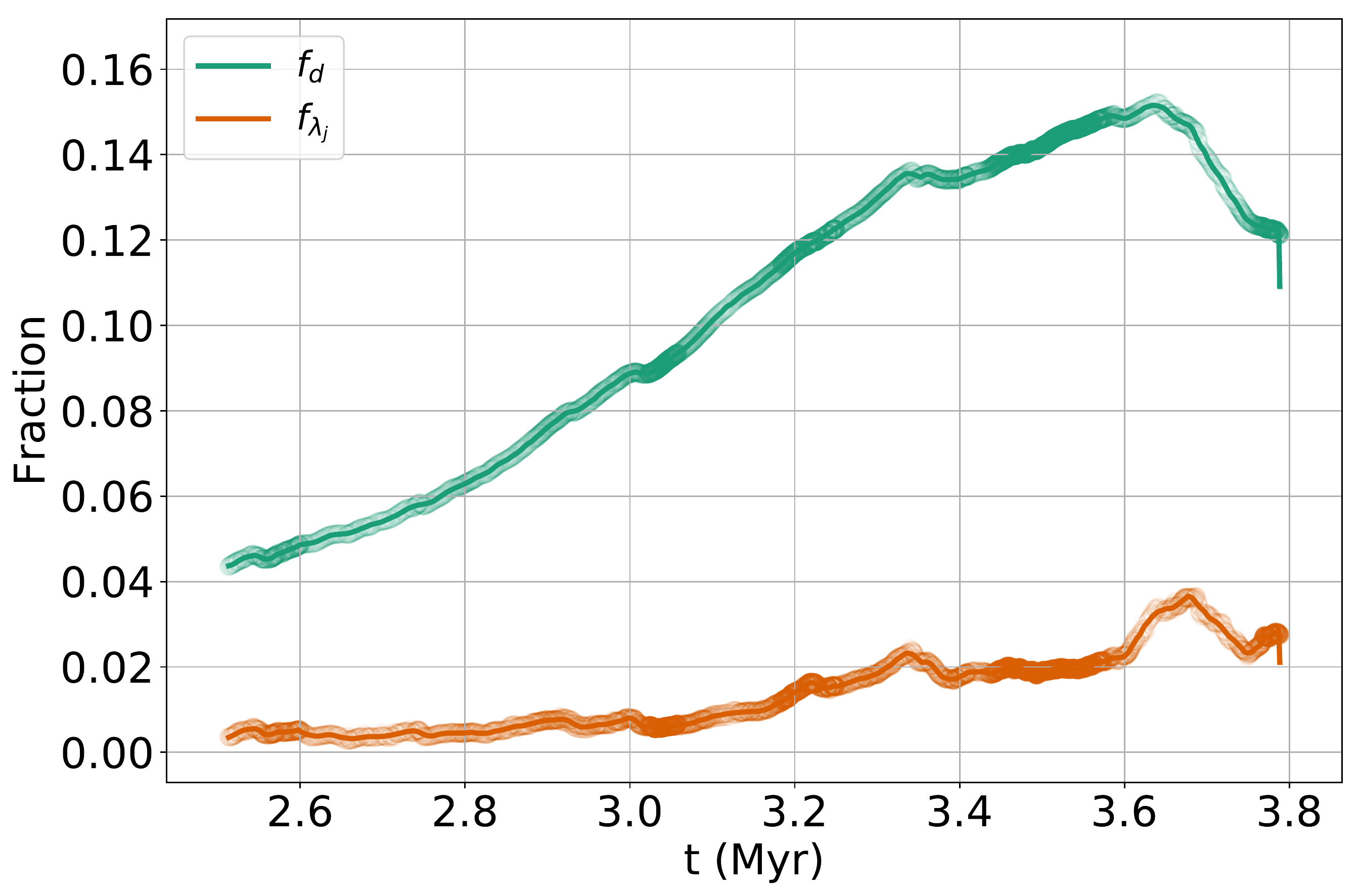} \\
		(a) & (b) \\[6pt]
		\includegraphics[width=0.5 \textwidth]{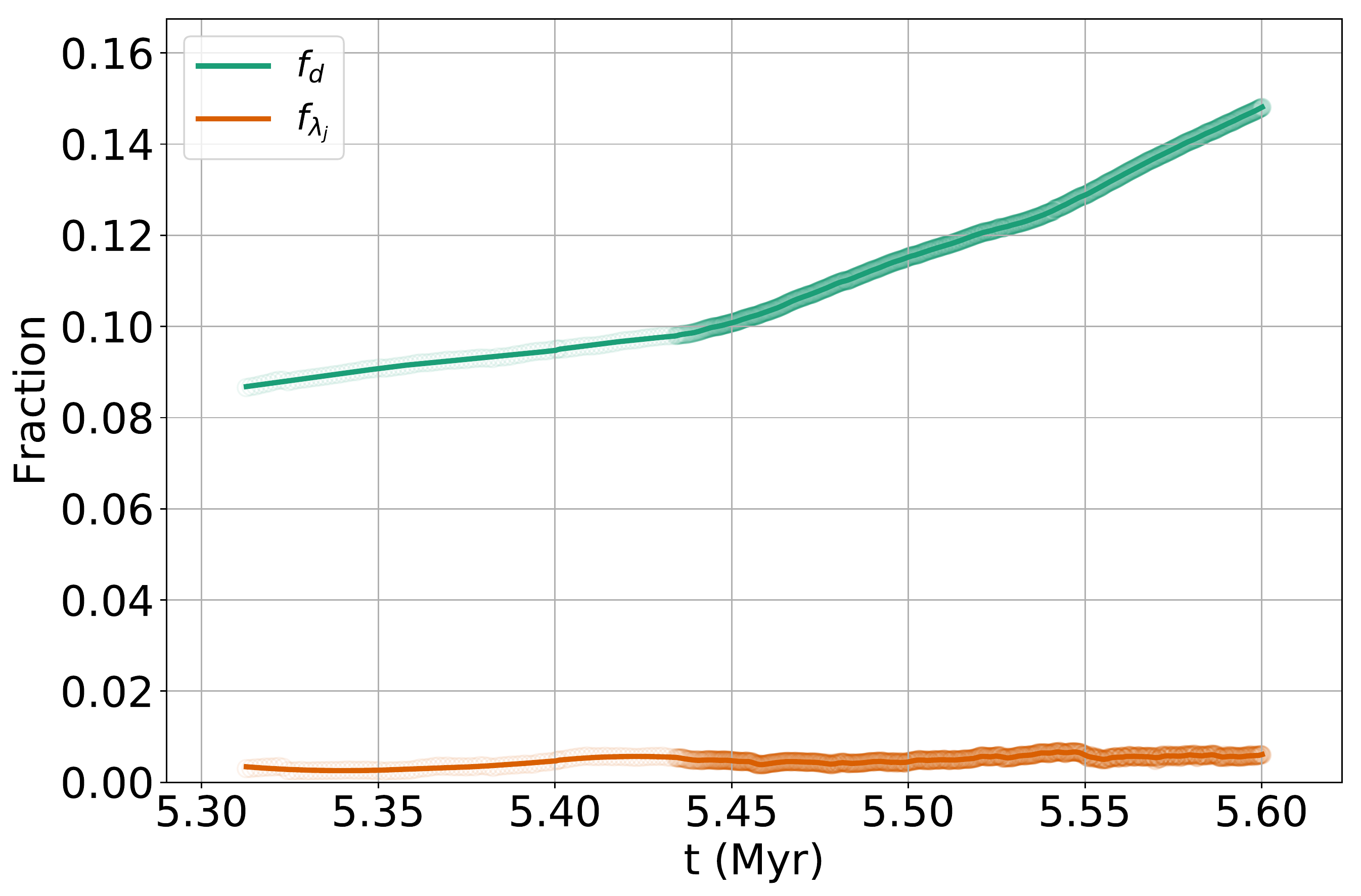} &
		\includegraphics[width=0.5 \textwidth]{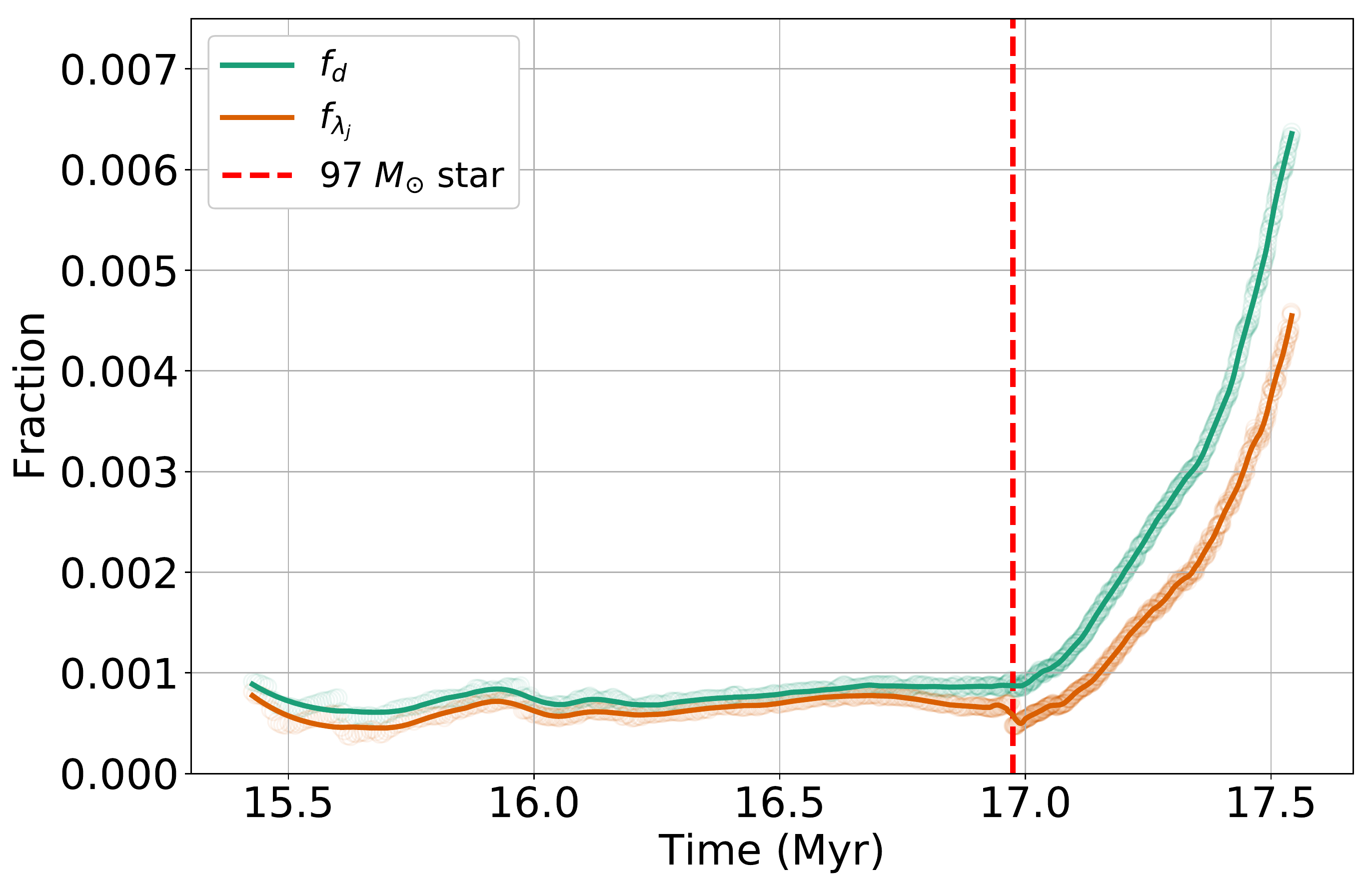} \\
		(c) & (d) \\[6pt]
	\end{tabular}
	\caption{Total dense and Jeans unstable gas fractions for runs
          (a) M3 (b) M3f (c) M3f2 and (d) M5f.  
          \label{fig:frac_totals}}
\end{figure*}

The amount of gas available to form stars is presented in
Figure~\ref{fig:frac_totals} where we show both the fraction (by mass)
of dense gas \citep[for which $n_{\rm H} > \SI{e4}{cm^{-3}}$, the
limit generally considered for gas to be star
forming;][]{Lada_SFR_clouds_2010} and the fraction of Jeans unstable
gas. For all the runs except M5f, the amount of Jeans unstable gas is
much smaller, generally by a factor of two or more, than the amount of
dense gas. This agrees with results found by
\citet{Dale_early_evo_clusters}, who concluded that dense gas is a
necessary but not sufficient condition for star formation. Also in
agrement with \citet{Dale_early_evo_clusters} is our finding that
effective stellar feedback, which occurs in runs M3f2 and M5f,
actually produces more dense gas, rather than reducing it. Only in run
M5f does feedback also seem to increase the amount of Jeans unstable
gas, thereby leading to increased star formation near the center of
feedback. The difference between the feedback in M3f2 and M5f is that
the stellar wind from the extremely massive star in M5f can sweep up a
shell dense enough to trap its own H~{\sc ii} region, allowing some
triggered star formation in the shell (see Sect.~\ref{m5f_sf}).

\subsection[Group Structure]{Stellar Group Structural Evolution}

\label{fb_struct}

We next examine the effectiveness, or lack thereof, of stellar
feedback on the evolution of the groups that contain massive
stars. Given the expectation that 90\% of all clusters are disrupted
\citep{Lada_and_Lada_2003}, presumably by gas expulsion, we might
expect any feedback that completely ejects the natal gas to destroy
the cluster it formed from by removing the dense gas potential helping
to bind the cluster
\citep[e.g.][]{tutukov1978,elmegreen1983,parmentier2008,goodwin2009,rahner2017,rahner2019}.

\subsubsection{Energetics}

\label{subsub:energy}
\begin{figure*}
	\begin{tabular}{cc}
          	\includegraphics[width=0.5 \textwidth]{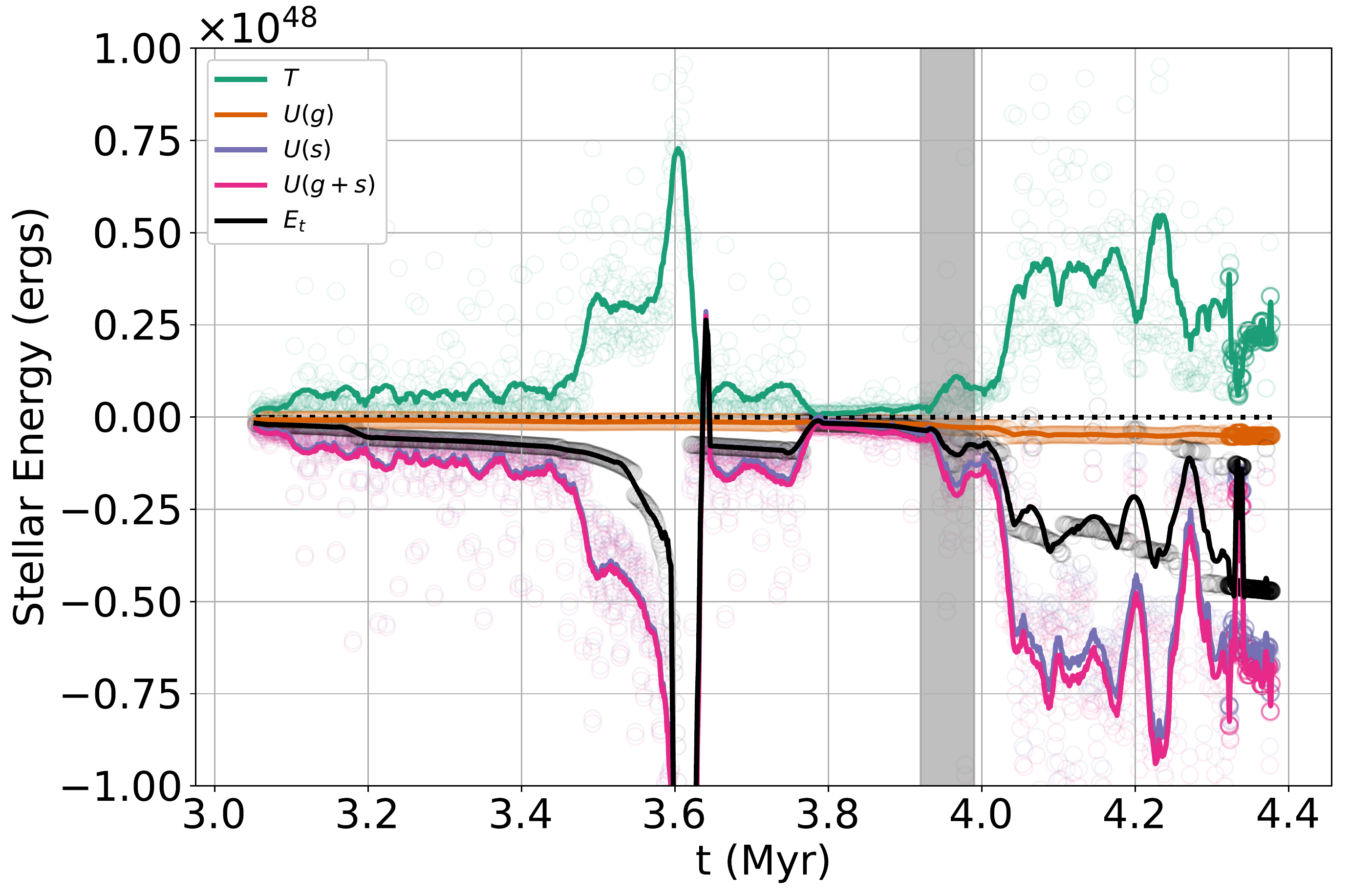} &
		\includegraphics[width=0.5 \textwidth]{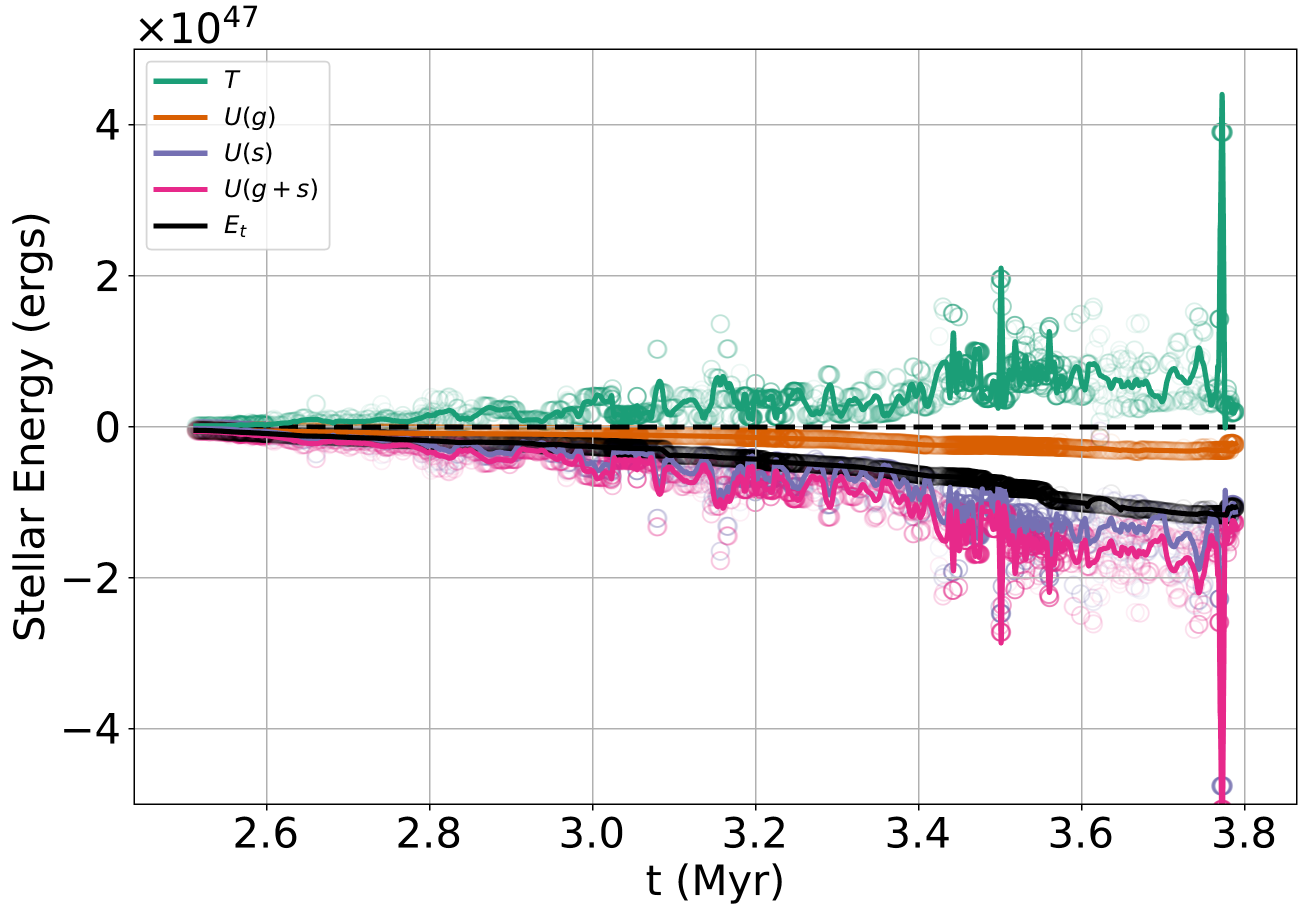} \\
		(a) & (b) \\[6pt]
		\includegraphics[width=0.5 \textwidth]{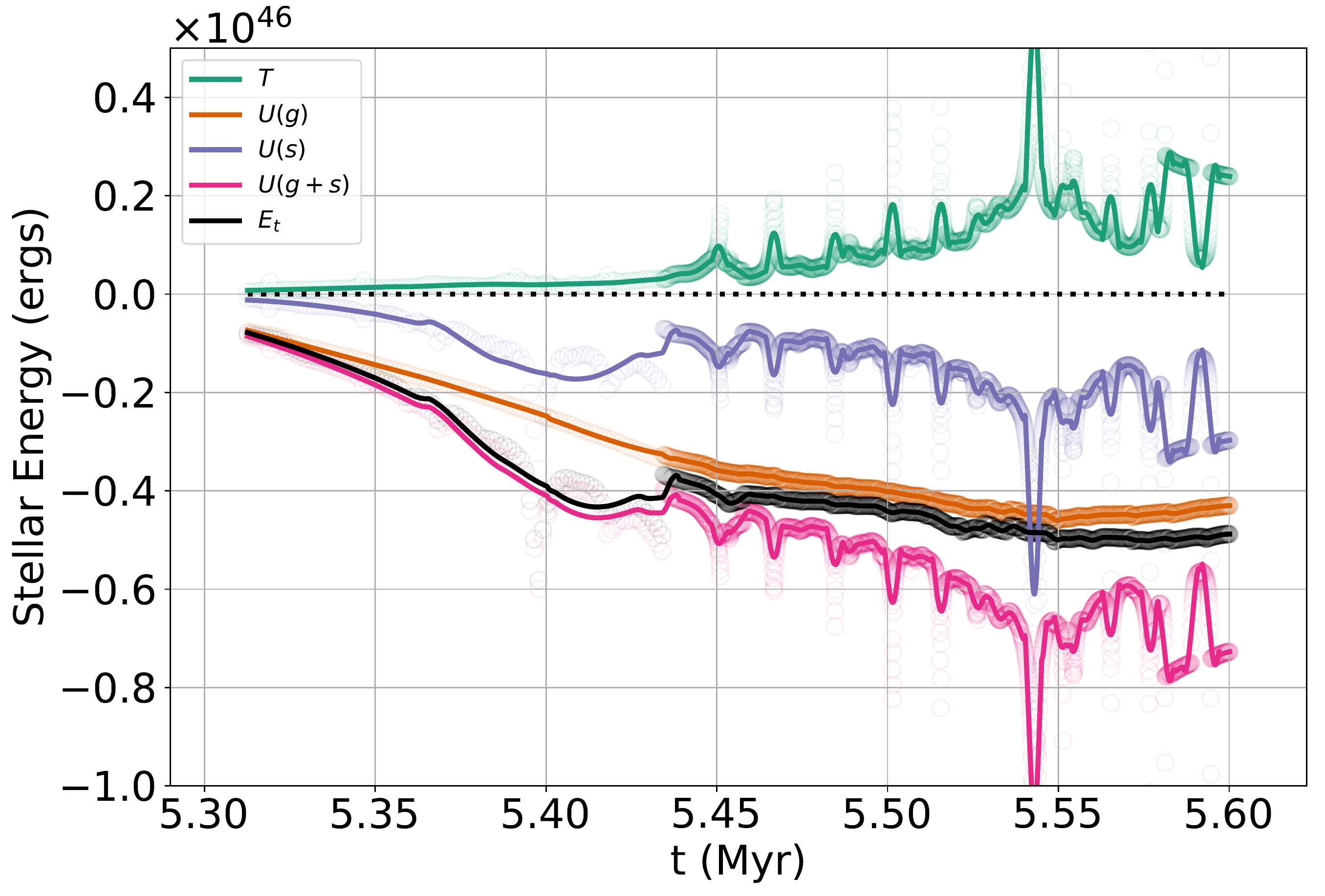} &
		\includegraphics[width=0.5 \textwidth]{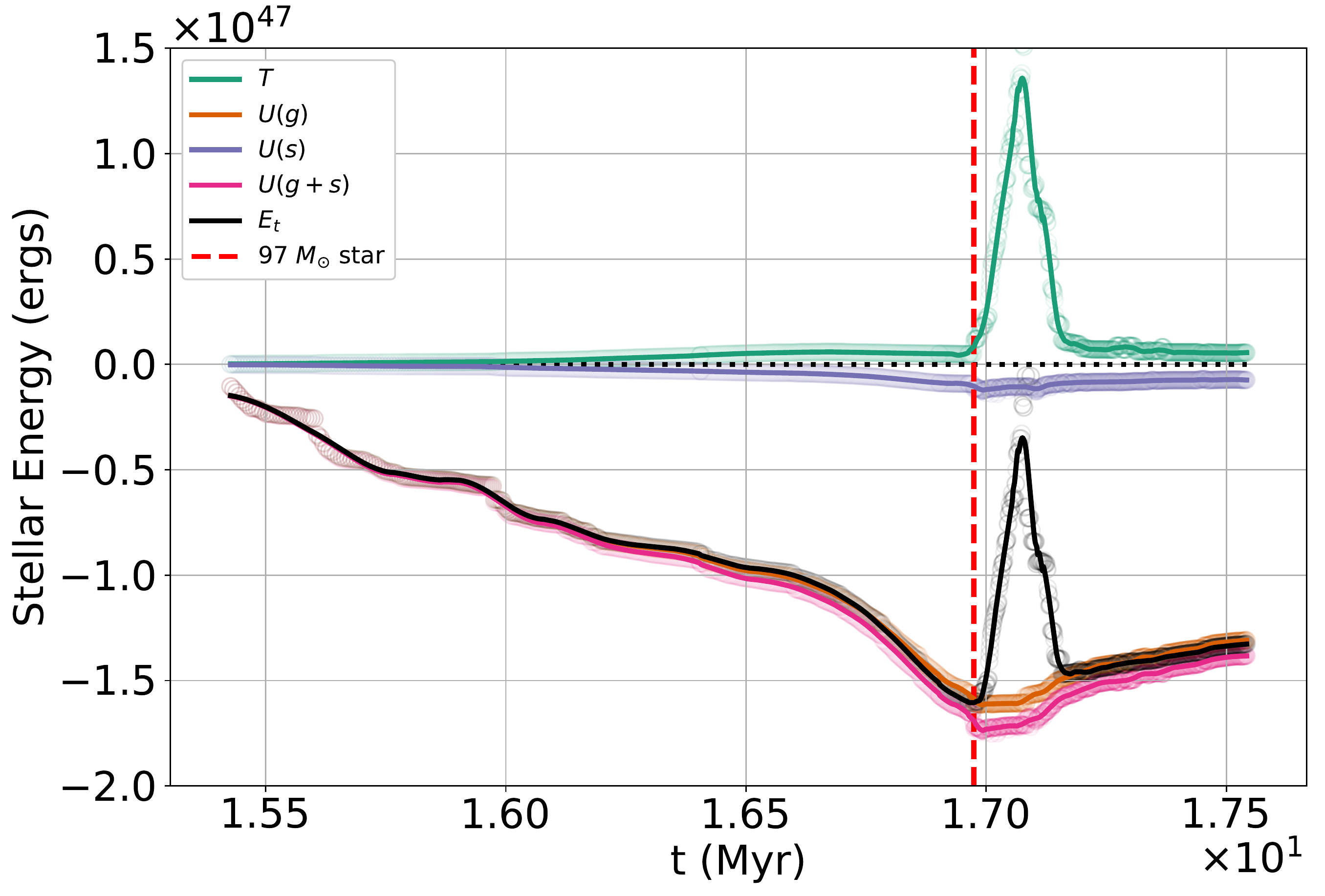} \\
		(c) & (d) \\[6pt]
	\end{tabular}
	\caption{Total energy of the stars $E_{\rm t}$ in the main
          stellar groups for runs (a) M3 (b) M3f (c) M3f2 and (d)
          M5f, showing that they all end bound. Also shown are stellar
          kinetic energy $T$, potential energy 
          due to gas $U(g)$ and stars $U(s)$, and their sum
          $U(g+s)$. The grey
          shaded area in (a) shows time of subgroup merger, while the red
          dashed vertical line in (d) shows the formation of $A_*$, the
          \SI{97}{\msun} star. Varying energy ranges come from
          varying compactness of the main group in each case. \label{fig:star_energy}}
\end{figure*}
\begin{figure*}
	\begin{tabular}{cc}
		\includegraphics[width=0.5 \textwidth]{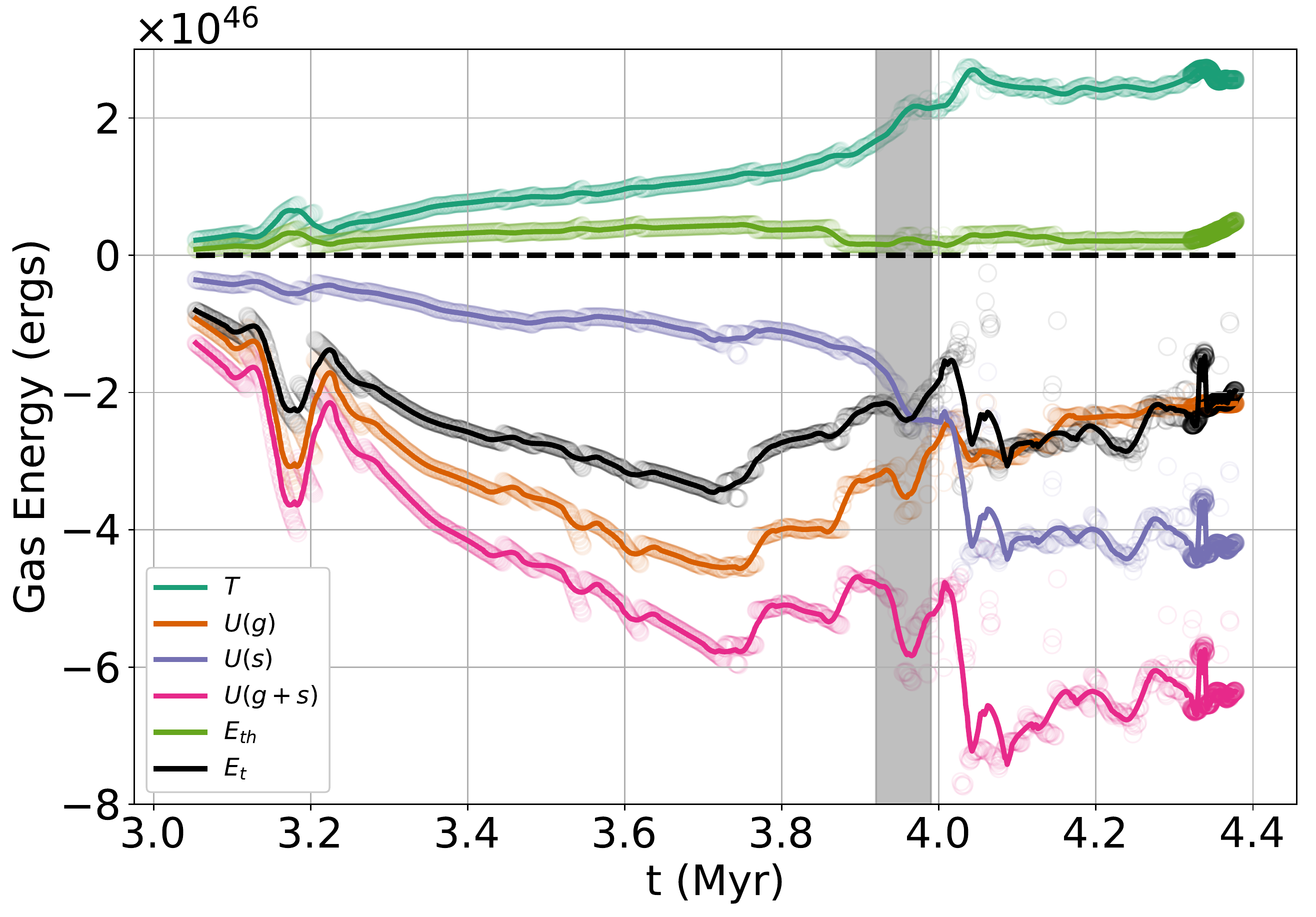} &
		\includegraphics[width=0.5 \textwidth]{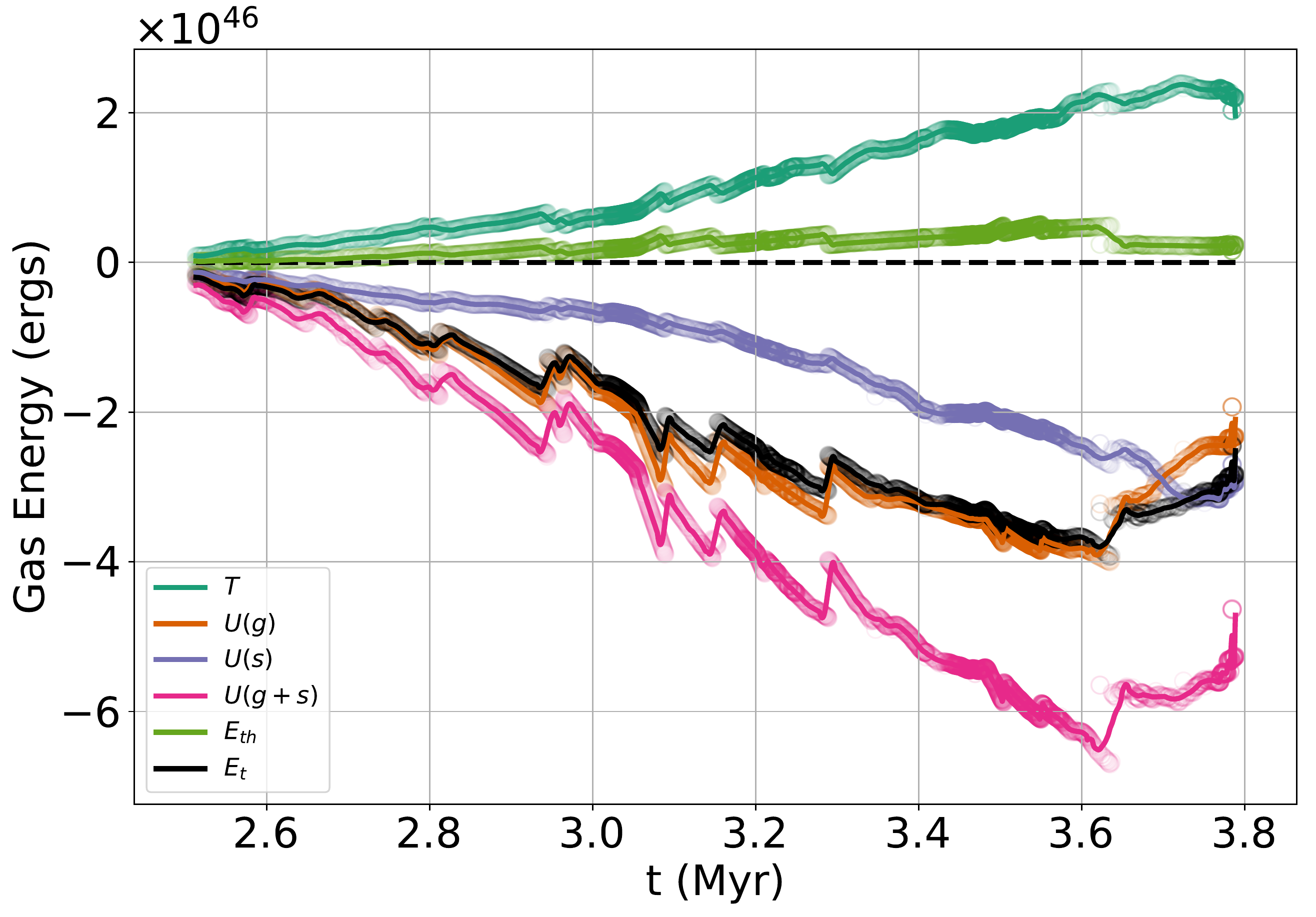} \\
		(a) & (b) \\[6pt]
		\includegraphics[width=0.5 \textwidth]{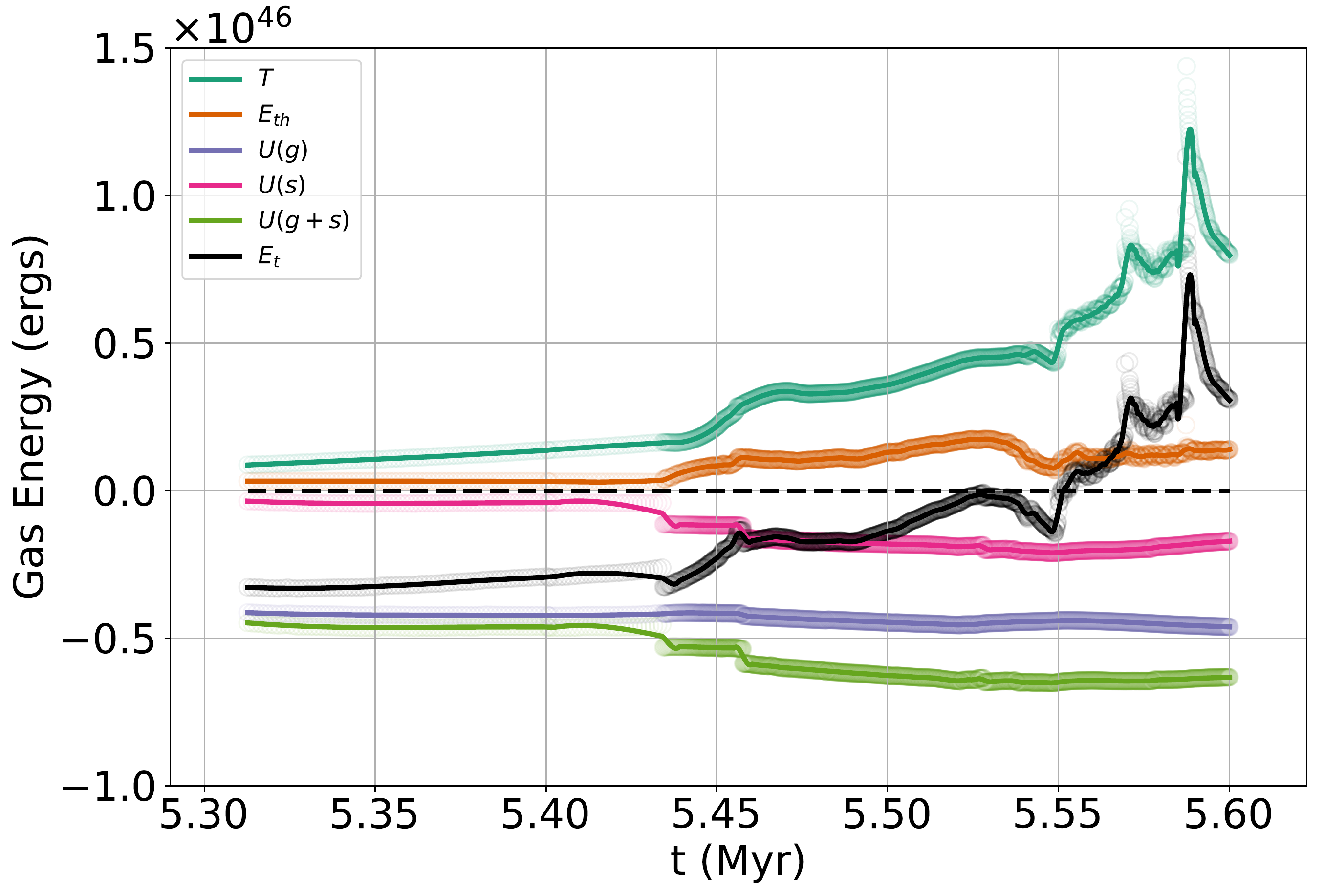} &
		\includegraphics[width=0.5 \textwidth]{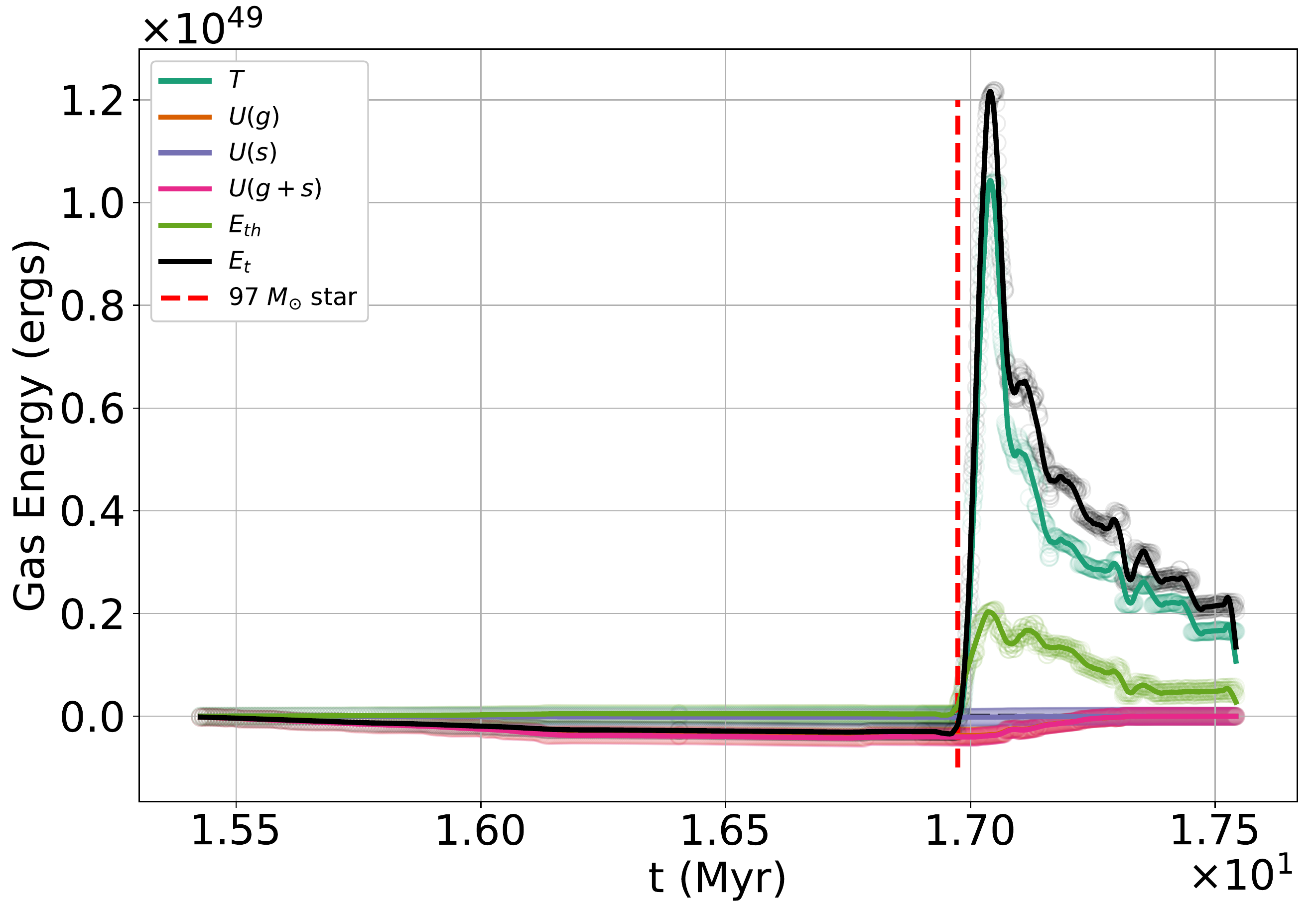} \\
		(c) & (d) \\[6pt]
	\end{tabular}
	\caption{Total energy of the gas $E_{\rm t}$ for the main groups in runs
          (a) M3 (b) M3f (c) M3f2 and (d) M5f.  Also shown are gas
          kinetic energy $T$, thermal energy $E_{\rm th}$, and potential energy 
          due to gas $U(g)$ and stars $U(s)$, and their sum
          $U(g+s)$. The grey
          shaded area in (a) shows time of subgroup merger, while the red
          dashed vertical line in (d) shows the formation of $A_*$, the
          \SI{97}{\msun} star. \label{fig:gas_energy}}
\end{figure*}
Figures~\ref{fig:star_energy}
and~\ref{fig:gas_energy} show the energy in stars and gas respectively
contained within the radius of the main group for each run. These
figures show that only \added{the main groups in} runs M3f2 and M5f actually eject their 
natal gas, when their total gas energy becomes positive. However both
stellar groups remain bound with negative total energies after the gas
is removed, identifying them as true clusters.

\begin{figure*}
	\begin{tabular}{cc}
		\includegraphics[width=0.5 \textwidth]{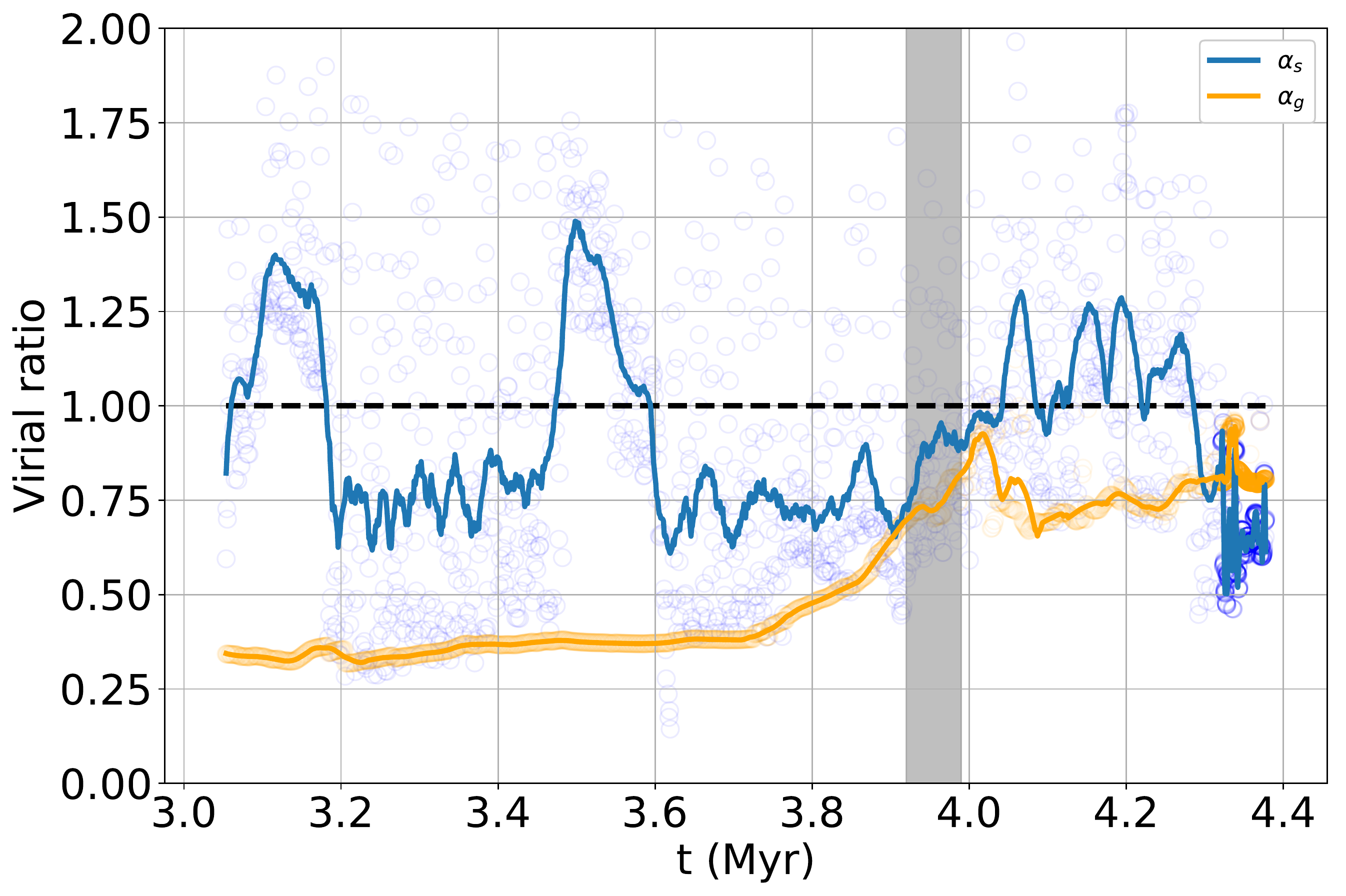} &
		\includegraphics[width=0.5 \textwidth]{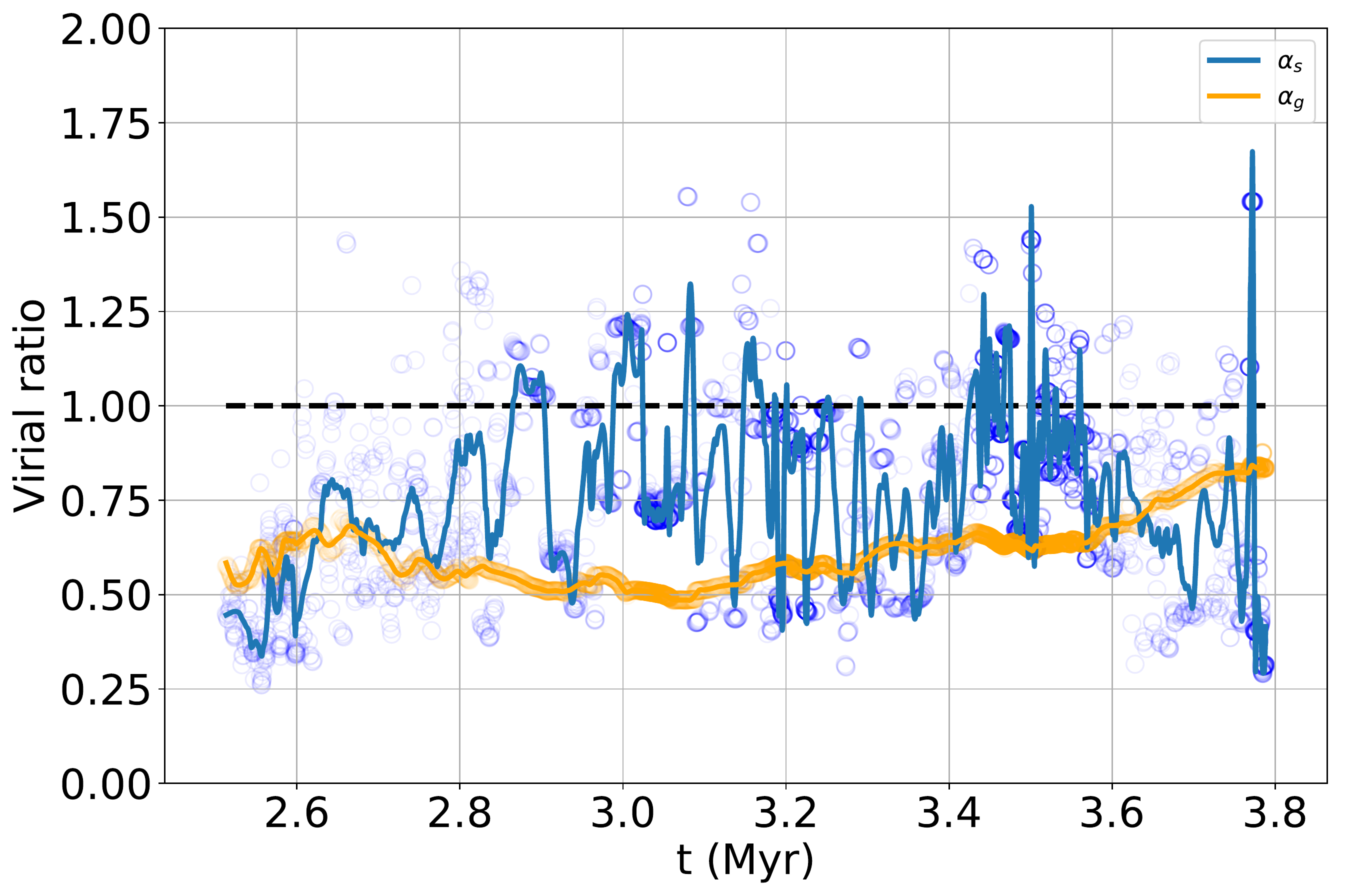} \\
		(a) & (b) \\[6pt]
		\includegraphics[width=0.5 \textwidth]{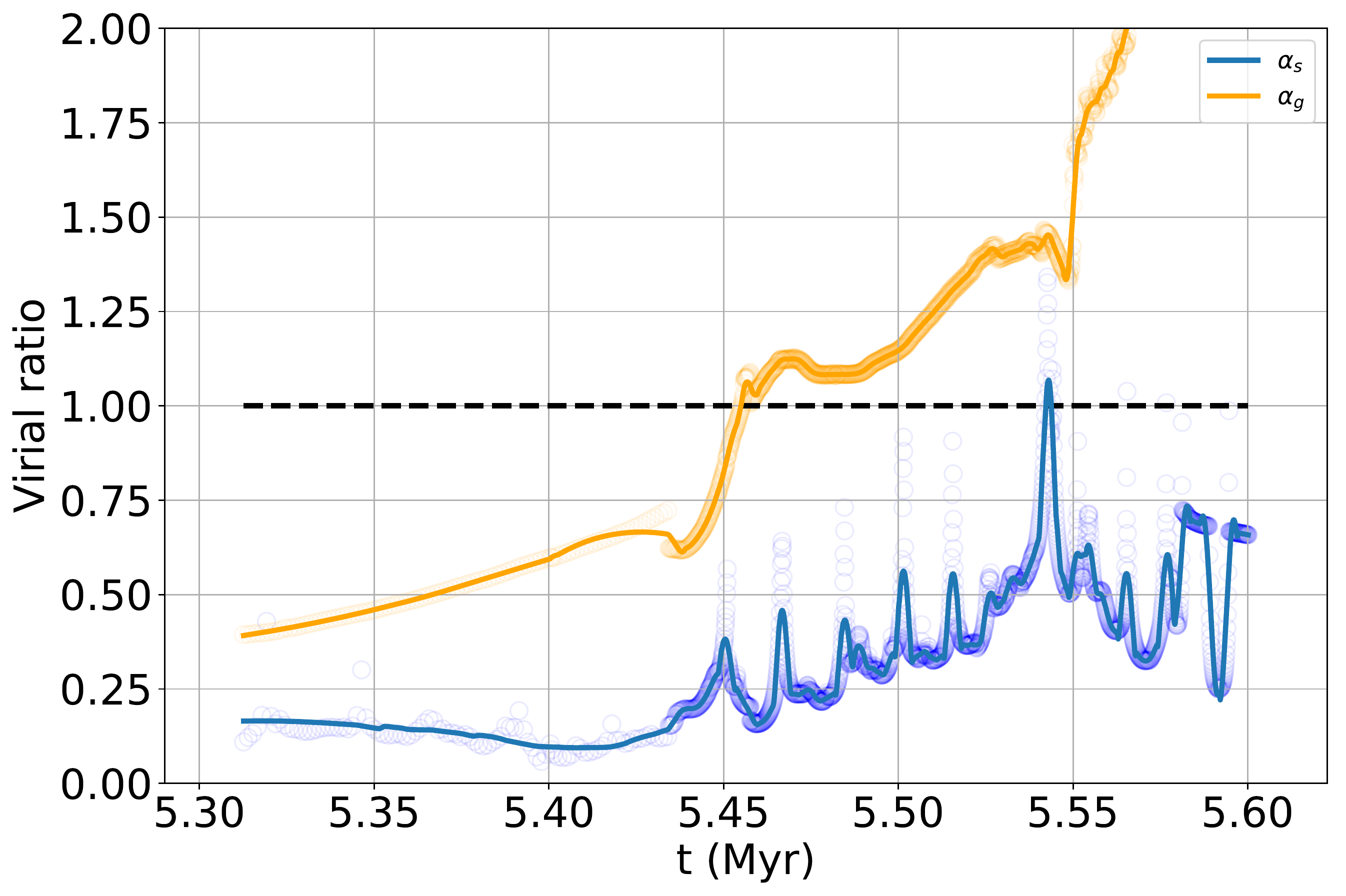} &
		\includegraphics[width=0.5 \textwidth]{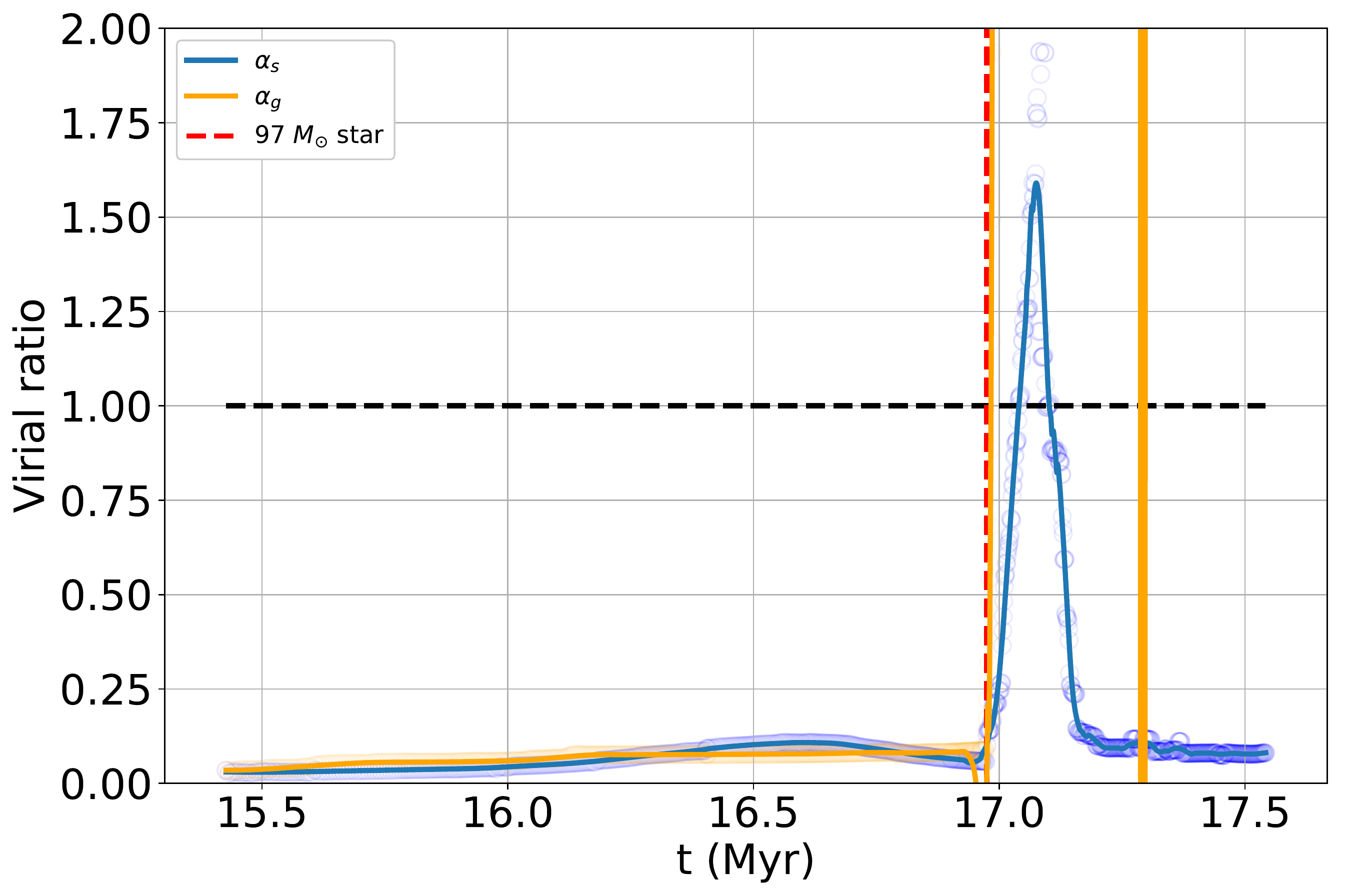} \\
		(c) & (d) \\[6pt]
	\end{tabular}
	\caption{Virial ratios for the stars $\alpha_s$ and gas
          $\alpha_g$ in the main groups for runs (a) M3 (b) M3f (c)
          M3f2 and (d) M5f. The grey
          shaded area in (a) shows time of subgroup merger, while the red
          dashed vertical line in (d) shows the formation of $A_*$, the
          \SI{97}{\msun} star. \label{fig:vir_ratio}}
\end{figure*}

This is further confirmed by looking at the virial ratios $\alpha =
2T/U$ of the gas and stars in these groups
(Fig.~\ref{fig:vir_ratio}), where the group as a whole appears briefly
unbound during the time that some of the outer stars escape following
gas ejection, but the overall \replaced{cluster}{group} survives and returns to a bound
virial ratio in both cases. Indeed, the \replaced{cluster}{group} in run M5f is subvirial
at the time of the final snapshot, and the other \replaced{3}{three} \replaced{cluster}{group}s are close
to being virialized, regardless of the current state of the gas in the
region defined by the \replaced{cluster}{group}.
We do see mass segregation in our runs, as detailed below, which might
contribute to their being observed as supervirial, something we will
examine in more detail in future work.  Our \replaced{cluster}{group}s appear likely to survive gas ejection, and mass segregation may assist in their survival, since increasing stellar to gas density ratios increases the likelihood of surviving the gas ejection stage \citep{Kruijssen_clusters_structure_2012}.


%
%

\subsubsection{Mass}

\label{subsub:mass}
\begin{figure*}[h]
	\begin{tabular}{cc}
		\includegraphics[width=0.5 \textwidth]{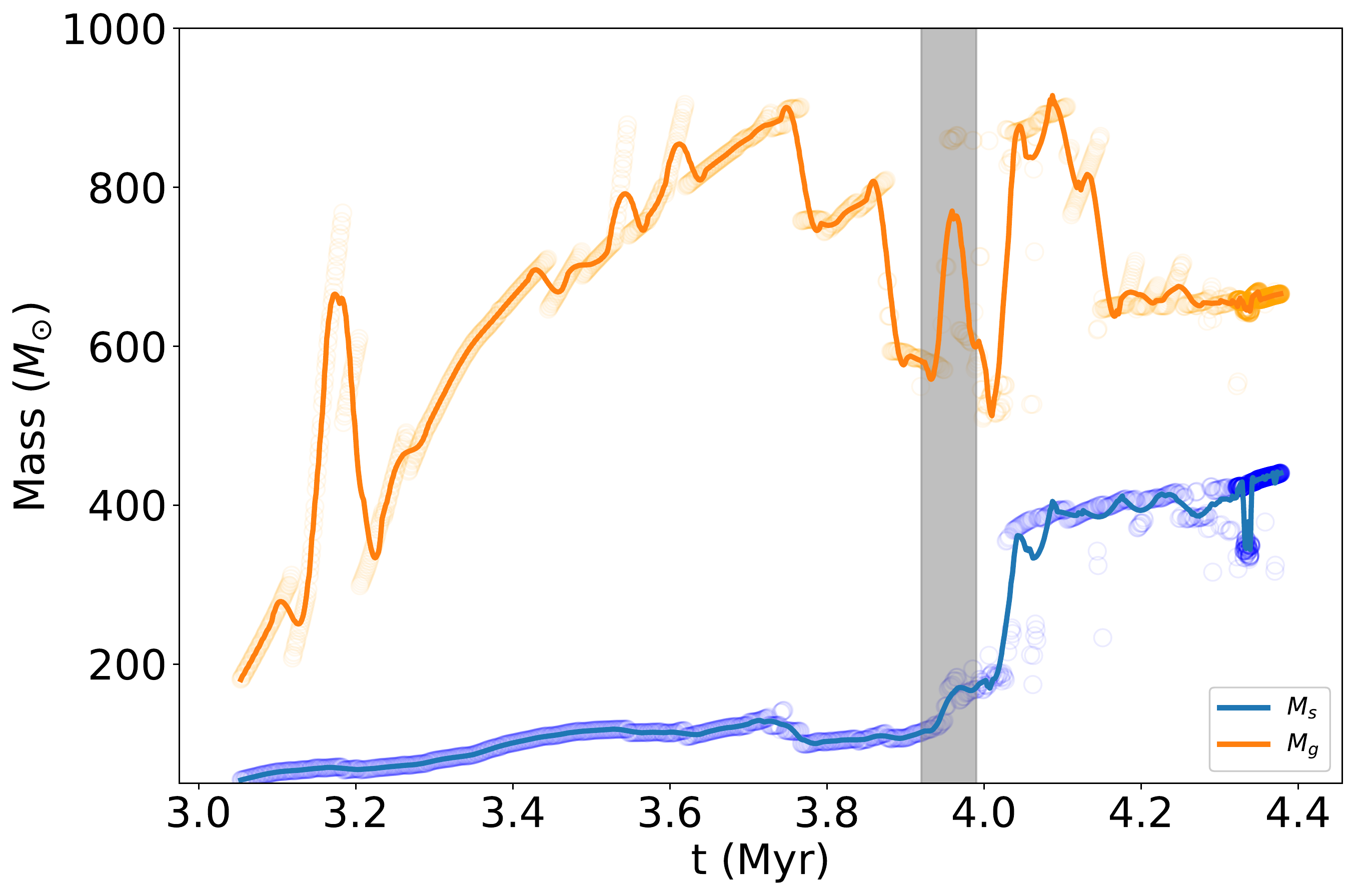} &
		\includegraphics[width=0.5 \textwidth]{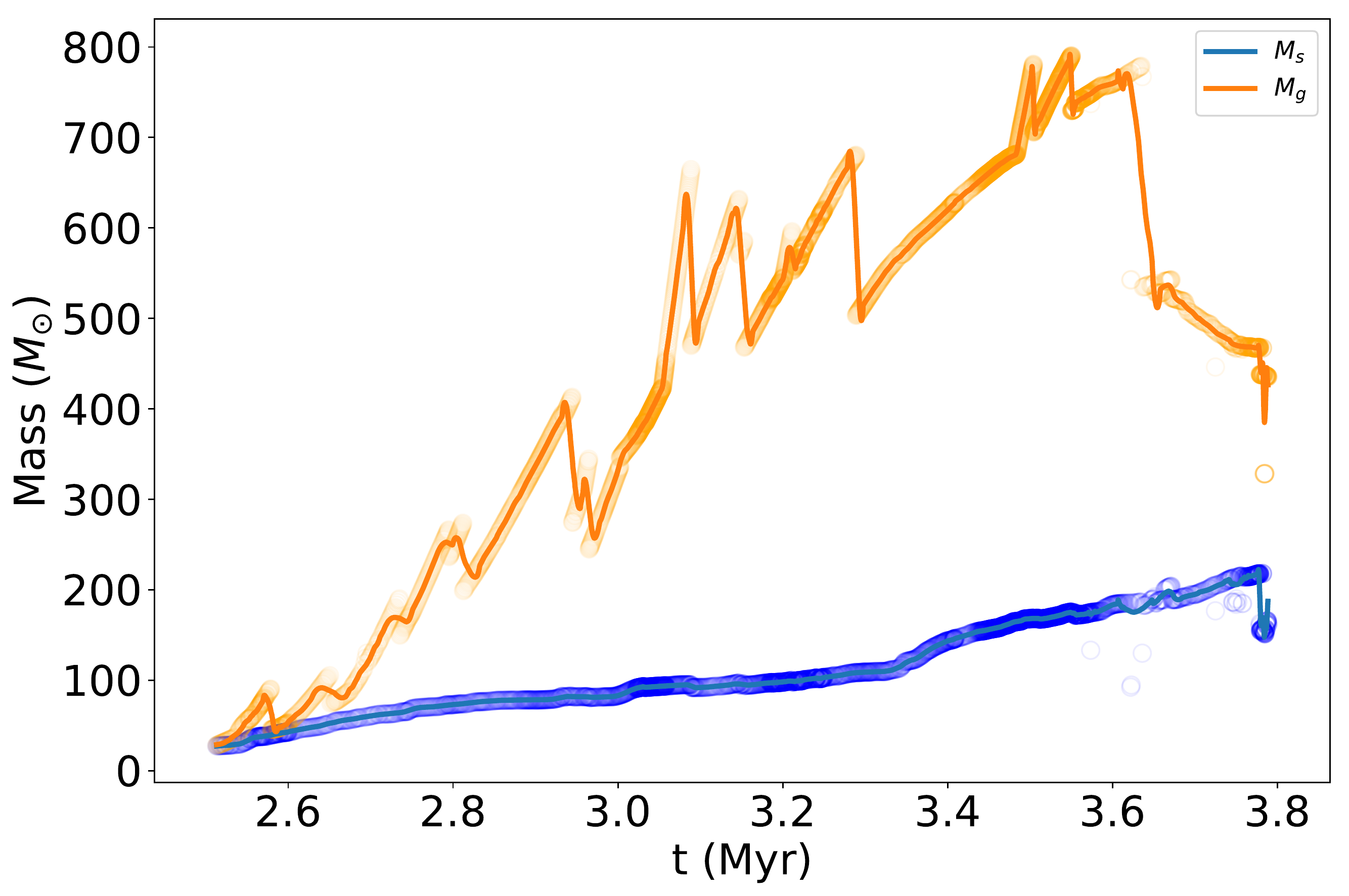} \\
		(a) & (b) \\[6pt]
		\includegraphics[width=0.5 \textwidth]{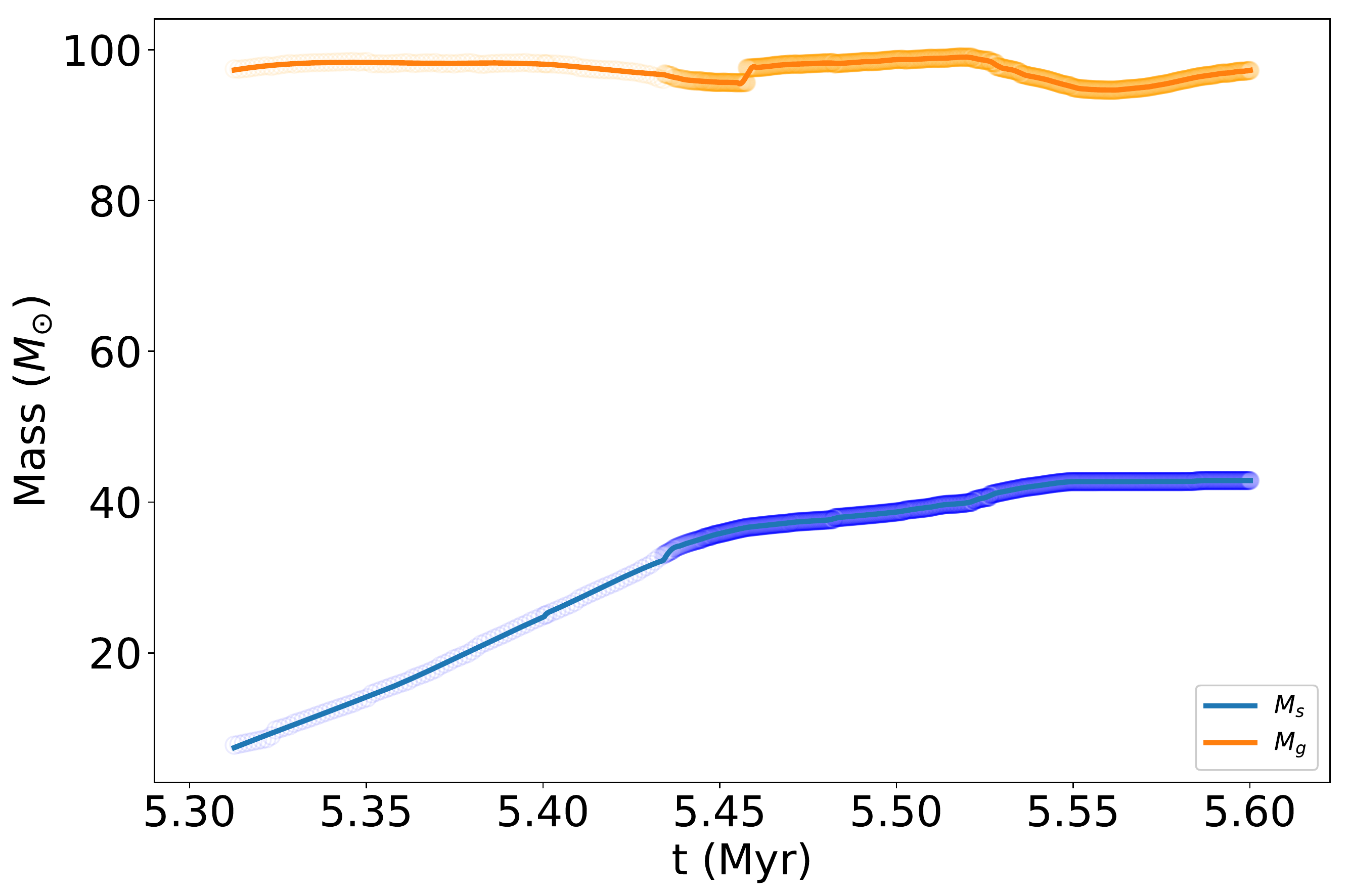} &
		\includegraphics[width=0.5 \textwidth]{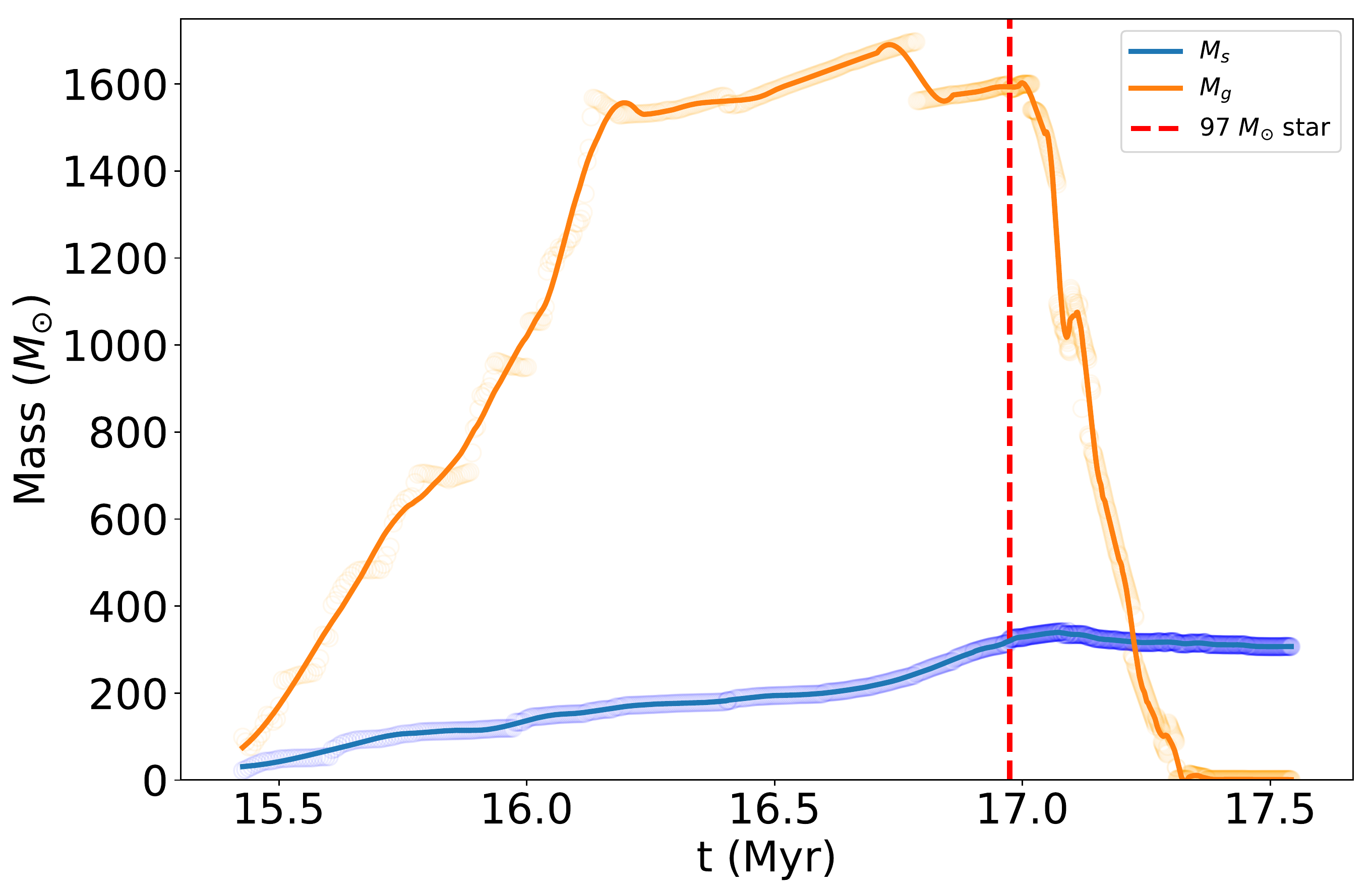} \\
		(c) & (d) \\[6pt]
	\end{tabular}
	\caption{Total mass in stars and gas in the main groups for
          runs (a) M3 (b) M3f (c) M3f2 and (d) M5f. The grey
          shaded area in (a) shows time of subgroup merger, while the red
          dashed vertical line in (d) shows the formation of $A_*$, the
          \SI{97}{\msun} star.\label{fig:mass_totals}}
\end{figure*}

\begin{figure*}[h]
	\begin{tabular}{cc}
		\includegraphics[width=0.5 \textwidth]{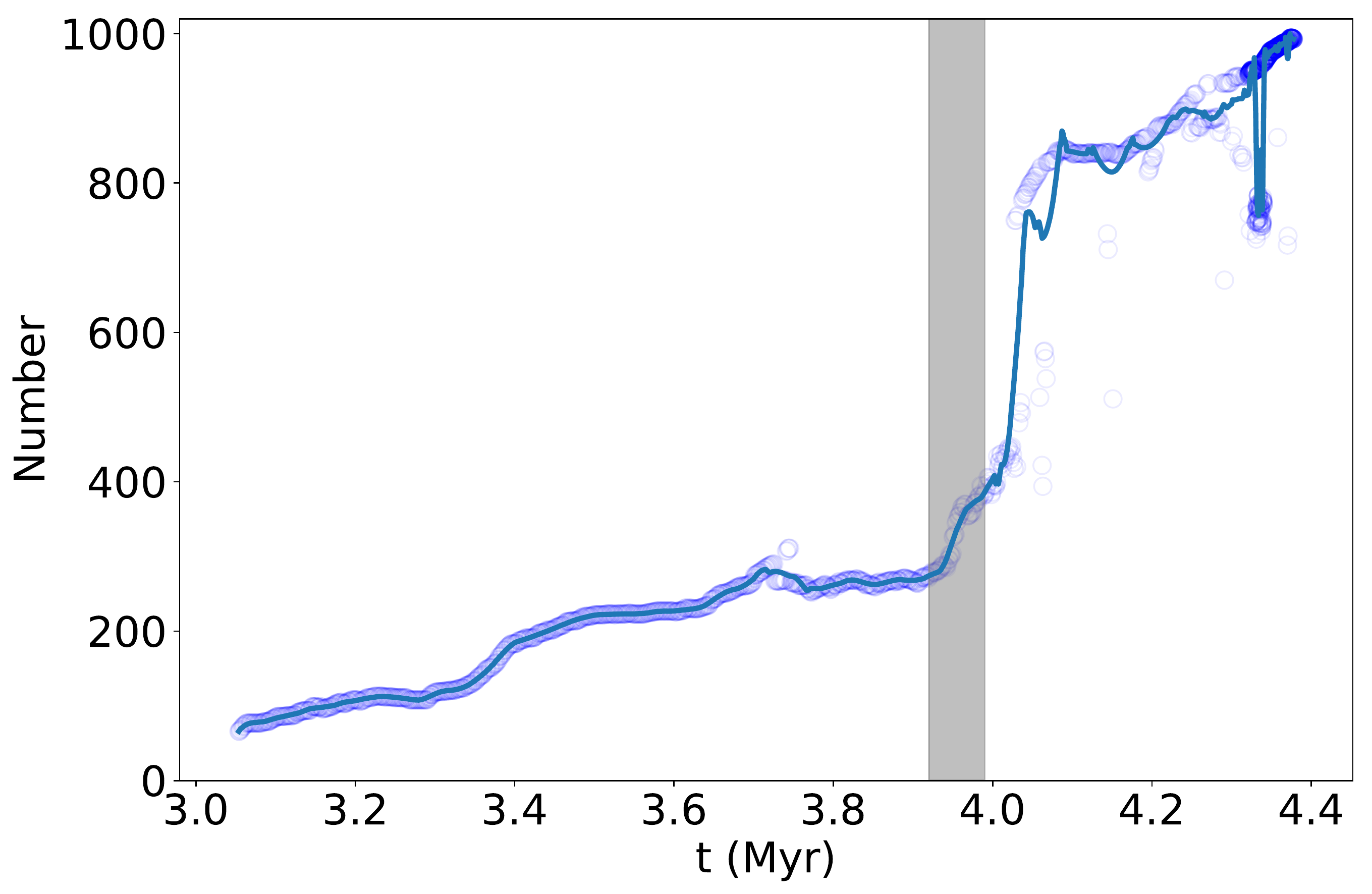} &
		\includegraphics[width=0.5 \textwidth]{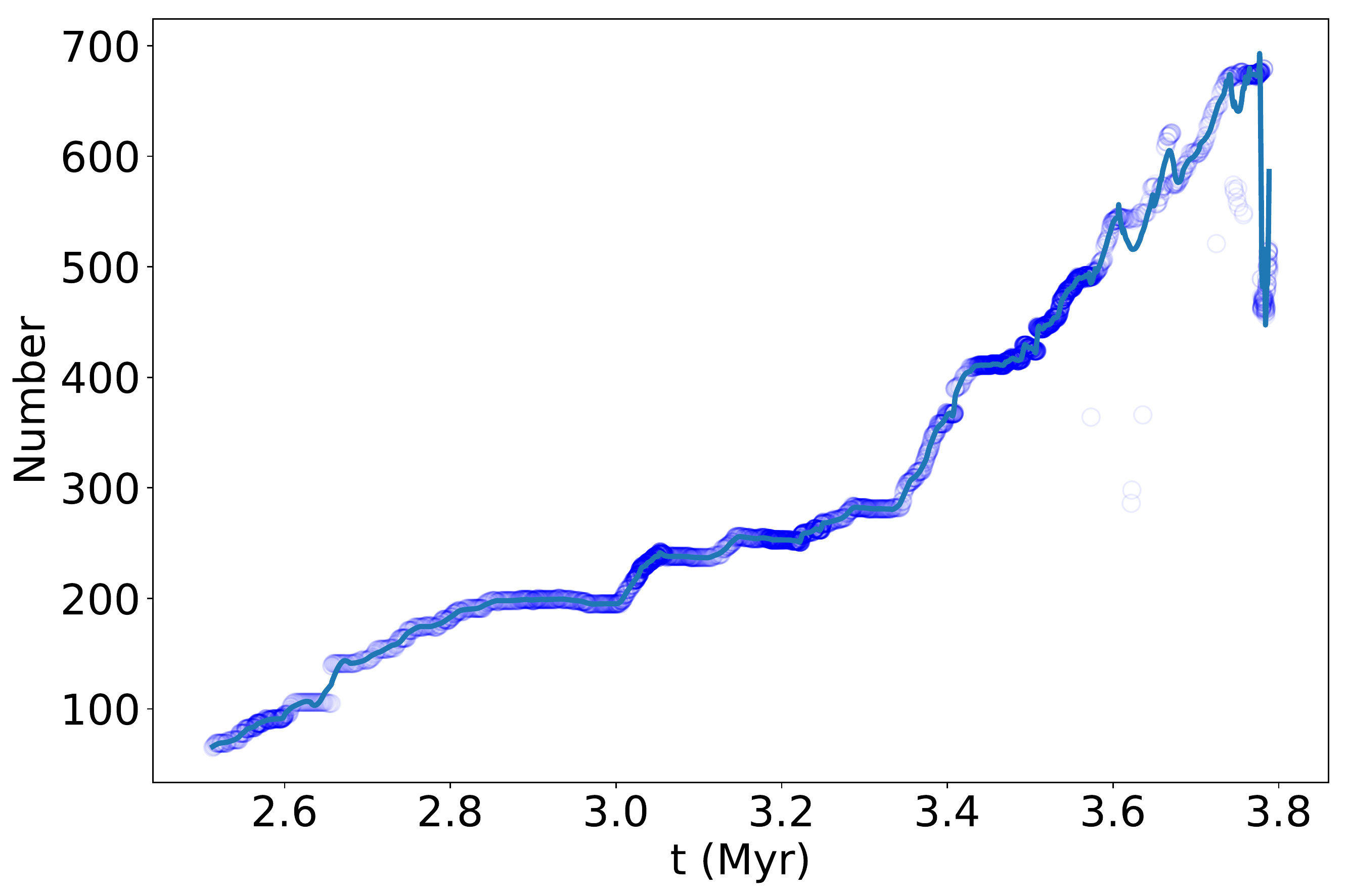} \\
		(a) & (b) \\[6pt]
		\includegraphics[width=0.5 \textwidth]{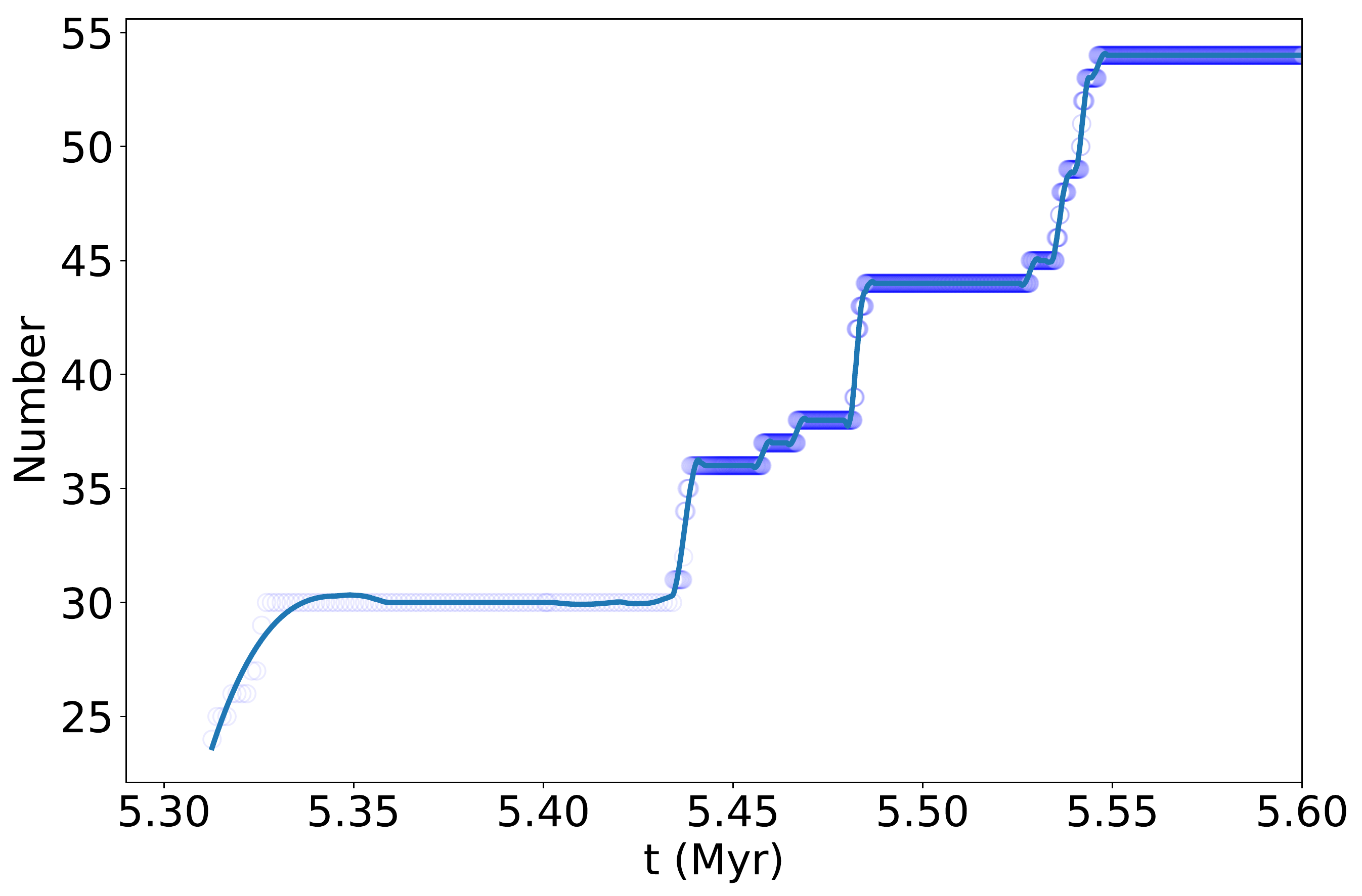} &
		\includegraphics[width=0.5 \textwidth]{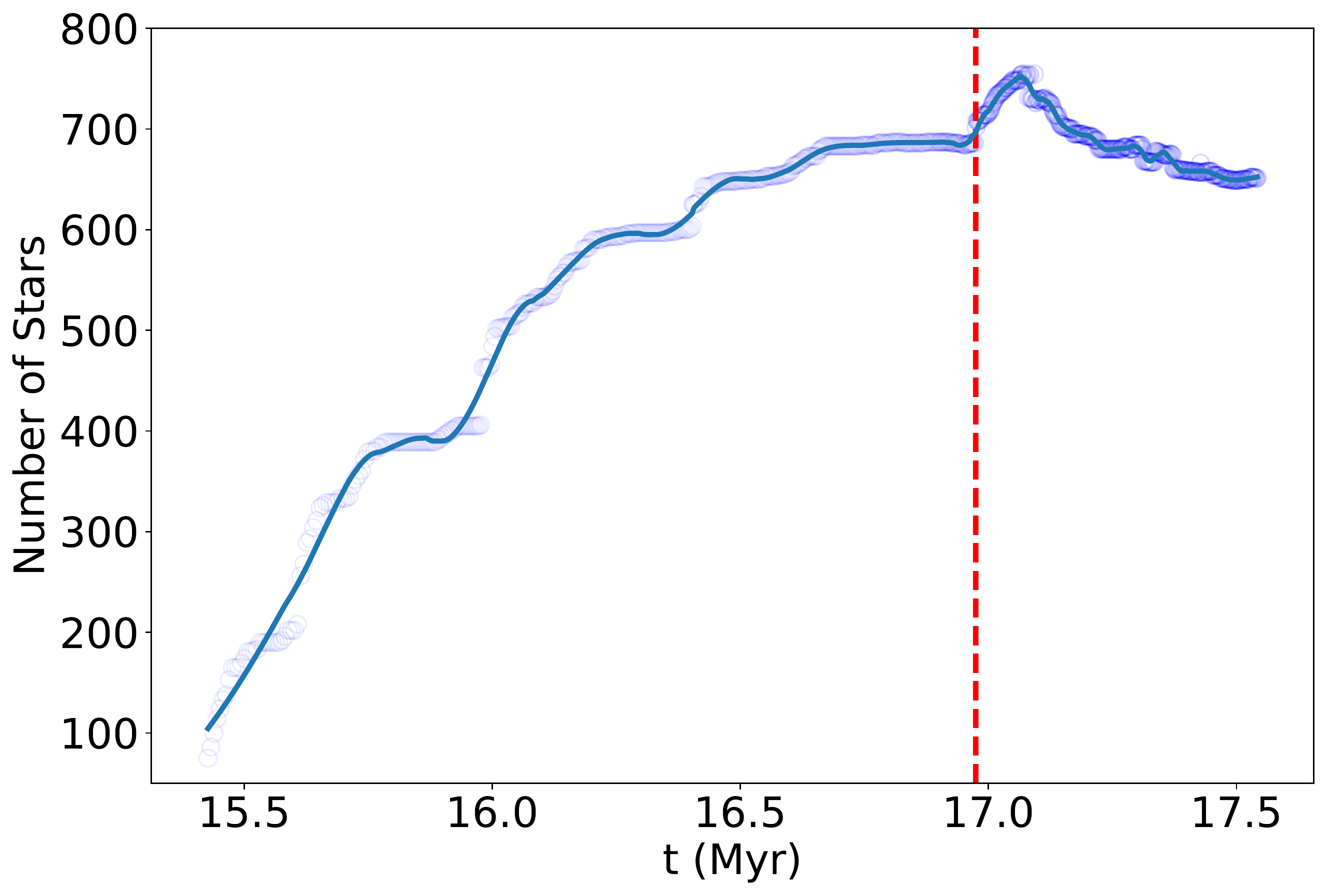} \\
		(c) & (d) \\[6pt]
	\end{tabular}
	\caption{Total number of stars in the main groups for runs (a)
          M3 (b) M3f (c) M3f2 and (d) M5f. The grey
          shaded area in (a) shows time of subgroup merger, while the red
          dashed vertical line in (d) shows the formation of $A_*$, the
          \SI{97}{\msun} star.\label{fig:num_stars}}
\end{figure*}

Figure~\ref{fig:mass_totals} shows that the mass in gas dominates the
mass in all the groups for most of their evolution, only being driven
completely out of the group under the intense feedback of M5f's
massive star. Even in M3f2, where the gas actually has positive
energy, it has yet to be driven from the group entirely. Indeed, dense
gas is growing in the group (Fig.~\ref{fig:frac_totals} c) as it
continues to fall in from the filament and build up along the edge of
the H~{\sc ii} region.
This infall itself may lead to more star
formation, although at the end of the run there has been no similar
increase in the amount of Jeans unstable gas, and the total number of
stars in the group has been steady for the last \SI{5e4}{yr}, as shown
in Figure~\ref{fig:num_stars}. This is similar to our other
\SI{e3}{\msun} simulation M3f, where feedback near the main group
during the previous \SI{e5}{yr} has also stabilized the number of
stars. In M5f, feedback eventually leads to the loss of some of the
least bound stars in the main group as the gas is ejected, shown by
the correlation in the drop in gas mass with the decline in the number
of stars in the group.

\subsubsection{Radius}

\label{subsub:radius}

\begin{figure*}[h]
	\begin{tabular}{cc}
		\includegraphics[width=0.5 \textwidth]{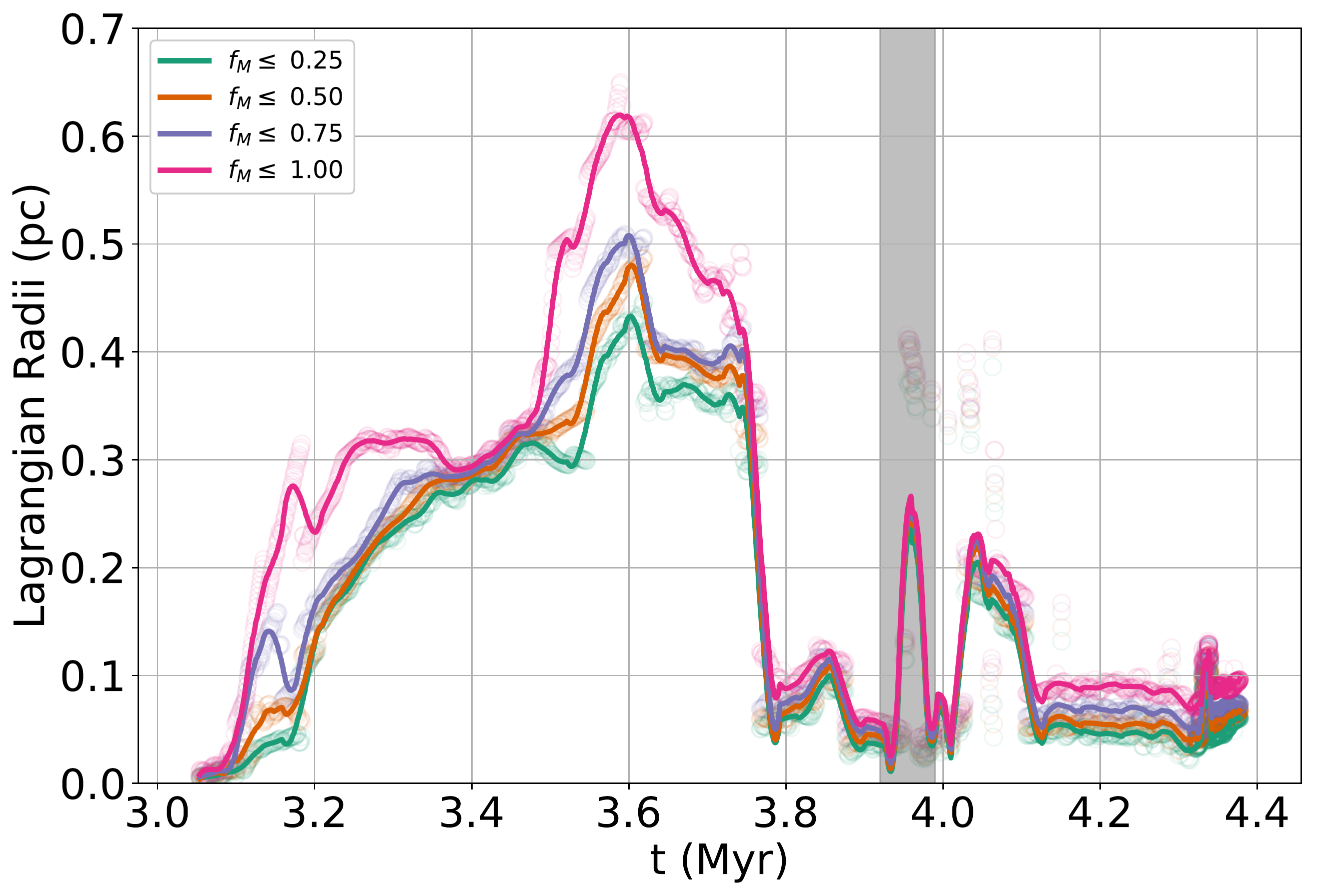} &
		\includegraphics[width=0.5 \textwidth]{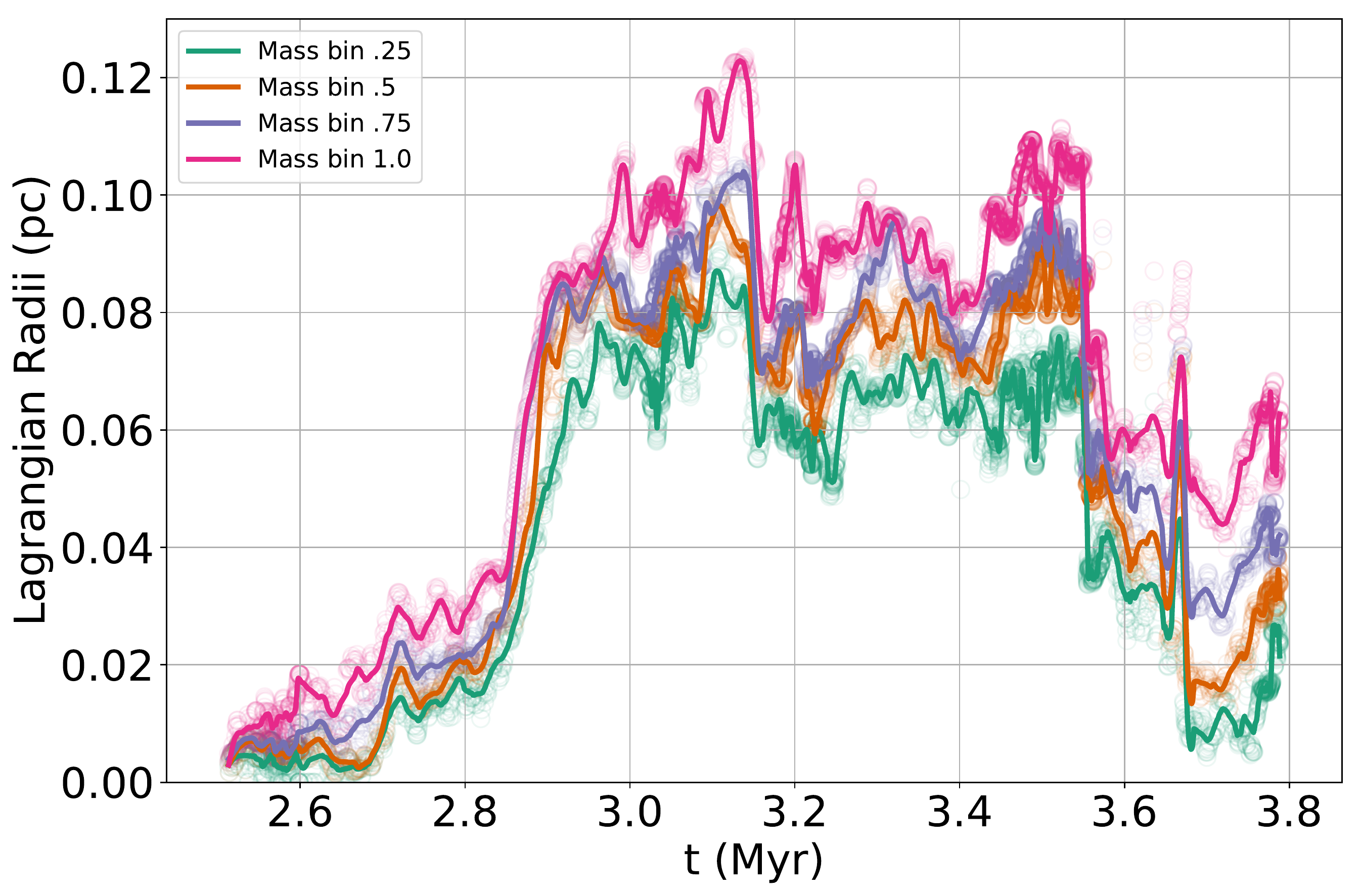} \\
		(a) & (b) \\[6pt]
		\includegraphics[width=0.5 \textwidth]{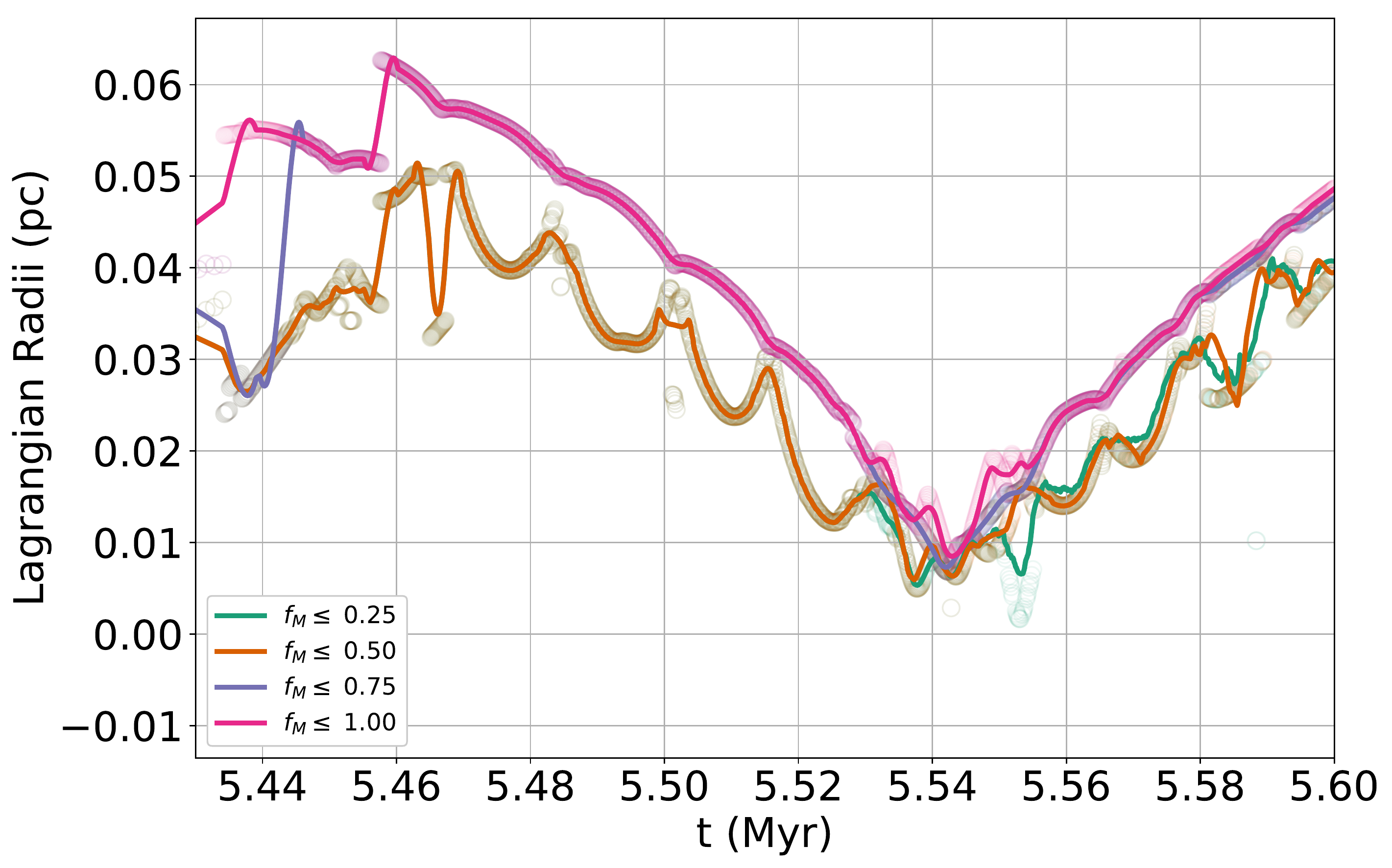} &
		\includegraphics[width=0.5 \textwidth]{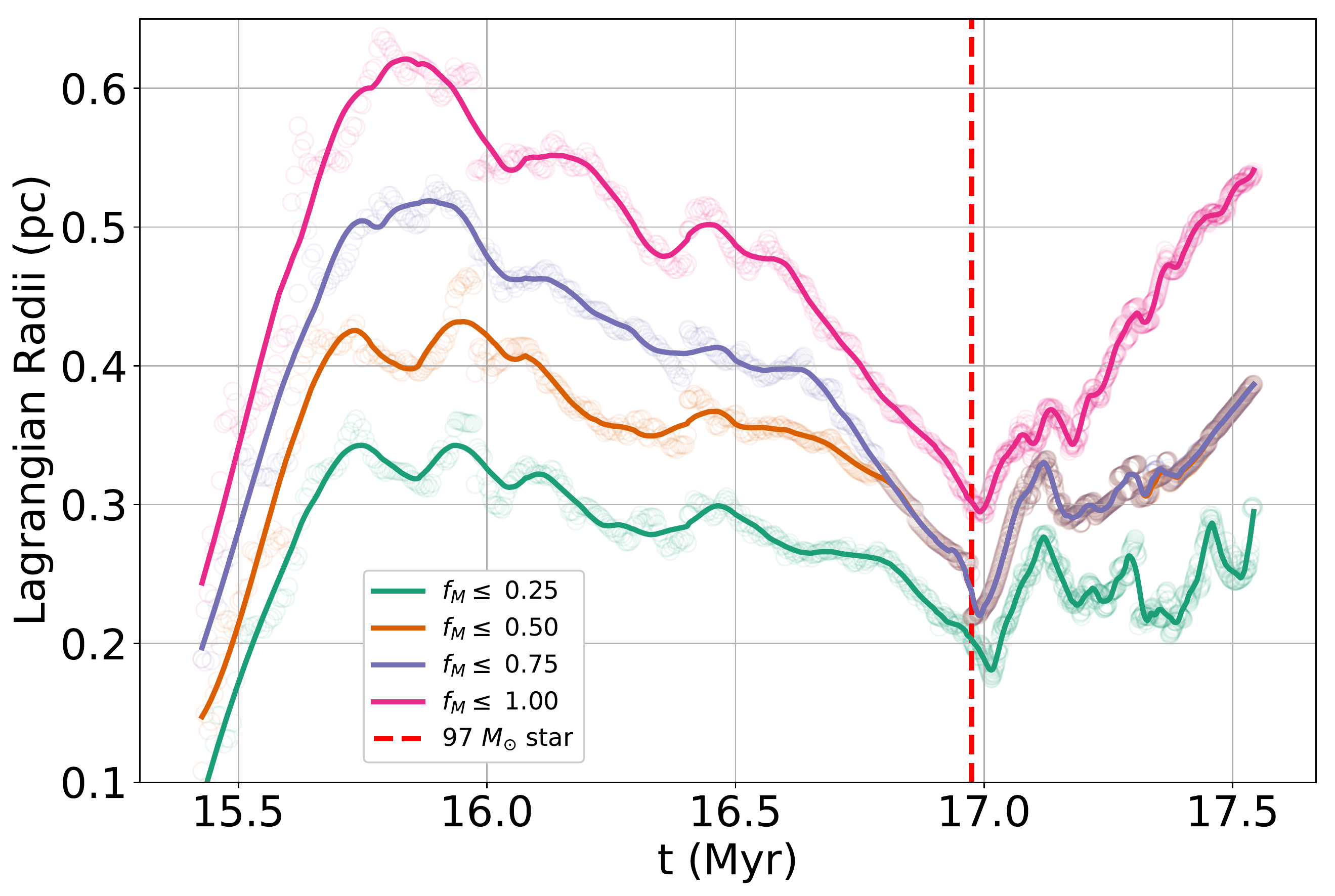} \\
		(c) & (d) \\[6pt]
	\end{tabular}
	\caption{Lagrangian radii of all the stars in the main groups
          for runs (a) M3 (b) M3f (c) M3f2 and (d) M5f. The grey
          shaded area in (a) shows time of subgroup merger, while the red
          dashed vertical line in (d) shows the formation of $A_*$, the
          \SI{97}{\msun} star.\label{fig:lag_radii}}
      \end{figure*}

In Figure~\ref{fig:lag_radii} we show the Lagrangian radii for evenly
spaced mass bins.  In M3 (Fig.~\ref{fig:lag_radii}a), lacking
feedback, the group radius drops, aside from a brief bounce when two
sub\replaced{cluster}{group}s merge (grey bar). Gas ejection leading to loss of the least bound stars can be seen in
the fast rate of growth of the Lagrangian radii of the central
groups for the two runs in which feedback expelled significant amounts
of gas: M3f2 (Fig.~\ref{fig:lag_radii}c), and M5
(Fig.~\ref{fig:lag_radii}d).  In M5f, the 25\%
Lagrangian radius of the group only grows by $\sim$33--50\%, but the
the outer (100\%) and half-mass (50\%) radii almost double after the onset of stellar
feedback. 




\begin{figure*}[h]
	\begin{tabular}{cc}
		\includegraphics[width=0.5 \textwidth]{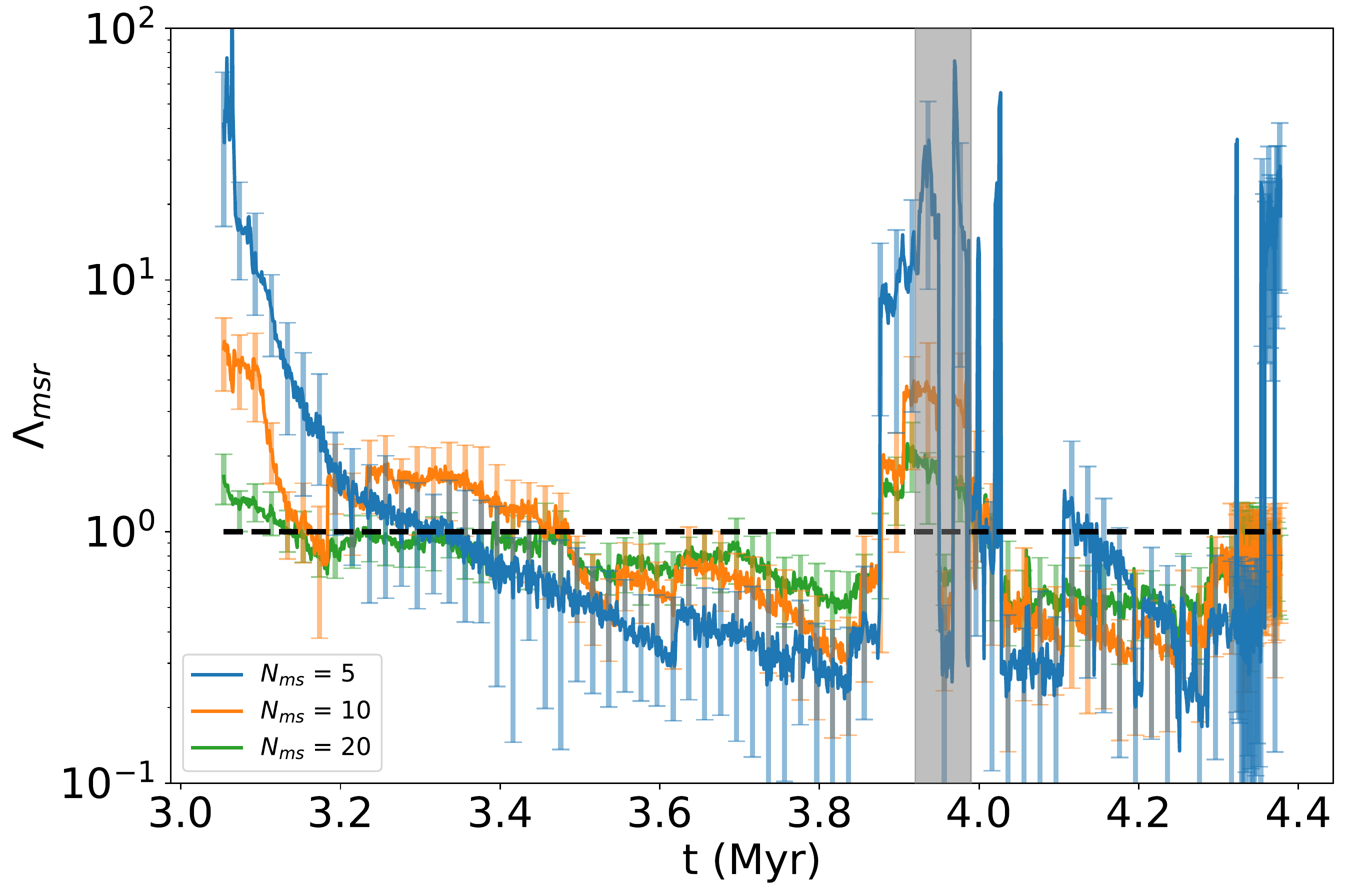} &
		\includegraphics[width=0.5 \textwidth]{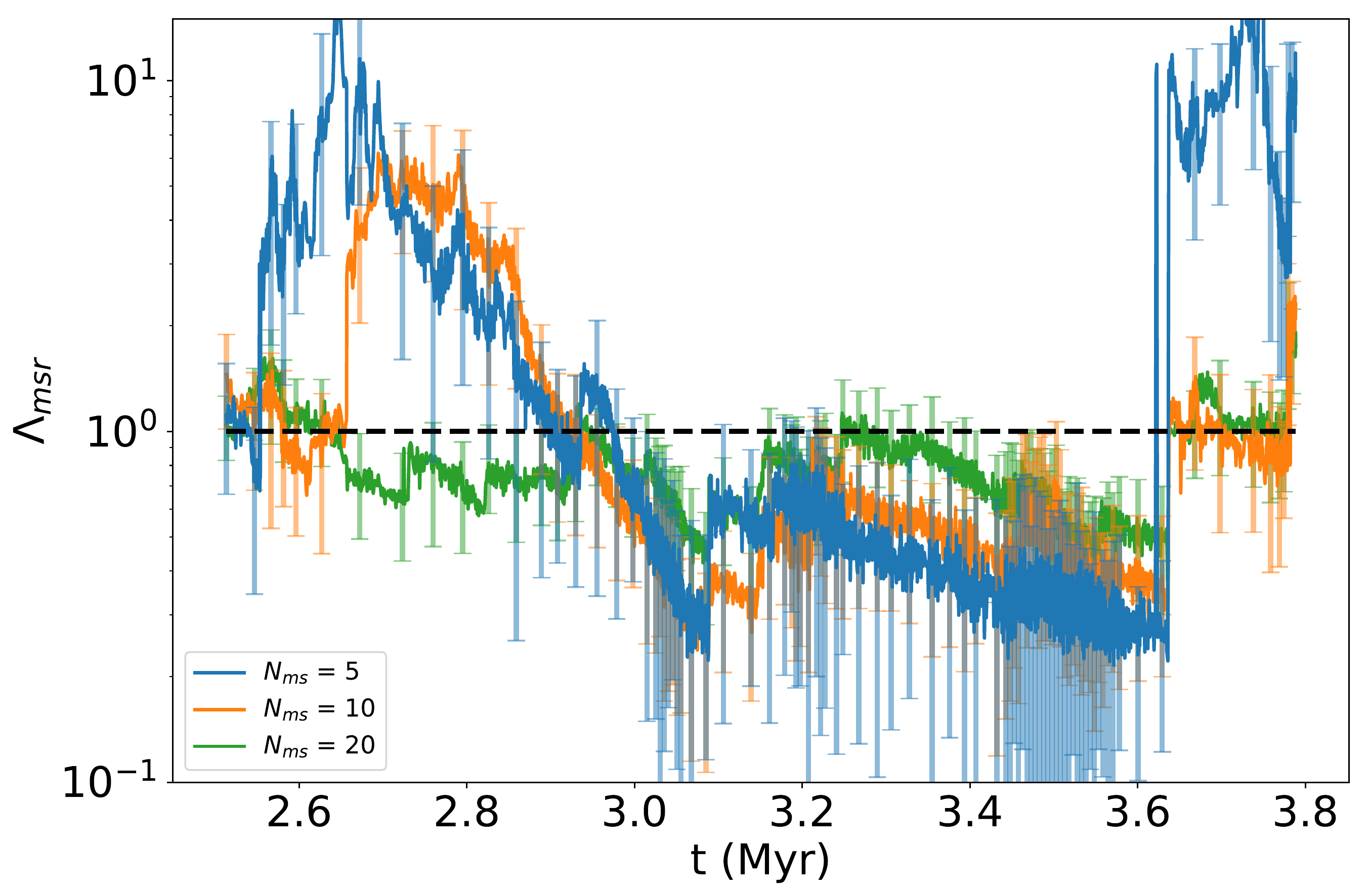} \\
		(a) & (b) \\[6pt]
		\includegraphics[width=0.5 \textwidth]{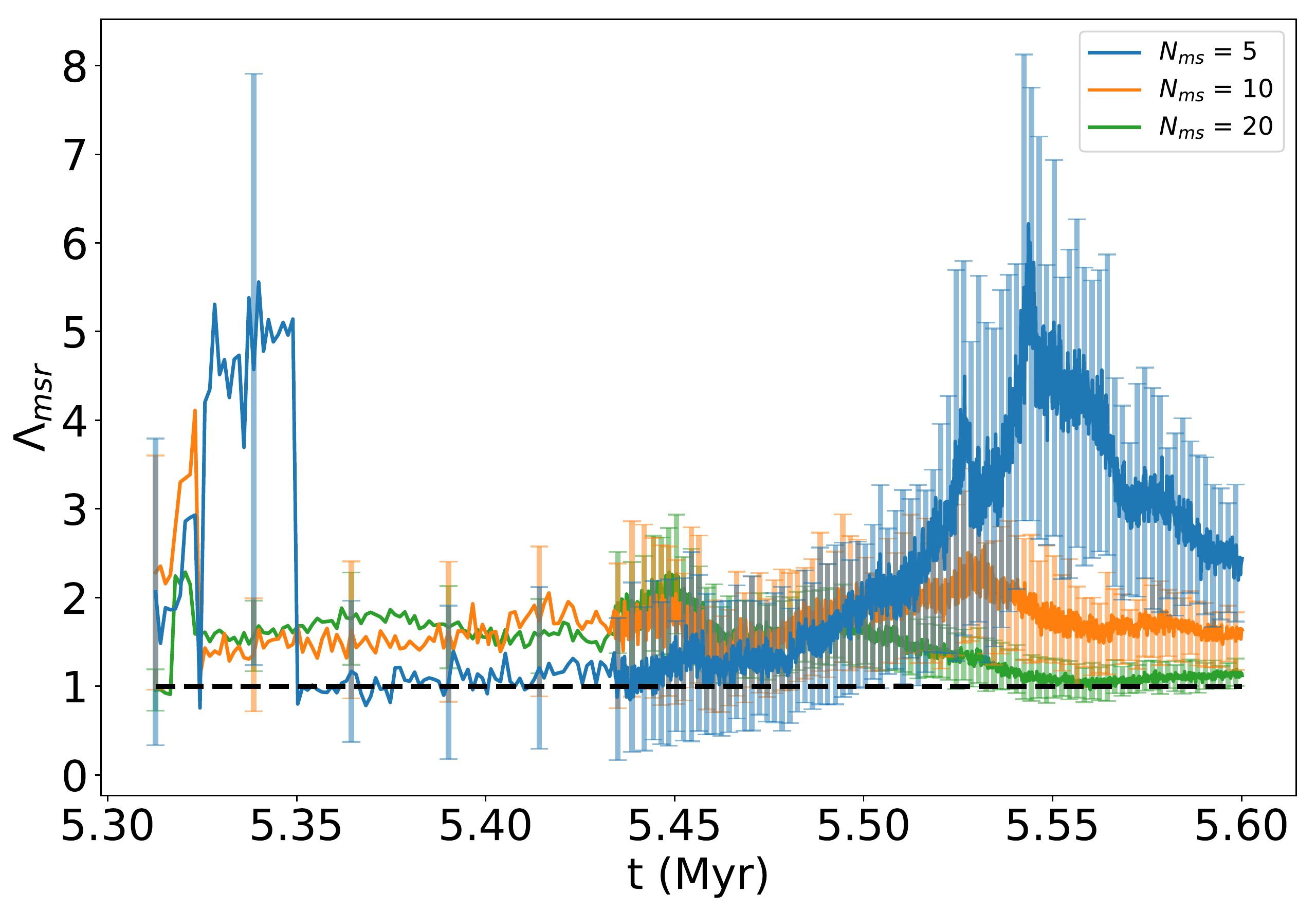} &
		\includegraphics[width=0.5 \textwidth]{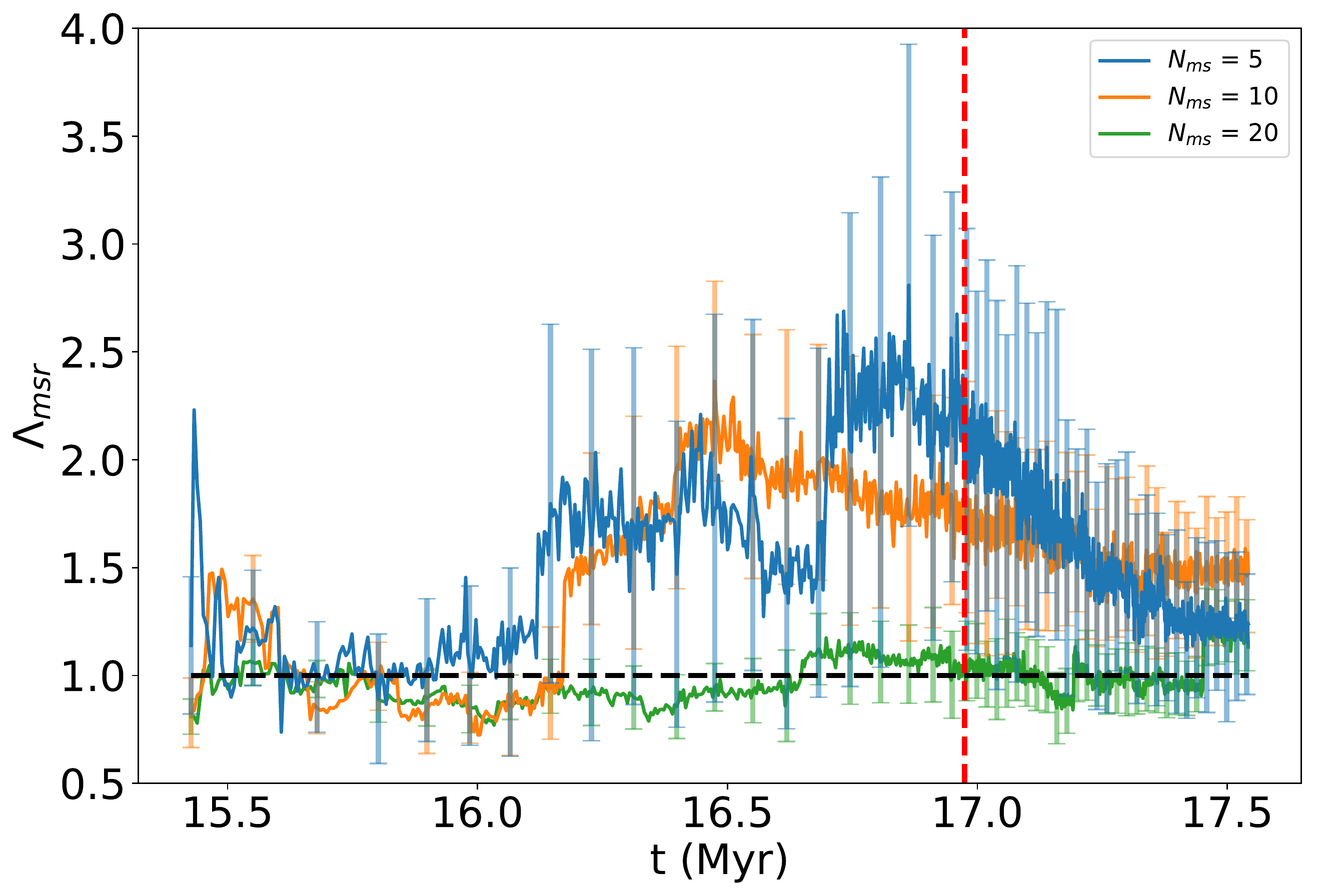} \\
		(c) & (d) \\[6pt]
	\end{tabular}
	\caption{Mass segregation ratio $\Lambda_{\rm msr}$
          (Eq.~\ref{eq:Lambda_msr}) for all 
          the stars in the main groups for runs (a) M3 (b) M3f (c)
          M3f2 and (d) M5f. Mass segregation produces $\Lambda_{\rm
            msr} > 1$. The grey band shows the merger of
          the two sub\replaced{cluster}{group}s in M3, while the red, dashed line shows
          the formation of $A_*$. \label{fig:msr}}
\end{figure*}

\subsection{Mass Segregation}

\label{subsec:ms}
Next we consider the mass segregation of the groups formed in our
simulations. Several methods exist to quantify mass segregation,
including looking at the half-mass radii $R_{\rm hm}$ of different
mass bins in the group
\citep{Mcmillan_dynamical_mass_seg_2007,McMillan_Rapid_Mass_Seg_2015},
the mean or median radius of a subset of massive stars
\citep{Bonnell_Davies_mass_seg_1998}, and calculating the Gini
coefficient of the group
\citep{Converse_and_Stahler_mass_seg_pleadies, Pelupessy_embedded_SC}.

\citet{Allison_MST_mass_seg_2009} pointed out several issues with
these methods of computing mass segregation, including how binning can
affect the results, the reliance on properly finding the group center, and difficulty in comparison to observations. They presented a new method, based on calculations of $N_{\rm random}$ minimum spanning trees (MSTs) of the group and the MST of the $N_{\rm ms}$ most massive stars contained in the same group as a model independent way of determining the amount of mass segregation. They compared the ratios of the norm of lengths of the random trees, $\langle l_{\rm norm} \rangle$, to the length of the massive star tree, $l_{\rm ms}$, to obtain the mass segregation ratio 
\begin{equation} \label{eq:Lambda_msr}
\Lambda_{\rm msr} = \frac{\langle l_{\rm norm} \rangle}{l_{\rm ms}} \pm \frac{\sigma_{\rm norm}}{l_{\rm ms}},
\end{equation}
where $\Lambda_{\rm msr}>1$ indicates mass segregation in the group. 

This method is independent of any determination of the group center,
always returns the same tree lengths (even if the tree is drawn in a
different order), and is simple to implement in both two and three
dimensions. Further, since the number of random trees calculated
provides a standard deviation of tree length, error for the
calculations are straight forward to obtain. We show our calculated
values of the three-dimensional value of $\Lambda_{\rm msr}$ in
Figure~\ref{fig:msr} using $N_{\rm ms} = \lbrace 5,10,20 \rbrace$ and
50 random samples drawn from each group for the comparison trees. We
only start tracking groups once they reach 64 stars in size, with the
exception of M3f2 where we start following the group at 24 stars.
\added{We note that even in M5f, where the most massive star is an
  outlier, the fifth, tenth, and twentieth most massive stars have
  masses that are still well above the average value, with masses of
  4.9 $M_{\odot}$, 3.4 $M_{\odot}$, and  2.4 $M_{\odot}$.} 

Two points can be made with this data. First, all of our runs become
mass segregated at early times. This presumably occurs because our
groups of $N$ stars, with initially short crossing times $t_{\rm
  cr}=R_{\rm hm}/\sigma_v$, have likewise short half-mass relaxation
times \citep{binney2011galactic} $t_r=0.1 N t_{\rm cr}/\ln N$.  The
wide range in stellar masses then accelerates the dynamical evolution
of the \replaced{cluster}{group} to a fraction of the half-mass relaxation time scale
$t_{\rm seg} \sim \left(\langle m \rangle / \langle m_{\rm hm} \rangle
\right)t_r$, where $\langle m \rangle$ and $\langle m_{\rm hm}
\rangle$ are the mean mass of all and of the high mass stars
respectively \citep{2002ApJ...576..899P}.  The clumpiness of the
stellar distribution helps to preserve this primordial mass segregation
throughout the assembly of more massive stellar conglomerates, as was
predicted by \citet{Mcmillan_dynamical_mass_seg_2007} from simulations
of small merging subgroups.

Second, feedback seems to be correlated with mass segregation in all
of the runs including it. We attribute this to gas expulsion having a
stronger effect on low mass, loosely bound stars, causing their
orbital radii and kinetic energy to increase more than massive stars
and leading naturally to an increase in mass segregation even as the
whole group expands. Also, all runs that experience significant mass
segregation sustain that segregation over their ten most massive stars
(Fig.~\ref{fig:msr}c and d), even if the five most massive stars have
strong interactions that reduce their ratio $\Lambda_{\rm msr}$. This
supports the view \citep[e.g.][]{Girichidis_ICs_and_statistics_2_frag_ind_starve,Girichidis_ICs_and_statistics_3_proto_cluster_struct} that subgroups will start more mass segregated than dynamics alone can account for if they can survive the ejection of their natal gas, as seen in many observations of young stellar groups \citep{Hillenbrand_Hartmann_mass_seg_ONC_1998,de_Grijs_mass_seg_LMC_2002,Gouliermis_mass_seg_SMC_LMC_2004,Converse_and_Stahler_mass_seg_pleadies}.

%

\subsection[Triggered Star Formation]{Triggered Star Formation}\label{m5f_sf}

A modest level of triggered star formation has been seen both
observationally
\citep{Thompson_triggered_sf_statistics,ALMA_triggered_seq_sf_2017}
and numerically
\added{\citep{gonzalez-samaniego2020}}
\citep{Dale_triggered_sf_2012,gonzalez-samaniego2020},
although some care 
must be exercised in observations since the time evolution of the
system is not available as it is in simulations, making it difficult
to disentangle triggered star formation from formation that would have
otherwise occurred naturally due to gravitational collapse
\citep{Dale_trigger_happy_2015}. \citet{Dale_triggering_2007b} and
\citet{Dale_trigger_happy_2015} divide triggering into two categories;
{\it weak triggering} where star formation that would already occur
due to normal collapse is accelerated, but without increasing either
the overall star formation efficiency or the number of stars created;
and {\it strong triggering} where collapse is induced in previously
stable gas that increases the total star formation efficiency, the
number of stars, or both.

\begin{figure}
	\includegraphics[width=0.75 \linewidth]{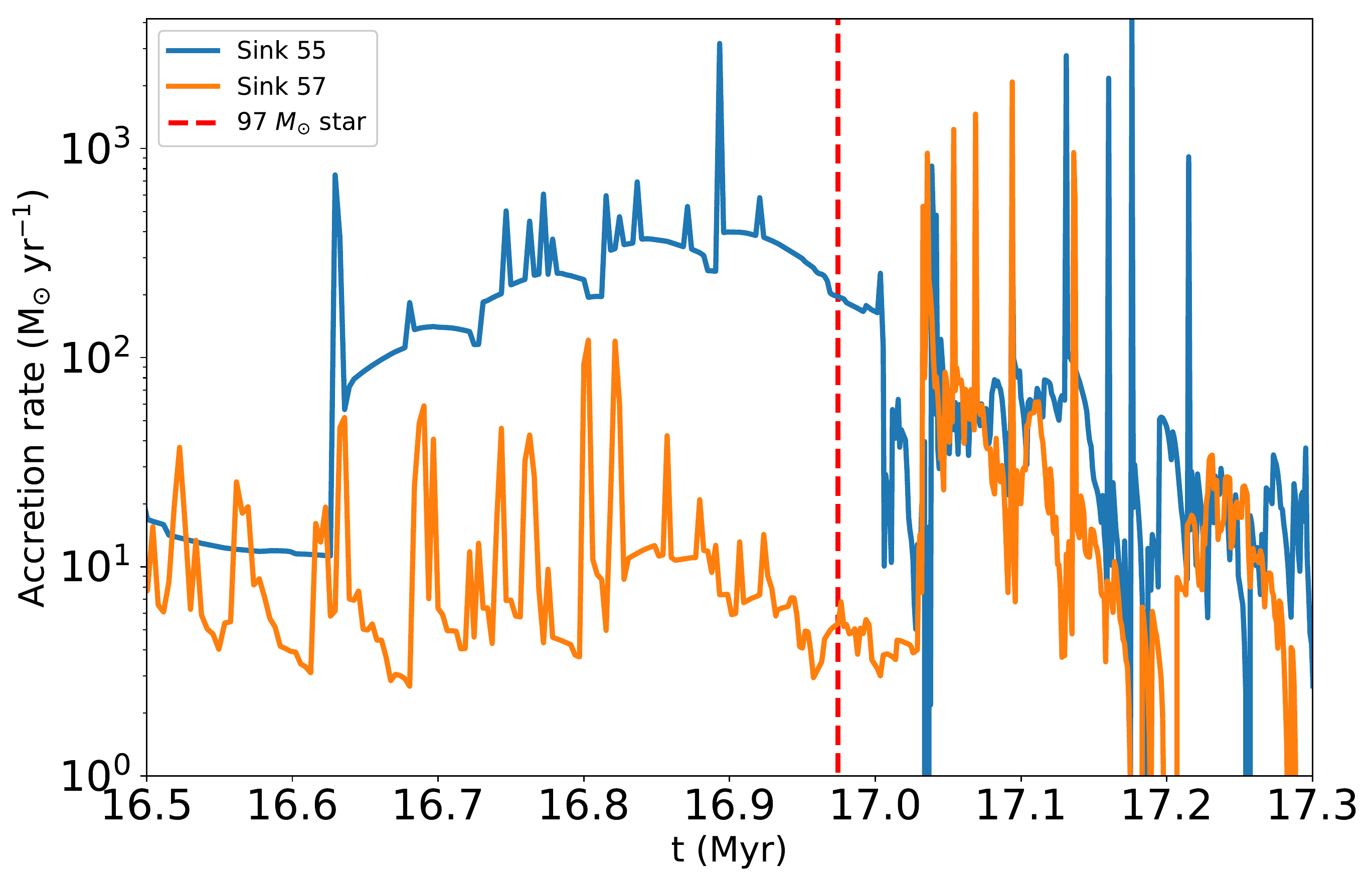}
	\caption{The individual mass accretion rates for the two sinks
          ejected from the main group due to feedback.  The grey
          shaded area in (a) shows time of subgroup merger, while the red
          dashed vertical line in (d) shows the formation of $A_*$, the
          \SI{97}{\msun} star. \label{fig:sink_accr_rate}}
\end{figure}

Apparent triggered star formation occurs in
run M5f, which has the strongest
feedback. At the time of formation of the
\SI{97}{\msun} star (hereafter referred to as $A_*$) in the main group, there are three
star-forming sinks present.
The first sink (sink \# 56)
is the one that actually forms $A_*$, and its accretion immediately
shuts down.  The other two sinks
(\# 55 and 57) continue accreting gas
for another 0.3~Myr 
from gravitationally unstable regions in the swept up shell driven by
the stellar wind from $A_*$ (Fig.~\ref{fig:sink_accr_rate}).  This appears to be a case of triggering
maintaining the global star formation rate temporarily  (see Fig.~\ref{fig:sfr} d) even in the presence of
strong negative feedback.




In the case of sink 55, star formation was already proceeding at a
vigorous rate before $A_*$ appeared. Therefore this cannot be
considered even weakly triggered star formation, however it seems that
the intense feedback was unable to do more than briefly slow the
production of stars for about $10^4$~yr before the sink returned to
producing stars at a rate exceeding \SI{4e-5}{\msun yr^{-1}}. In the
case of sink 57, though, star formation was at less than \SI{1e-5}{\msun\
  yr^{-1}} when the stellar wind bubble compressed gas in which the sink was embedded. Once this occurred, the rate of star formation grew rapidly. For this to be considered triggered star formation, we should also be able to observe an increase in both the amount of dense gas {\it and} Jeans unstable gas in the region surrounding the group occurring as the feedback impacts the region. We show that this increase indeed occurs in Figure~\ref{fig:frac_totals} (d).

During their ejection, both sinks reached peak speeds of $\sim
\SI{30}{km~s}^{-1}$ with respect to the group center as they
followed the expanding gas driven by feedback.
This velocity is noteworthy, since it defines the boundary velocity
for massive O and B stars that are considered runaway stars
\citep{Gies_OB_runaways_1986}. Generally the production of OB runaways
has been considered the result of kicks from a binary partner that
goes supernova
\citep{Blaauw_SN_kicks_for_OB_runaway,2000ApJ...544..437P} or due to
dynamical interactions with binaries
\citep{Leonard_dyn_origin_OB_runaway_1988,Fujii_SPZ_OB_runaways_2011}. However
since our sinks (and therefore also the gas they accrete) reach
velocities comparable to that of OB runaways, triggered star formation
in our simulation shows a third, and not previously considered, method
for producing OB runaways. In this case we produced many lower mass
stars, but the total mass produced by the two sinks during this time
was over \SI{30}{\msun}, therefore the lack of formation of an OB star
was simply due to the random selection of our star formation method.

The high gas velocity is clearly connected to the feedback of $A_*$,
but which physical process contributes the most? The radius and velocity of the D-type front are \citet{Spitzer_ISM}
\begin{eqnarray}
R &=& R_{\rm St} \left(1 + \frac{7}{4} \frac{c_s t}{R_{\rm St}}\right)^{-3/4}, \\
\frac{dR}{dt} &=& c_s \left(\frac{R}{R_{\rm St}}\right)^{-3/4},
\end{eqnarray}
The velocity of the D front has a maximum value of $v_{\rm d} \sim
\SI{15}{km~s}^{-1}$, too slow for our gas, ruling out compression by
radiation. Note this also likely rules out radiation driven implosion
\citep{Sanford_Whitaker_Klein_RDI_1982} as a primary trigger. We also considered a champagne flow as a possible method of driving the gas velocities, but as shown in \citet{Bodenheimer_TT_Yorke_champagne_1979} and similar to the case of radiation driven implosion, champagne flows only accelerate the lower density gas to high velocities. Even in their case (5), where they allowed a D-type front to move past a small dense cloud, the dense gas was compressed but the dense gas velocities never exceeded \SI{8}{km~s}$^{-1}$.

This leaves the effect of the winds as the main factor accelerating
the gas flow. Normally, the wind bubble would evolve while trapped
within the H~{\sc ii} region \citep{Weaver_1977}. This means for
moderately massive O stars the H~{\sc ii} region dominates the
dynamics, since the D-type front strikes the ambient gas first
\citep{McKee_Van_Buren_Lazareff_winds_in_cloudy_1984}. However in the
case of winds moving rapidly into a region of dense gas the H~{\sc ii}
regions can become trapped within the wind shells
\citep{Van_Buren_Mac_Low_UCHII_bow_shocks_1990,Mac_Low_Van_Buren_UCHII_bow_shocks_1991}. To
calculate the speed of the shell from the wind, we obtained the
luminosity and temperature of $A_*$ using \seba and then calculated
the wind luminosity using our stellar wind code as described in
Sect.~\ref{winds}, finding $L_w = \SI{2.47e37}{erg}$.
The shell velocity of a stellar wind bubble in a uniform medium is \citep{Weaver_1977}
\begin{align}
V_2(t) = 16n_o^{-1/5} L_{36}^{1/5}t_6^{-2/5}~\si{km~s}^{-1},
\end{align}
with $n_o = \SI{e3}{cm^{-3}}$ 
to find the wind shell velocity at the peak time of the sink
velocities, which is $\sim \SI{3e4}{yr}$ after the formation of
$A_*$. This gives us a shell velocity of $V_2 = \SI{31}{km~s}^{-1}$,
consistent with the maximum sink velocity of \SI{30}{km~s}$^{-1}$. 

\begin{figure}
	\includegraphics[width=0.75 \linewidth]{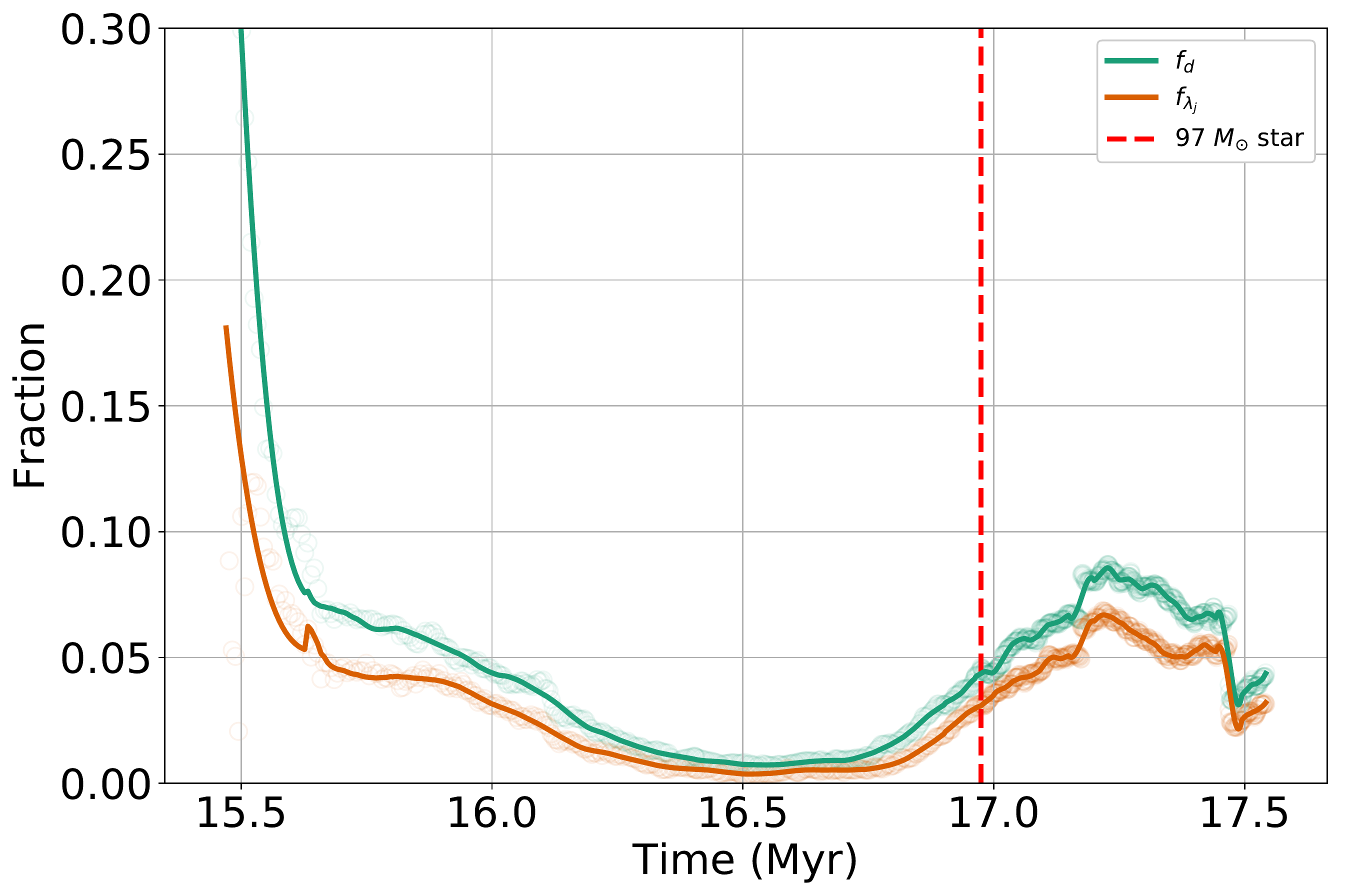}
	\caption{Fractions of dense ($n_{\rm H} > \SI{e4}{cm^{-3}}$)
          and Jeans unstable gas in the region defined by group 2. The grey
          shaded area in (a) shows time of subgroup merger, while the red
          dashed vertical line in (d) shows the formation of $A_*$, the
          \SI{97}{\msun} star. \label{fig:frac_clust2}}
\end{figure}

Eventually all the star forming filaments that are in close proximity
to the main group are disrupted by the feedback of $A_*$. At this
point ($\sim \SI{17.3}{Myr}$) the overall star formation rate rapidly
drops, as all gas in the region becomes warm, low density H~{\sc ii}
gas or hot wind shocked gas. Some small amount of star formation still
occurs in a smaller secondary group containing sink \# 22, but formation here is slow due to the overall smaller fraction of dense gas present in the group, as shown in Figure~\ref{fig:frac_clust2}.

\section{Summary}\label{fb_conclusions}

\replaced{In this paper we have described the implementation of stellar feedback
methods in \flash, whose inclusion within the \amuse software
framework was described in \citetalias{paper1}. We have shown that our}{This paper describes
the implementation of stellar feedback methods in the Torch software
package, which incorporates \flash into the \amuse software
framework \citepalias{paper1}. Our}
implementations reproduce standard benchmarks for ionizing radiation,
stellar winds, and supernovae. We also include heating from cosmic rays
and non-ionizing radiation from both individual stars and the galactic
background and radiative cooling from both gas in collisional
equilibrium ionization and dust.
\replaced{These implementations are part of a
larger effort to model the formation and early evolution of star
clusters, where we combine}{The implementation of feedback in the
Torch package allows its use to model the formation and early evolution of star
clusters by combining}
magnetohydrodynamics using \flash, collisional N-body
dynamics using \phfour, binary and
higher-order multiple dynamics using \multiples, and stellar evolution using \seba, with the
stellar feedback described here.

We have begun to use this
framework to study \added{the structure and dynamics of} newly formed
stellar groups and clusters.
We here report on four proof-of-concept simulations with initial gas masses
of either $10^3\, \msun$ or $10^5\, \msun$. \added{Because the focus
  of this paper is on the feedback implementations, we use a single
  approximation for the choice of the mass, position, and velocity of
  stars formed from sink particles. Future work will study whether
  our results are sensitive to variations in how this choice is made.}

\replaced{From these models we conclude the following:}{Our four
  models lead to the following tentative conclusions:} 
\begin{enumerate}
\item Stellar feedback \replaced{effectively controls the SFR}{can effectively
    terminate star formation in the region around a stellar group}
  (Sect.~\ref{subsec:SF}). The details of cloud structure do matter,
  however, both for the overall star formation rate and because dense
  shells swept up by feedback can trigger small amounts of additional
  star formation.
\item Stellar feedback tends to increase the amount of dense gas
  present in the star forming region, agreeing with
  \citet{Dale_early_evo_clusters}. Contrary to them, however, we do find a case in
  which feedback even increases the amount of Jeans unstable gas.
\item Our stellar groups generally form subvirial and end marginally
  virialized (Sect.~\ref{subsub:energy}). Both groups that ejected their gas (M3f and M5f) went
  through a period of supervirial expansion but ended subvirial, even
  while they continue to expand.
\item Feedback results in the ejection of gas from our \replaced{cluster}{group}s, but
  did not disrupt any of the stellar groups created in the runs presented
  here, although the least bound stars were lost
  (Sects.~\ref{subsub:mass} and \ref{subsub:radius}).
\item Our stellar groups quickly become mass segregated
  (Sect.~\ref{subsec:ms}).  Feedback-driven gas removal further
  stratifies the stars according to their current binding
  energy. After expulsion of their gas, the groups remain mass
  segregated, consistent with observations.
\end{enumerate}

\acknowledgments We acknowledge A. Tran for
documentation and development of the open source Bitbucket repository,
A. van Elteren, I. Pelupessy and S. Rieder for assistance with AMUSE,
R. Banerjee and D. Seifried for providing the base code for dust and
molecular cooling, \added{W. Farner for measuring total stellar
  masses,} R. W\"unsch for providing a helper script for 
the initial conditions, and M. Davis, C. Federrath, S. Glover,
A. Hill, J. Moreno, and E. Pellegrini for useful discussions.  This work was supported by NASA grant
NNX14AP27G, the Netherlands Research School for Astronomy
(NOVA), NWO (grant \# 621.016.701 [LGM-II]), NSF grants AST11-0395
and AST18-15461, an award to M-MML by the Alexander-von-Humboldt
Stiftung, the Deutsche Forschungsgemeinschaft (DFG) via SFB 881 ``The Milky Way
System'' (sub-projects B1, B2 and B8), and SPP 1573 ``Physics of the
ISM'', and the European Community’s Seventh Framework Programme via
the ERC Advanced Grant ``STARLIGHT'' (project number 339177). The
Dutch National Supercomputing Center SURFSara grant 15520 
provided computing resources for our simulations.  
\software{AMUSE \citep{AMUSE,AMUSE13}, Flash \citep{FLASH}, ph4 \citep{ph4},
  SeBa \citep{Portegies_SeBa}, yt \citep{yt}, numpy \citep{numpy},
  matplotlib \citep{hunter2007}, HDF \citep{HDF}}



\bibliographystyle{aasjournal}

\bibliography{./Wall.bib}

\end{document}